\documentclass[fleqn,usenatbib]{mnras}

\usepackage{newtxtext,newtxmath}
\usepackage[T1]{fontenc}

\DeclareRobustCommand{\VAN}[3]{#2}
\let\VANthebibliography\thebibliography
\def\thebibliography{\DeclareRobustCommand{\VAN}[3]{##3}\VANthebibliography}

\usepackage{graphicx}	% Including figure files
\usepackage{amsmath}	% Advanced maths commands
\usepackage{multirow}

\title[Dynamical evolution of IMB]{Dynamical evolution of the inner asteroid belt}

\author[S. F. Dermott et al.]{
Stanley F. Dermott,$^{1}$\thanks{E-mail: \href{mailto:sdermott@ufl.edu}sdermott@ufl.edu}
Dan Li,$^{2}$
Apostolos A. Christou,$^{3}$
Thomas J. J. Kehoe,$^{4}$
\newauthor
Carl D. Murray,$^{5}$
and J. Malcolm Robinson$^{1}$
\\
\\
$^{1}$Department of Astronomy, University of Florida, Gainesville, FL 32611, USA\\
$^{2}$NSF’s National Optical-Infrared Astronomy Research Laboratory, Tucson, AZ 85719, USA\\
$^{3}$Armagh Observatory and Planetarium, College Hill, Armagh, BT61 9DG, UK\\
$^{4}$Department of Physical Sciences, Embry-Riddle Aeronautical University, 600 S. Clyde Morris Blvd, Daytona Beach, FL 32114\\
$^{5}$School of Physics and Astronomy, Queen Mary University of London, 327 Mile End Road, London E1 4NS, UK
}

\date{Accepted XXX. Received YYY; in original form ZZZ}

\pubyear{2021}

\begin{document}
\label{firstpage}
\pagerange{\pageref{firstpage}--\pageref{lastpage}}
\maketitle

\begin{abstract}
A determination of the dynamical evolution of the asteroid belt is difficult because the asteroid belt has evolved since the time of asteroid formation through mechanisms that include: (1) catastrophic collisions, (2) rotational disruption, (3) chaotic orbital evolution and (4) orbital evolution driven by Yarkovsky radiation forces. The timescales of these loss mechanisms are uncertain and there is a need for more observational constraints. In the inner main belt, the mean size of the non-family asteroids increases with increasing inclination. Here, we use that observation to show that all inner main belt asteroids originate from either the known families or from ghost families, that is, old families with dispersed orbital elements. We estimate that the average age of the asteroids in the ghost families is a factor of 1/3 less than the Yarkovsky orbital evolution timescale. However, this orbital evolution timescale is a long-term average that must allow for the collisional evolution of the asteroids and for stochastic changes in their spin directions. By applying these constraints on the orbital evolution timescales to the evolution of the size-frequency distribution of the Vesta asteroid family, we estimate that the age of this family is greater than 1.3 $Gyr$ and could be comparable with the age of the solar system. By estimating the number of ghost families, we calculate that the number of asteroids that are the root sources of the meteorites and the near-Earth asteroids that originate from the inner main belt is about 20.
\end{abstract}


%%%%%%%%%%%%%%%%%%%%%%%%%%%%%%%%%%%%%%%%%%%%%%%%%%

%%%%%%%%%%%%%%%%% BODY OF PAPER %%%%%%%%%%%%%%%%%%

\section{Introduction}
Understanding the origin and dynamical evolution of the asteroid belt may be key to understanding the formation and the dynamical evolution of the inner rocky planets and the initial solid cores of the major planets. We need to know why the total mass of the belt ($5\times10^{-4}M_{\earth}$ \citep{DeMeoCarry2013}) is so small and why the belt is now dynamically excited with orbital eccentricities and inclinations so high that the asteroids could not possibly have formed on their present orbits. There is a marked radial separation of the different asteroid types in the main belt, but this separation is far from complete, implying that there has been large-scale radial mixing since the time of asteroid formation \citep{gradieTedesco1982}. Measurements of the isotopic abundances of some unstable elements in a wide range of chondritic and iron meteorites show that the asteroids were formed in two distinct reservoirs of either carbonaceous (CC) or non-carbonaceous (NC) material that were separated in the solar nebula by location and by time of formation \citep{Kruijer.et.al2017}. Numerical investigations of the formation and associated orbital evolution of the major planets suggest that the asteroids that accreted in these two reservoirs were scattered by planetary perturbations into the present belt and thus that the larger asteroids are the remnants of the original building-blocks of both the terrestrial planets and the initial solid cores of the major planets \citep{Walsh.et.al2011}. It is this particular scenario of asteroid formation that forms the starting point of this paper. We assume that after all planetary migration and the scattering that resulted from that migration ceased, further evolution of the dynamically excited belt was driven by: (1) the collisional and (2) the rotational destruction of asteroids \citep{Dohnanyi1969,Jacobson.et.al2014}; (3) chaotic orbital evolution \citep{Wisdom1985,Farinella.et.al1994,MorbidelliNesvorny1999,MintonMalhotra2010}; and (4) Yarkovsky-driven transport of small asteroids to the escape hatches located at orbital resonances \citep{Migliorini.et.al1998,FarinellaVokrouhlicky1999,VokrouhlickyFarinella2000}.

Depletion of the asteroid belt due to collisional evolution was first discussed by \citet{Dohnanyi1969} in terms of an equilibrium cascade resulting from destructive collisions between solid asteroids with strengths independent of their size. However, since that early work, measurements of the low mean densities and, by implication, the high porosities of some asteroids suggest that most small asteroids are not coherent solid bodies, but unconsolidated rubble-piles \citep{Davis.et.al1985,holsapple2002}. These observations dictate that in discussing and quantifying the collisional lifetimes of small (diameter, $D<10$ $km$) main belt asteroids, we may need to  distinguish between the average time needed to shatter an asteroid and the average time needed to disperse the gravitationally bound fragments. At present, both of these times are uncertain \citep{holsapple2002}.

Orbital evolution due to Yarkovsky radiation forces accounts for the V-shaped distribution of family asteroids in semimajor axis and inverse diameter ($1/D$) space and these observations have been used to date some of the families (see, for example, \citealt{Spoto.et.al2015}). However, these ages are not absolute ages because, while we have observations of the rates of orbital evolution of a large number of very small near-Earth asteroids \citep{Greenberg.et.al2020}, the rates of orbital evolution of the larger main-belt asteroids have not been measured. Thus, the ages of the  families are currently uncertain, partly because the thermal properties of small, porous asteroids are uncertain, but also because other parameters, particularly the mean densities, are uncertain. We note that if most of the major families have ages less than about half the age of the solar system, as estimated by \citet{Spoto.et.al2015}, this raises the question of where are the major families that must have been formed at earlier times?  

The chief aim of this paper is to introduce new observational constraints on the long-term dynamical evolution of the asteroid belt. In particular, we analyze the observed correlations between the mean proper orbital elements and the mean diameters of the non-family asteroids, and the non-linear, log-log SFDs of the small asteroids in the major families. The number of asteroids in our data set is both large and observationally complete and we do not use subsets of the data for which uncertain observational selection corrections have to be applied. Previously, we argued that the asteroid size-orbital element correlations of the non-family asteroids in the inner main belt (IMB) are evidence for the existence of ghost families \citep{Dermott.et.al2018}). While recognizing the validity of that argument, and the likely existence of ghost families, here we show that the resonant structure of the IMB must also have had a role in the production of the observed correlations. The escape hatches that bound the IMB are the $\nu_6$ secular resonance and the 3:1 Jovian mean motion resonance \citep{Wisdom1985,Farinella.et.al1994}. We show that because of the unique resonant structure of the IMB, orbital evolution driven by Yarkovsky radiation forces results, inevitably, in the relative depletion of small asteroids from the high-inclination orbits. We also show that analysis of the observed asteroid size and orbital inclination correlation of the non-family asteroids allows us to separate Yarkovsky-driven orbital evolution from the other asteroid loss mechanisms and to constrain the timescale of that loss mechanism. 

This is a paper on the statistics of the distributions of asteroid sizes and proper orbital  elements. Given that we are looking for correlations, we must allow not only for the existence of asteroid families that have correlated orbital elements, but also for the likelihood that a large fraction of the asteroids that are currently classified as non-family are halo asteroids, that is, they are actually family asteroids and also have correlated orbital elements. In Section 2, in an attempt to circumvent this problem, we isolate a set of asteroids in the IMB that is devoid of the correlations associated with the major families and their halos. By making this choice, we confine our analysis to a small patch of the main belt that contains only about 1\% of the main belt asteroids.  Fortunately, the number of asteroids in this confined space that have known orbital and physical properties and are currently classified as non-family, proves to be sufficient for our purpose and, fortuitously, this space is of interest because of its unique dynamical structure. In Section 2, we discuss the observations that constrain our models of asteroid evolution. In Section 3, we discuss the mechanisms that result in asteroid loss from the main belt. Our models of asteroid evolution that are constrained by the observations discussed in Section 2 are presented in Section 4. These models allow us to estimate the timescale of orbital evolution of main belt asteroids due to Yarkovsky radiation forces. In Section 5, we present evidence for the existence of ghost families. In Section 6, we discuss the observed correlation between the sizes and orbital eccentricities of the IMB non-family asteroids. Finally, in Section \ref{section7}, we show how Yarkovsky forces largely define the shape of the SFD of the small asteroids in the Vesta family and we use that shape to constrain the age of the family.

\section{Asteroid size distributions}

\begin{figure}
    \centering
	\includegraphics[width=0.8\columnwidth]{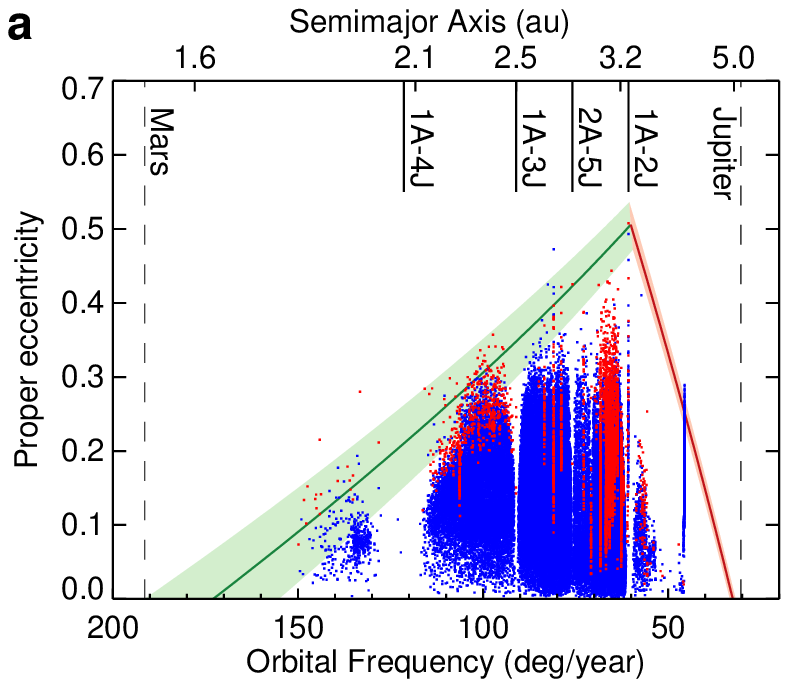}
	\includegraphics[width=0.8\columnwidth]{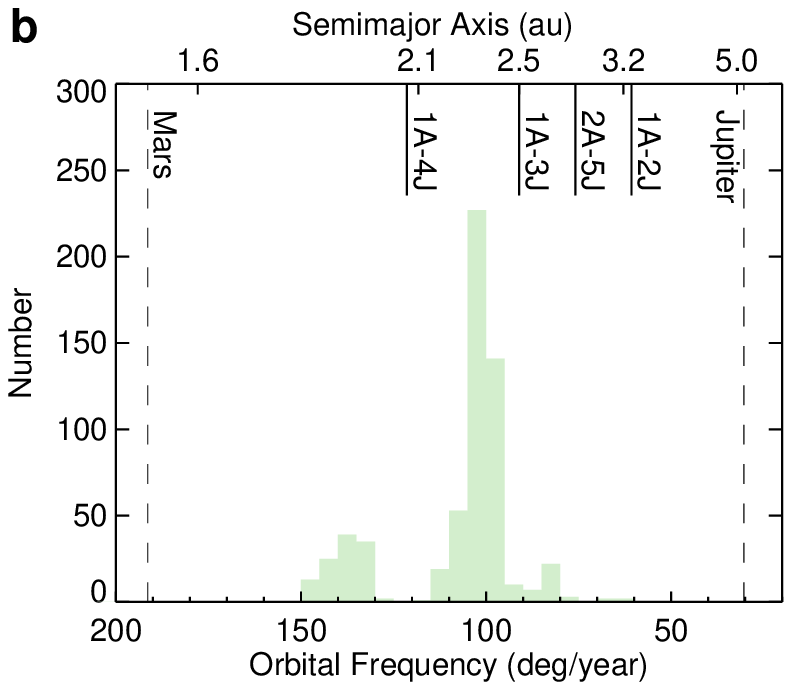}
    \caption{Panel A: A scatter plot of proper eccentricity $e$ and semimajor axis, $a$ of all the asteroids in the IMB with absolute magnitude $H<15$. The colours indicate the maximum Lyapunov Characteristic Exponents, mLCE of the orbits, as calculated by \citep{KnezevicMilani2007}, with red being the most chaotic (mLCE $>$ 0.00012 year$^{-1}$) and black the most stable (mLCE $<$ 0.00012 year$^{-1}$). In the Mars-crossing zone, which is shaded green, the orbit of an asteroid can cross that of Mars. The width of this zone depends on the orbital eccentricity of Mars which varies from near zero to $\sim$0.14 on a 2 $My$ timescale \citep{MurrayDermott1999}. The upper bound of the zone corresponds to Mars in a circular orbit. The brown shaded region is the Jupiter-crossing zone. Panel B: A histogram of the semimajor axes of the asteroids in the Mars-crossing zone. }
    \label{fig:1}
\end{figure}

\begin{figure*}
    \centering
	\includegraphics[width=0.32\textwidth]{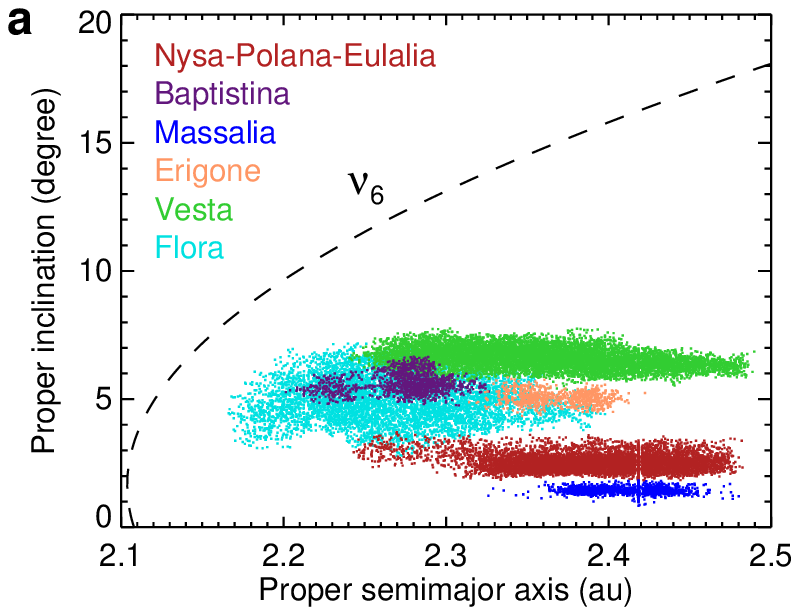}
	\includegraphics[width=0.32\textwidth]{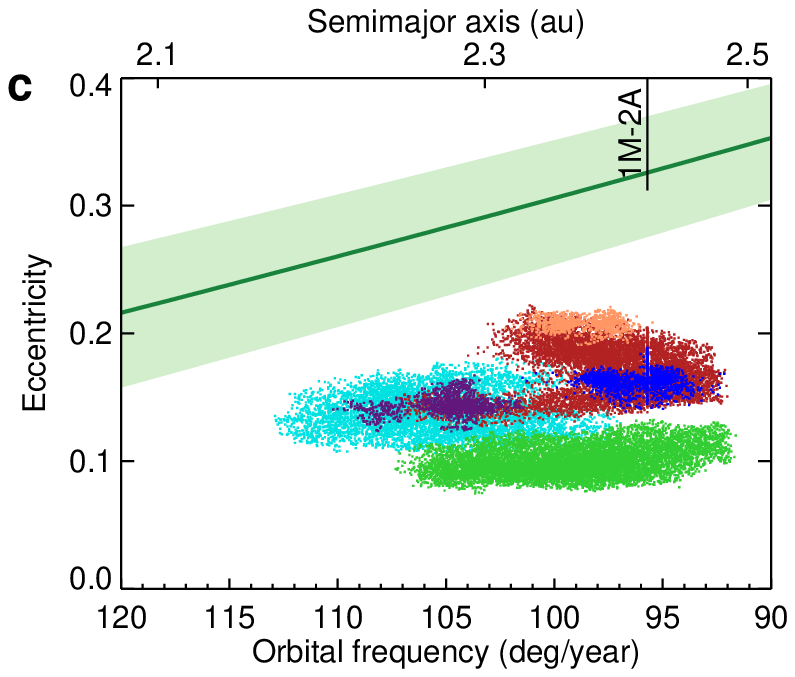}
	\includegraphics[width=0.32\textwidth]{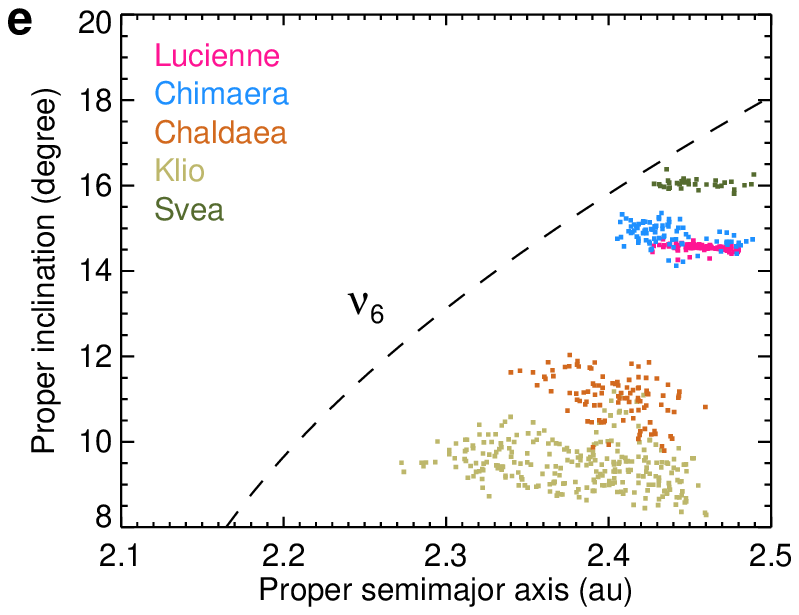}
	\includegraphics[width=0.32\textwidth]{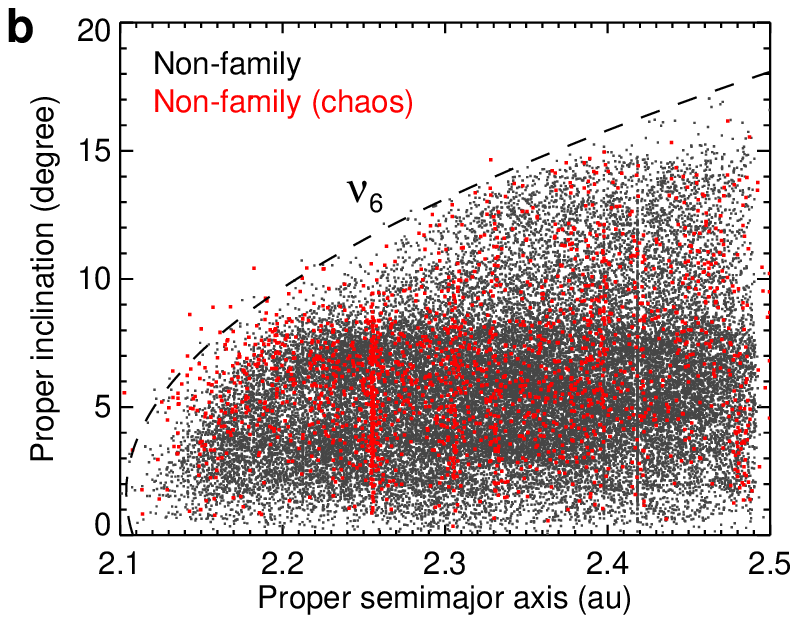}
	\includegraphics[width=0.32\textwidth]{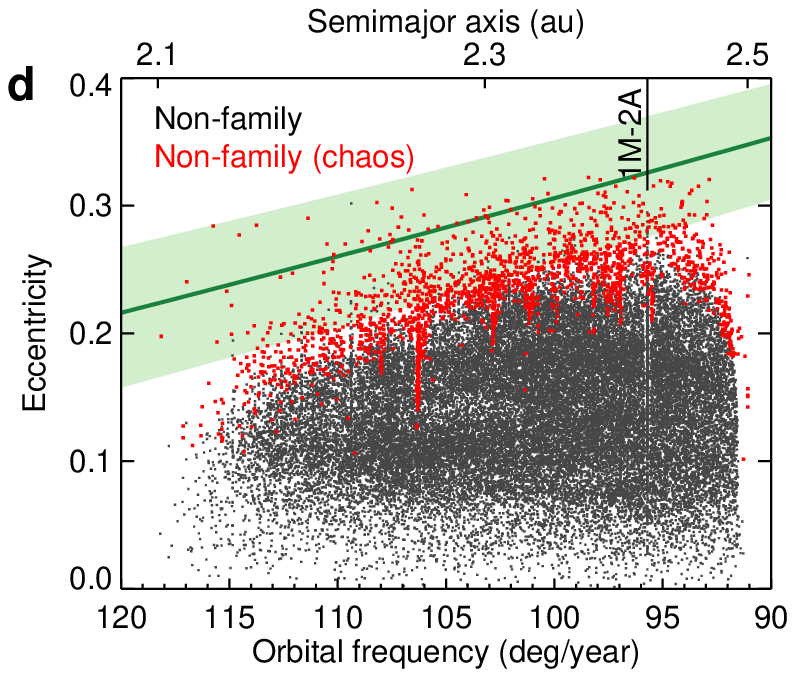}
	\includegraphics[width=0.32\textwidth]{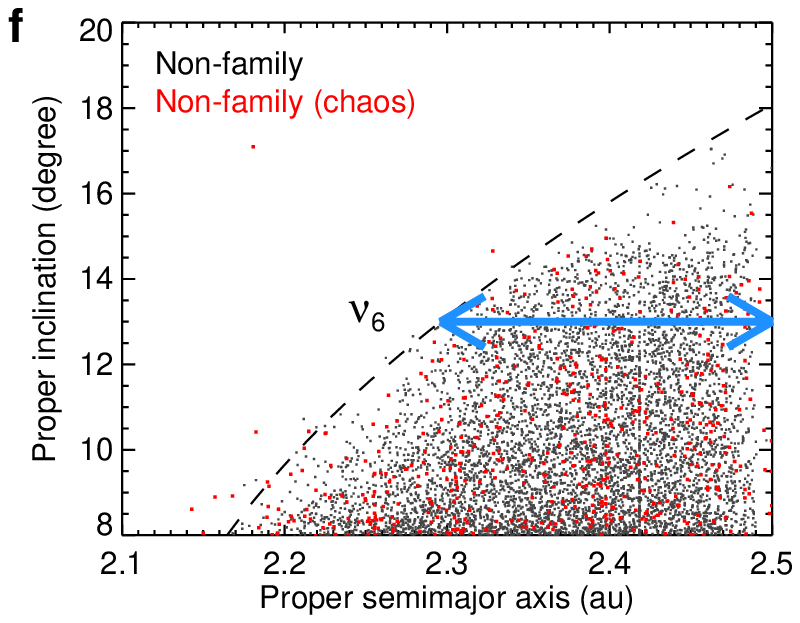}
    \caption{Panel A: A scatter plot of the proper inclination $I$ and semimajor axis, $a$ of all the asteroids in the IMB with absolute magnitude, $H<16.5$. The major families, as defined by \citet{Nesvorny.et.al2015}, are shown colour-coded. The dashed curve is the $\nu_6$ secular resonance. Panel B: An $a,I$ scatter plot of the non-family asteroids with points colour-coded according to their mLCEs, red being the most chaotic as in Fig. \ref{fig:1}. Panel C: The major families in the IMB. The green-shaded region is the Mars-crossing zone.  Panel D: An $a,e$ scatter plot of the non-family asteroids with points colour-coded according to their mLCEs, red being the most chaotic. Panel E: The minor families with high inclinations. Panel F: Non-family asteroids with high inclinations. The blue arrow indicates that Yarkovsky forces transport small asteroids to the bounding resonances. }
    \label{fig:2}
\end{figure*}

Fig. \ref{fig:1} shows the distribution of the proper eccentricities, $e$ and the semimajor axes, $a$ of all the asteroids in the main belt ($2.10<a<3.28$ $au$) with absolute magnitudes, $H<15$ and less than the observational completeness limit of the main belt as a whole \citep{hendler.et.al2020}. The eccentricity distribution is capped by the Mars-crossing zone, implying that Mars has a role in the removal of asteroids from the main belt and in their delivery to the inner solar system. The orbits in Fig. \ref{fig:1} are colour-coded according to their chaotic orbital evolution timescales, as calculated by \citet{KnezevicMilani2007}, with red being the most chaotic. The asteroids in the outer main belt ($2.96<a<3.28$ $au$) with moderate eccentricities are highly chaotic because of the overlap of high-order, Jovian mean motion resonances and these Jovian resonances have a role in the delivery of asteroids to the outer Mars-crossing-zone \citep{DermottMurray1983}. The asteroids in the IMB ($2.1<a<2.5$ $au$) with moderate eccentricities are also highly chaotic (Fig. \ref{fig:2}d), but in this case the chaos arises from a dense web of high-order, 2-body, Martian and Jovian mean motion resonances and various secular and 3-body resonances \citep{Milani2014}. These resonances also have a role in the delivery of asteroids to the Mars-crossing-zone \citep{MorbidelliNesvorny1999}. 

In Fig. \ref{fig:1}b, we show that the majority of the asteroids in the Mars-crossing zone are in the IMB. The scattering lifetimes of asteroids in the Mars-crossing zone decrease with decreasing semimajor axis and, on that basis, we might expect a deficiency of Mars-crossing asteroids in the IMB. The fact that we observe the opposite implies that the IMB Mars-crossing zone must have abundant, ongoing sources of replenishment implying, in turn, that the IMB is a major source of both near-Earth asteroids (NEAs) and meteorites, a conclusion that has previously been reached through an assessment of the efficacy of the likely escape routes \citep{Gladman1997,Granvik2017,Granvik2018}. One aim of this paper is to present new observational evidence that may allow us to determine and quantify how the IMB asteroids are transported to the inner Mars-crossing zone.

The asteroids in the IMB are bound in $a-I$ space by the resonant escape hatches at the 3:1 Jovian mean motion resonance at $a=2.5$ $au$, where the orbital periods of Jupiter and an asteroid are in the ratio 3:1, and at the $\nu_6$, eccentricity-type, secular resonance at semimajor axes, $\nu_6(I)$ given by
\begin{equation}
    \nu_6(I)=4.99332\mathrm{sin}^2I-0.28734\mathrm{sin}I+2.10798 \,au.
	\label{eq:1}
\end{equation}
\citep{Delbo.et.al2019}. These asteroids have proper (or long-term average) orbital elements in the ranges: $2.1<a< 2.5$ $au$, $0 < e < 0.325$, and $0 < I < 17.8$ $deg$, where $a$, $e$ and $I$ are, respectively, the proper semimajor axis, the proper eccentricity and the proper inclination \citep{KnezevicMilani2007}.

\begin{table*}
	\centering
	\caption{\citet{Nesvorny2015} families in the IMB with $H<16.5$ ordered by inclination $I$.}
	\label{tab:1}
	\begin{tabular}{clccccccc}
		\hline
		\# & Family & N & $a_{min}$ ($au$) & $a_{max}$ ($au$) & $e_{min}$ & $e_{max}$ & $I_{min}$ ($deg$) & $I_{max}$ ($deg$) \\
		\hline
		410 & 27 Euterpe & 177 & 2.2969 & 2.4578 & 0.1751 & 0.2014 & 0.45 & 1.22\\
		404 & 20 Massalia & 1,450 & 2.3260 & 2.4801 & 0.1390 & 0.1888 & 0.84 & 1.82\\
		405 & 44 Nysa-Polana-Eulalia & 7,975 & 2.2433 & 2.4825 & 0.1245 & 0.2220 & 1.82 & 3.88\\
		402	& 8 Flora & 7,226 & 2.1664 & 2.3978 & 0.1034 & 0.1802 & 2.79 & 7.17\\
		407 & 302 Clarissa & 36 & 2.3871 & 2.4145 & 0.1037 & 0.1098 & 3.24 & 3.56\\
		417 & 108138 2001 GB11 & 1 & 2.4649 & 2.4649 & 0.1525 & 0.1525 & 3.93 & 3.93\\
		406 & 163 Erigone & 777 & 2.3143 & 2.4224 & 0.1913 & 0.2207 & 4.34 & 5.85\\
		403 & 298 Baptistina & 1,305 & 2.2014 & 2.3263 & 0.1234 & 0.1672 & 4.76 & 6.66\\
		408 & 752 Sublimities & 175 & 2.4013 & 2.4882 & 0.0817 & 0.1023 & 4.79 & 5.68\\
		412 & 21509 Lucascavin & 1 & 2.2812 & 2.2812 & 0.1269 & 0.1269 & 5.23 & 5.23\\
		411 & 1270 Datura & 3 & 2.2347 & 2.2349 & 0.1534 & 0.1535 & 5.30 & 5.30\\
		401	& 4 Vesta & 9,631 & 2.2407 & 2.4897 & 0.0747 & 0.1323 & 5.37 & 7.75\\
		& Family ($I>9$ $deg$) & & & & & & & \\
		413 & 84 Klio & 259 & 2.2726 & 2.4601 & 0.1741 & 0.2114 & 8.29 & 11.18\\
		415 & 313 Chaldea & 101 & 2.3399 & 2.4596 & 0.2117 & 0.2432 & 9.79 & 12.03\\
        409 & 1892 Lucienne & 98 & 2.4270 & 2.4801 & 0.0841 & 0.1049 & 14.26 & 14.74\\
        414 & 623 Chimaera & 87 & 2.4054 & 2.4887 & 0.1331 & 0.1625 & 14.12 & 15.36\\
        416 & 329 Svea & 35 & 2.4280 & 2.4895 & 0.0797 & 0.1023 & 15.81 & 16.38\\
		\hline
	\end{tabular}
\end{table*}

\begin{table*}
	\centering
	\caption{\citet{Nesvorny2015} family and non-family asteroids in the IMB with $H<16.5$.}
	\label{tab:2}
	\begin{tabular}{clcc}
		\hline
		 & IMB asteroids with $H<16.5$ & Number & \% \\
		\hline
		$a=b+c+d+f$ & All & 63,692 & 100.0\\
		$b$ & Family ($I>9$ $deg$) & 580 & 0.9 \\
		$c$ & Non-family ($I>9$ $deg$) & 4,400 & 6.9 \\
		$d$ & Family ($I<9$ $deg$) & 28,757 & 45.1 \\
		$e$ & Non-family ($I<9$ $deg$) & 29,955 & 47.0 \\
		$f=c\times2.3$ & Estimate of non-family ($I<9$ $deg$) that are non-halo & 10,124 & 15.9\\
		$g=e-f$ & Halo asteroids ($I<9$ $deg$) & 19,831 & 31.1 \\
		$h=d+g$ & Family + halo asteroids ($I<9$ $deg$) & 48,588 & 76.2 \\
		\hline
	\end{tabular}
\end{table*}

The families in the IMB, as defined by \citep{Nesvorny.et.al2002,Nesvorny.et.al2015} using the Hierarchical Clustering Method (HCM), are listed in Table \ref{tab:1} and shown in Fig. \ref{fig:2}. HCM is a clustering algorithm \citep{Zappala.et.al1990} that defines a critical velocity difference between neighbouring orbits in $a-e-I$ space that is chosen to avoid family overlap. This critical velocity difference has no dynamical significance and the HCM cannot be used to attach the asteroids in the extensive family halos to their parent families, or to separate the family-halo asteroids from the non-family asteroids. This is a major problem for the analysis described in this paper. To minimize that problem, we confine our analysis to a small patch of the IMB that is devoid of asteroids originating from the major families and from the halos of those families. To avoid observational bias, we also confine our analysis to asteroids with $H < 16.5$ and less than the IMB completeness limit \citep{Dermott.et.al2018,hendler.et.al2020}.

\begin{figure}
    \centering
	\includegraphics[width=0.8\columnwidth]{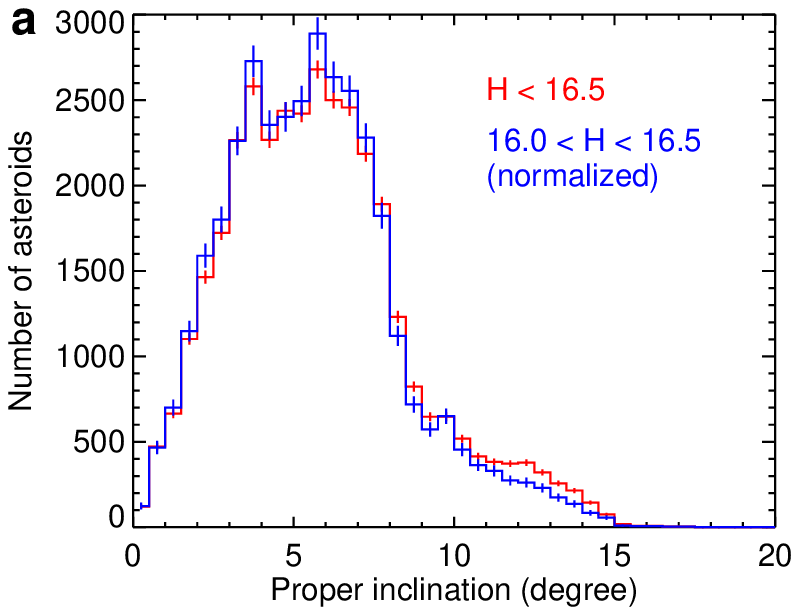}
	\includegraphics[width=0.8\columnwidth]{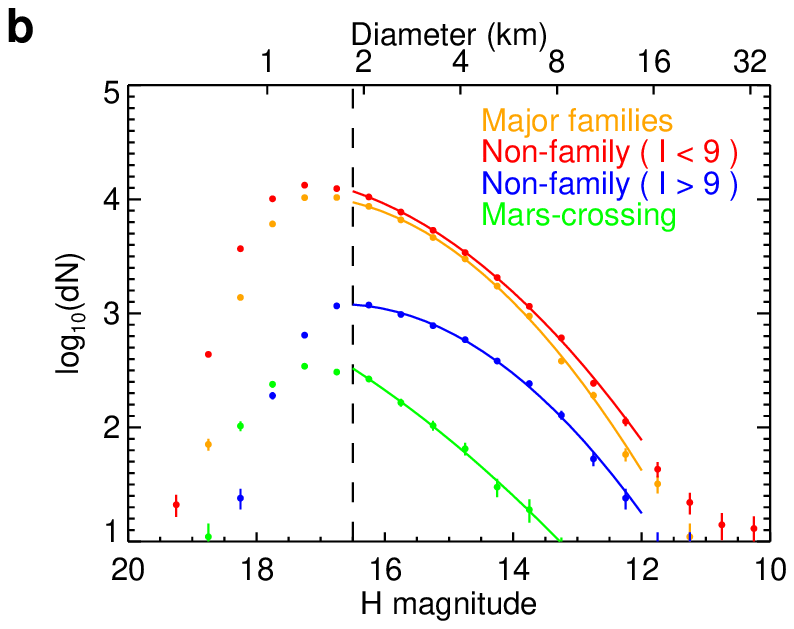}
    \caption{Panel A: The red histogram shows all the \citet{Nesvorny.et.al2015} non-family asteroids with $H < 16.5$, the IMB observational completeness limit. The blue histogram, that has been normalized to have an equal area, shows the small ($16.0 < H < 16.5$) non-family asteroids. Comparison of these two histograms shows that there is a marked lack of small asteroids with high inclinations. Panel B: Quadratic polynomial fits to the SFDs (with $dH=0.5$) for the asteroids in the separate major families (yellow), the non-family ($I<9$ $deg$) asteroids (magenta), the non-family ($I>9$ $deg$) asteroids (blue) and the asteroids in the Mars-crossing zone (green). Nominal asteroid diameters have been calculated from H assuming an albedo of 0.13.}
    \label{fig:3}
\end{figure}

\begin{figure}
    \centering
	\includegraphics[width=0.8\columnwidth]{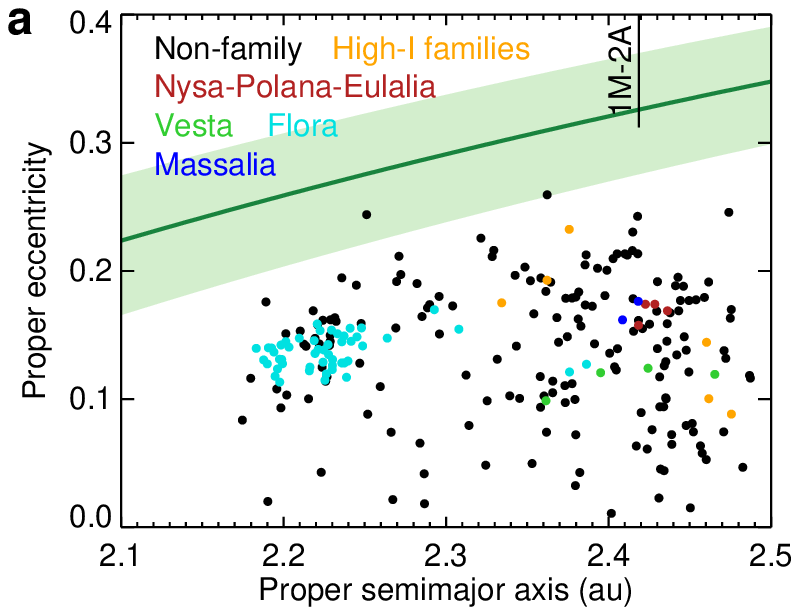}
	\includegraphics[width=0.8\columnwidth]{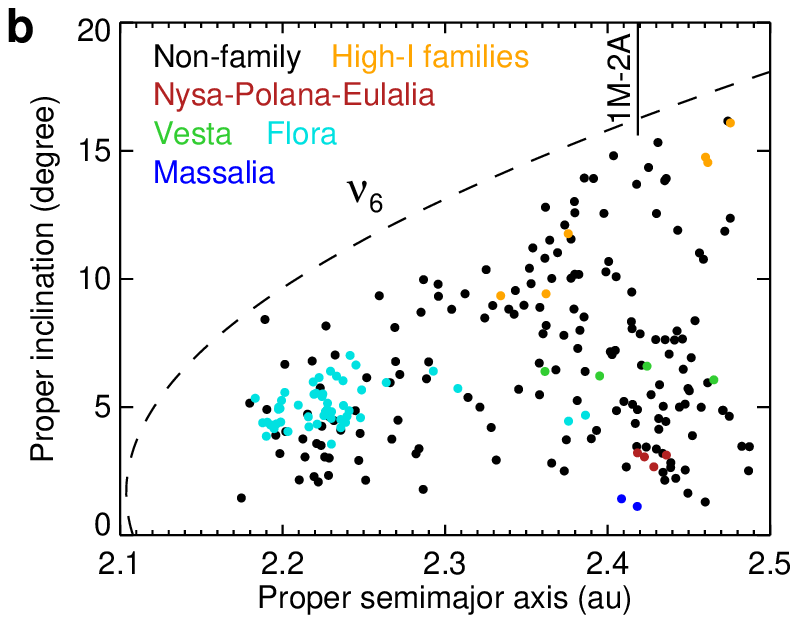}
    \caption{Panel A:  An $a,e$ scatter plot of all the asteroids in the IMB with absolute magnitude, $H<12$. Panel B: An $a,I$ scatter plot of all the asteroids in the IMB with absolute magnitude, $H<12$. Non-family asteroids are coloured black. The colours of the family asteroids correspond to those shown in Fig. \ref{fig:2}. Asteroids in the prominent Flora family are coloured cyan. A CC asteroid with $H=12$ and albedo, $A=0.07$ has a diameter, $D=20$ $km$. An NC asteroid with $H=12$ and albedo, $A=0.24$ has a diameter, $D=11$ $km$.}
    \label{fig:4}
\end{figure}

The distribution of the inclinations of the non-family asteroids are shown in Figs. \ref{fig:2} and \ref{fig:3}. The IMB is special in that there are no major asteroid families with proper $I>9$ $deg$ and above that inclination there is also a marked decrease in the asteroid number density in $a-I$ space (Fig. \ref{fig:3}a). Numerical investigations of the stability of the orbital inclinations show that any transport of asteroids between the $I<9$ $deg$ and the $I>9$ $deg$ asteroid groups by chaotic orbital evolution can be discounted (Appendix \ref{appendixA}). This group of 4,400 high-inclination ($I>9$ $deg$), non-family asteroids (Table \ref{tab:2}), that we assume is devoid of all major family asteroids and the asteroids in the halos of those families, is the focus of our investigation.

In Table \ref{tab:2}, we divide the 63,692 asteroids in the IMB with $H<16.5$ into groups according to their probable origins. The majority (92.1\%) of the asteroids in the IMB have $I<9$ $deg$ and, following \citet{Nesvorny.et.al2002,Nesvorny.et.al2015}, we divide these low-inclination asteroids into family (45.1\%) and non-family (47.0\%). However, the low-inclination ($I<9$ $deg$), non-family asteroids consist of both halo asteroids originating from the families and non-family asteroids of unknown origin. In Fig. \ref{fig:4}, we show the orbital elements of all the larger asteroids in the IMB with $H<12$. For reference, an asteroid with $H=12$ and albedo, $A = 0.07$ (CC) or 0.24 (NC) has a diameter, $D=20$ $km$ or 11 $km$, respectively. It is noticeable that a majority of these larger asteroids, particularly those in the range $2.3<a<2.5$ $au$, are not associated with any known family. An important question is whether these large non-family asteroids are primordial, or whether they are also the products of catastrophic collisions and members of ghost families, that is, old families with dispersed orbital elements. 

If we assume that the non-family asteroids that are not halo asteroids are distributed uniformly in $a-I$ space, then using equation~(\ref{eq:1}) and assuming that the average inner edge of the 3:1 Jovian mean motion resonance is at $a=2.478$ $au$, we calculate from equation~(\ref{eq:1}) that the ratio of the areas in $a-I$ space available to the $I>9$ $deg$ and the $I<9$ $deg$ asteroid groups is 0.43461. Then, using the observation that 4,400 non-family asteroids have $I>9$ $deg$, we estimate that the number of non-family asteroids with $I<9$ $deg$ that are not halo asteroids is 10,124 (Table \ref{tab:2}). It follows that the total number of family and family-halo asteroids with $H<16.5$ is 48,588 or 76.2\% of all the asteroids in the IMB. Thus, about three quarters of the asteroids in the IMB originate from just six asteroids (Massalia, Nysa-Polana-Eulalia, Flora and Vesta). In later sections, we present evidence that the remaining asteroids originate from a small number of ghost families. By estimating the number of these ghost families, we are able to estimate the number of source asteroids of the meteorites and the NEAs that originate from the IMB. We note here that the calculations listed in Table \ref{tab:2} also imply that all the major families listed by \citet{Nesvorny.et.al2015} must have significant numbers of interlopers that originate from ghost families. In addition, the large sizes of the non-family asteroids shown in Fig. \ref{fig:4} imply that the probability of an asteroid being an interloper increases with increasing asteroid size.

\begin{figure}
    \centering      
	\includegraphics[width=1.0\columnwidth]{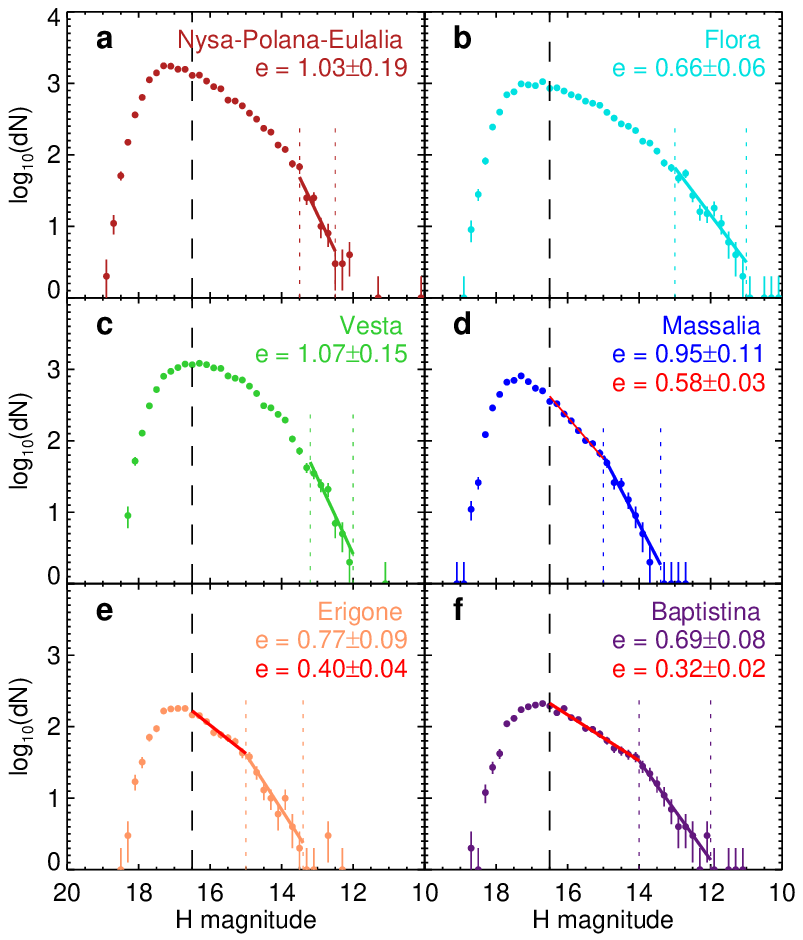}
    \caption{Panels A-F: The vertical dotted lines show the ranges of $H$ used in fitting the linear slopes of the SFDs of the large asteroids. Panels D-F: We also show linear fits to the small family asteroids in the range $16.5 > H > 13.5$. In all of the panels, the vertical dashed line is the observational completeness limit in the IMB, $H = 16.5$ \citep{Dermott.et.al2018}. }
    \label{fig:5}
\end{figure}

The incremental distribution of a collisional cascade  is described by
\begin{equation}
    \log dN=bH+c,
	\label{eq:2}
\end{equation}
where $dN$ is the number of asteroids in a box of width $dH$ and $c$ is a constant for asteroids with the same albedo. A cascade spanning all sizes must have $b<0.6$, otherwise most of the mass would be in the smallest asteroids and the total mass in the cascade would be infinite \citep{DurdaDermott1997}. \citet{Dohnanyi1969} calculated that an equilibrium cascade would have a slope, $b=0.5$. However, the slopes of the SFDs of the large asteroids in the major families are observed to be $> 0.6$ and in the case of the Vesta family and the Nysa-Polana-Eulalia family complex, the slopes are close to unity (Fig. \ref{fig:5}). These steep slopes are likely to be at least partly the result of either the total bombardment history of the precursor asteroids before their catastrophic disruption or the dynamics of crater formation.

\begin{figure}
    \centering
	\includegraphics[width=0.8\columnwidth]{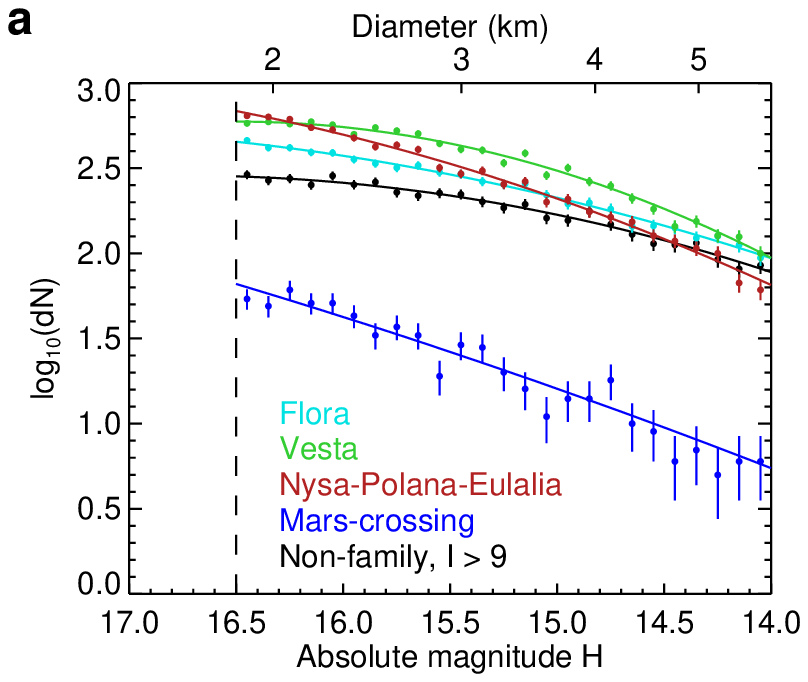}
	\includegraphics[width=0.8\columnwidth]{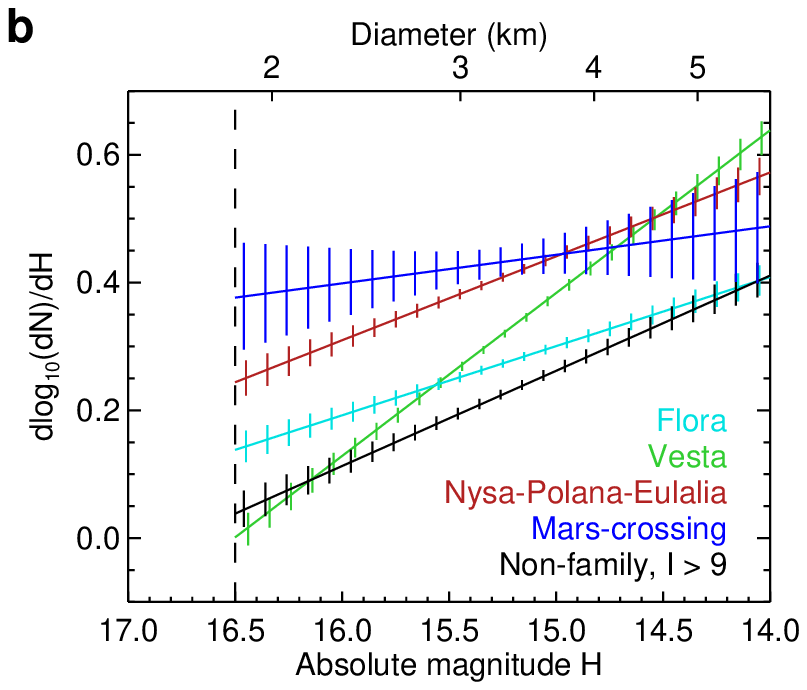}
    \caption{Panel A: Quadratic polynomial fits to the SFDs (with $dH = 0.1$) for the asteroids in the separate major families, for the non-family ($I>9$ $deg$) asteroids, and for the asteroids in the Mars-crossing zone. Nominal asteroid diameters have been calculated from $H$ assuming an albedo of 0.13. Panel B: Variation with absolute magnitude, $H$ of the slopes to the quadratic fits shown in panel A. }
    \label{fig:6}
\end{figure}

In Fig. \ref{fig:6}, we see that the slopes of the SFDs of the smaller asteroids ($H>14$) in the major families vary with $H$ and tend to zero for $H$ between 17 and 18. Given that our data set is observationally complete, this depletion of small asteroids cannot be ascribed to observational selection. The existence of large families is evidence that small asteroids are created by cratering events and by the catastrophic destruction of large asteroids. Therefore, we must expect collisional evolution to continuously change the SFDs of the smaller family asteroids. However, collisional evolution is not the only process that determines the observed SFDs. We argue here that the decrease of the slopes of the SFDs with increasing $H$ seen in Fig. \ref{fig:6}b is evidence that small asteroids are lost from the system by several mechanisms, including Yarkovsky-driven orbital evolution. We consider that Dohnanyi’s equilibrium cascade model that predicts a linear log-log SFD with slope $b=0.5$, must fail because once these small asteroids have left the system, an equilibrium collisional cascade cannot be established because there are too few small asteroids capable of destroying the larger asteroids.

In Fig. \ref{fig:3}a, we observe that the SFD of the major families, taken as a whole, and the SFD of low-inclination ($I<9$ $deg$), non-family asteroids are closely similar, supporting the argument that the low-inclination ($I<9$ $deg$), non-family asteroid group is dominated by the asteroids in the major-family halos that have SFDs similar to their parent families \citep{Dermott.et.al2018}. In contrast, in Fig. \ref{fig:3}b we observe that the high-inclination ($I>9$ $deg$), non-family group, which is devoid of both major-family and family-halo asteroids, has a markedly different SFD: these high-inclination, non-family asteroids are deficient in small asteroids. This depletion of small asteroids in high inclination orbits is also evident in Fig. \ref{fig:3}a. In contrast, the SFD of the asteroids in the Mars-crossing zone has an excess of small asteroids, suggesting that the asteroids that replenish that zone are predominantly small asteroids. 

\begin{figure}
    \centering
	\includegraphics[width=0.8\columnwidth]{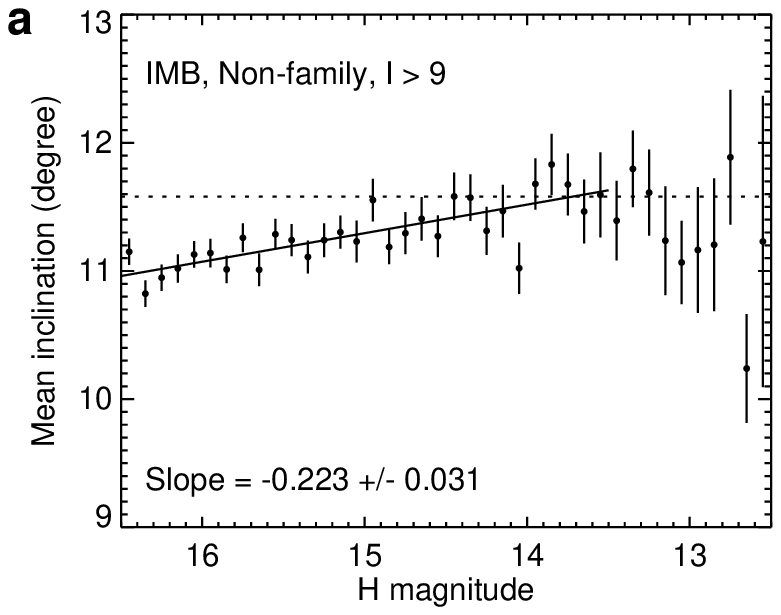}
	\includegraphics[width=0.8\columnwidth]{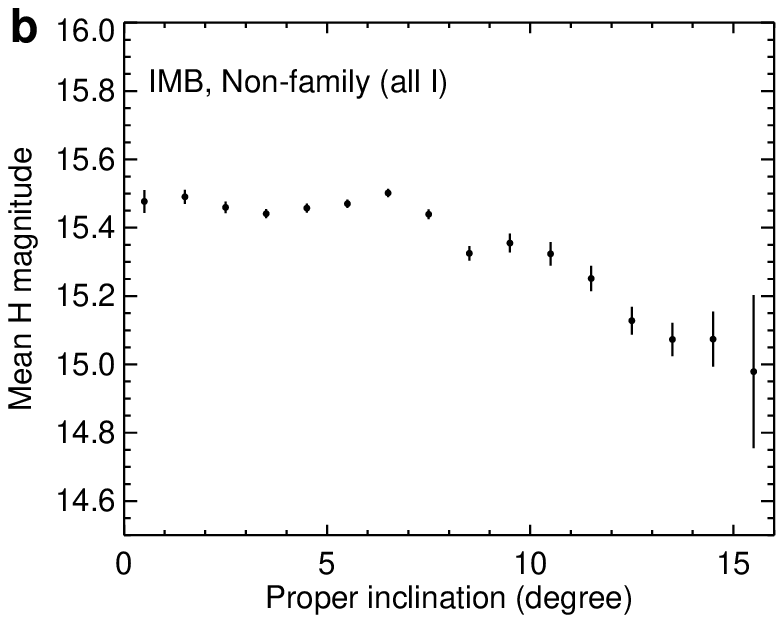}
    \caption{Panel A: Variation of the mean proper inclination of the high inclination ($I>9$ $deg$), non-family asteroids with absolute magnitude, $H$. The data is shown binned in $H$, but the slope has been determined from the individual points in the range $16.5<H<13.5$. Panel B: Variation of the mean absolute magnitude, $H$ of all the non-family asteroids in the IMB with proper inclination, $I$. In contrast to panel A, the data in panel B are shown binned in proper inclination. }
    \label{fig:7}
\end{figure}

The key observation that we discuss in this paper is the variation with $H$ of the mean inclination, $I$ of the high-inclination ($I>9$ $deg$), non-family asteroids shown in Fig. \ref{fig:7}. Plotting the data binned in $H$ shows that the mean $I$ increases with increasing asteroid size (or decreasing $H$) \citep{Dermott.et.al2018}. This correlation, that is significant at the 7$\sigma$ level, breaks down for asteroids with $H< 13.5$, a value that corresponds approximately to the changes in the slopes of the SFDs shown in Fig. \ref{fig:5}. By binning the same data in $I$, we observe that the decrease in mean $H$ with increasing $I$ is a property of the high-inclination ($I>9$ $deg$), non-family asteroids alone.  We argue that it is also significant that the space in the IMB above $I\simeq15$ deg and below the $\nu_6$ resonance is a desert that is almost devoid of asteroids (Fig. \ref{fig:2}f). We argue that both of these observations can be accounted for by the actions of Yarkovsky radiation forces.

\section{Asteroid loss mechanisms}

Mutual gravitational interactions force the eccentricities of the planetary orbits to vary periodically on timescales of between $\sim$50,000 and $\sim$500,000 years \citep{MurrayDermott1999}. An eccentricity-type secular resonance occurs when the average rate of change of the longitude of an asteroid’s pericenter equals one of the eigenfrequencies of the planetary system. In general, the location of an eccentricity-type secular resonance depends on  $a$, $e$ and $I$.  However, all of the asteroids in our IMB data set have $I<17.8$ $deg$ and for these inclinations the $\nu_6$ resonance is remarkable in that its location is largely eccentricity independent \citep{Morbidelli2002}. Yarkovsky forces transport small asteroids in the IMB to one of the bounding resonances, either the 3:1 Jovian resonance or the $\nu_6$ resonance, depending on the spin direction of the asteroid \citep{Bottke.et.al2002a}: witness the spread in the semimajor axes of the family asteroids shown in Fig. \ref{fig:2}. On encountering the $\nu_6$ resonance, numerical integrations by \citet{Farinella.et.al1994} have shown that resonant forces increase the eccentricity while the semimajor axis and the proper inclination remain unchanged with the result that the asteroid remains in resonance as the eccentricity increases to values approaching unity, allowing asteroids to fall into the Sun. Similar large increases in eccentricity occur when asteroids are captured in the 3:1 Jovian mean motion resonance \citep{Wisdom1985}. Once either of these resonances is encountered, asteroids are lost from the asteroid belt on timescales $\sim10^{6}$ $yr$ that are small compared with the transportation timescales.     

To account for the observed correlations between the mean inclinations and the mean sizes of the high-inclination ($I>9$ $deg$) asteroids, we need a loss mechanism that is dependent on the proper inclinations. Uniquely in the IMB, the transport of small asteroids to the escape hatches at the bounding resonances, driven by Yarkovsky radiation forces, provides such a mechanism. Thermal emission from a rotating asteroid is anisotropic with respect to the Sun-asteroid line and this anisotropy results in a weak force that, depending on the spin direction of the asteroid, expands or contracts the orbit. In the IMB, the separation in semimajor axis of the two escape hatches is determined by the proper inclination and this separation decreases almost linearly from $\sim$0.3 $au$ at $I=9$ $deg$ to zero at $I=17.8$ $deg$ (Fig. \ref{fig:2}f). The solar radiation intercepted by an asteroid increases as $D^2$, while the asteroid inertia increases as $D^3$ with the result that the asteroid acceleration increases as $1/D$, a result that has been confirmed by measuring the rates of change of the semimajor axes of very small near-Earth asteroids (NEAs) \citep{Greenberg.et.al2020}. The rates of change of the semimajor axes of the asteroids due to the Yarkovsky radiation forces do not depend on their inclinations, but in the IMB the lengths of the escape routes to the bounding resonances decrease with increasing inclination (Fig. \ref{fig:2}f). This results in a depletion of the smallest asteroids, with the highest rates of change of semimajor axis, from the high-inclination orbits.

We need to consider the loss of asteroids from the IMB due to the action of four mechanisms: collisional and rotational destruction; chaotic orbital evolution; and Yarkovsky-driven transport of small asteroids. Yarkovsky-driven transport is the only mechanism driving the asteroids to the escape hatches at the bounding resonances and we will show that Yarkovsky driven loss could be  the dominant loss mechanism in the size range that we investigate. \citet{Greenberg.et.al2020} used radar and optical observations of NEAs to estimate the strength of the Yarkovsky forces experienced by very small asteroids. They ignored the less important seasonal component of the Yarkovsky force and used the diurnal component to calculate that the average rate of change of the semimajor axis is given by
\begin{equation}
    \frac{da}{dt}= \pm  \xi  \frac{3}{4 \pi } \frac{1}{ \sqrt{a} } \frac{1}{1-e^2} \frac{L_ \odot }{c \sqrt{GM_ \odot } } \frac{1}{D \rho },
	\label{eq:3}
\end{equation}
where $L_{\odot}$ and $M_{\odot}$ are, respectively, the solar luminosity and the solar mass, $c$ is the speed of light, and $D$ and $\rho$ and are, respectively, the diameter and mean density of the asteroid. The Yarkovsky efficiency, $\xi$, depends on the spin pole obliquity and the thermal properties of the asteroid \citep{Bottke.et.al2002a}. On evaluating this equation, \citet{Greenberg.et.al2020} obtain 
\begin{equation}
    \frac{da}{dt}= \pm 7.2 \big( \frac{ \xi }{0.1} \big)  \sqrt{ \frac{1\,au}{a} } \big(\frac{1}{1-e^2}\big) \big( \frac{1\,km}{D} \big) \big( \frac{2000\,kg\,m^{-3}}{ \rho } \big) \frac{10^{-4}\,au}{Myr}.
	\label{eq:4}
\end{equation}
The range of semimajor axes in our models corresponding to $I=13.5$ $deg$ is 0.17 $au$. Over this small range, we ignore the variation of the Yarkovsky force and use an average value of $a=2.4$ $au$. We also ignore the eccentricity term. With these approximations, for asteroids in our non-family, high-inclination ($I>9$ $deg$) data set, equation~(\ref{eq:4}) reduces to
\begin{equation}
    \frac{1}{a} \frac{da}{dt}= \pm 2.0 \,\xi \big( \frac{1\,km}{D} \big) \big( \frac{2000\,kg\,m^{-3}}{ \rho } \big) Gyr^{-1}.
	\label{eq:5}
\end{equation}
While we expect $da/dt\propto1/D$, it is possible that other factors influence this dependence. For example, \citet{bolin2018a} considers that thermal inertia varies with asteroid size. Accordingly, in our models we do not assume that $\alpha=1$, rather we use the asteroid size-inclination correlation to determine the dependence of $da/dt$ on $D$. Following \citet{bolin2018a}), we use the more general expression
\begin{equation}
    \frac{1}{a} \frac{da}{dt}= \pm \big( \frac{1}{T_\mathrm{Y}} \big) \big( \frac{1\,km}{D} \big)^{\alpha},
	\label{eq:6}
\end{equation}
where $T_\mathrm{Y}$ is the Yarkovsky timescale, and determine $\alpha$ from the observations through modeling. The two largest uncertainties in $T_\mathrm{Y}$ are $\xi$, the Yarkovsky efficiency, and $\rho$, the bulk mean density. From observations of near-Earth asteroids, \citet{Greenberg.et.al2020} estimate that $\xi$ ranges from near-zero to over 0.7 with a median value of $0.12_{-0.06}^{+0.16}$. \citet{Carry2012} finds that the mean density of small, porous asteroids varies with diameter, $D$ and for small C-type and S-type asteroids could be, respectively, as small as 600 $kg$ $m^{-3}$ and 2,300 $kg$ $m^{-3}$. Our data set has equal numbers of CC and NC asteroids (see Fig. \ref{fig:14}) and \citet{Carry2012} lists the average density of C and S-type asteroids as 1,570 $kg$ $m^{-3}$ and 2,660 $kg$ $m^{-3}$ respectively, suggesting an average density of about 2,000 $kg$ $m^{-3}$. Assuming that $\xi=0.12$ and $\rho=2,000$ $kg$ $m^{-3}$, we estimate from the observations of \citet{Greenberg.et.al2020} that $T_\mathrm{Y}=4.0$ $Gyr$. However, we note that this estimate assumes that the spin rates and spin directions of the asteroids and also their masses are constant. Over the age of the solar system, this may not be the case and we need to distinguish between short-timescale evolution, as measured by the NEA observations, and the average timescale corresponding to orbital evolution over $\sim$$10^8$ $yr$ or longer.

Other loss mechanisms that do not depend on the orbital inclination include catastrophic disruption, rotational disruption and orbital evolution associated with the dense web of weak resonances that feeds asteroids into the Mars-crossing zone. \citet{Jacobson.et.al2014} argue that for the asteroids in our size range ($1 \lesssim D \lesssim 10$ $km$), the dominant loss mechanism is YORP-induced rotational disruption \citep{Rubincam2000} for which
\begin{equation}
    T_{\mathrm{YORP}}=10\big( \frac{D}{1\,km} \big)^2\, Myr.
	\label{eq:7}
\end{equation}
This timescale is even more uncertain than the Yarkovsky timescale, partly because of uncertainties in the thermal efficiencies and the mean densities, but also because the YORP forces depend on the unknown asymmetrical shapes of the asteroids. \citet{Jacobson.et.al2014} also state that the timescale of catastrophic disruption, $T_{disr}$ is given by  
\citep{Rubincam2000} for which
\begin{equation}
    T_{\mathrm{disr}}=450\big( \frac{D}{1\,km} \big)^{1/2}\, Myr
	\label{eq:8}
\end{equation}
and the timescale, $T_{rot}$ on which collisions change the asteroid spin directions is given by
\begin{equation}
    T_{\mathrm{rot}}=130\big( \frac{D}{1\,km} \big)^{3/4}\, Myr.
	\label{eq:9}
\end{equation}
However, a problem with the latter two estimates is the assumption that the SFD of the main belt asteroids is described by a Dohnanyi-type equilibrium collisional cascade and this needs to be questioned.

For heuristic purposes, let us consider a solid asteroid of diameter $D$ and impact strength $S$ disrupted by an asteroidal bullet of diameter $d$ and relative velocity $V$. If we assume that a fraction $f$ of the kinetic energy of the bullet is available to disrupt the larger asteroid, and that both asteroids have the same density, then if we ignore any variation of the impact strength with diameter, the critical ratio $d/D$ does not depend on $D$ and is given by 
\begin{equation}
    \frac{d}{D}>\big( \frac{2S}{fV^2} \big)^{1/3}.
	\label{eq:10}
\end{equation}
The collisional lifetime of the asteroid is then determined by the cross-sectional area of the target asteroid and the number of asteroid bullets with diameters greater than some critical diameter, a number that is determined by the SFD of the asteroid bullets. The purpose of this heuristic argument is to point out that one major source of uncertainty in this estimate is the uncertainty in the SFD of the asteroid bullets. If we assume, following \citet{Dohnanyi1969}, that the cumulative number of asteroids with diameter $>d$ is given by
\begin{equation}
    N=Cd^{-p},
	\label{eq:11}
\end{equation}
where $C$ is a constant and that in an equilibrium cascade $p=5b=5/2$ \citep{Dohnanyi1969,DurdaDermott1997}, then it follows that the collisional lifetime is $\propto d^{5/2}/D^2 \propto D^{1/2}$ (cf. equation~(\ref{eq:8})). The weakness of this analysis (apart from the facts that the effective strength of an asteroid and the magnitude of f are not independent of size) is that we have no strong evidence that the asteroid belt can be described by an equilibrium cascade. Furthermore, Yarkovsky forces and other loss mechanisms act to remove small asteroids from the main belt on timescales comparable with the collisional lifetimes, making it difficult for an equilibrium cascade with $b=0.5$ to be established. We do have good observations of the SFD of those asteroids with $H<16.5$, but these asteroids are disrupted by smaller asteroids for which $d/D\sim0.1$ and the SFD of these smaller asteroids, that are observationally incomplete, is currently uncertain.

If, again for heuristic purposes, we consider the possibly more relevant case of the disruption of a gravitationally bound, rubble-pile asteroid, then the critical mass, m of the bullet needed to overcome the gravitational forces binding an asteroid of mass $M$ is given by
\begin{equation}
    f \frac{1}{2}mV^2>\frac{12}{5} \frac{GM^2}{D}.
	\label{eq:12}
\end{equation}
In this case, assuming $V=5$ $km/s$, the critical diameter ratio is given by
\begin{equation}
    \frac{d}{D} > \big(\frac{4\pi G \rho D^2}{5fV^2}\big)^{1/3} = 0.005 \big( \frac{ 0.1 }{f} \big) \big( \frac{\rho}{2000\,kg\,m^{-3}} \big)^{1/3} \big( \frac{D}{1\,km} \big)^{2/3}.
	\label{eq:13}
\end{equation}
Hence, the critical bullet size increases as $D^{5/3}$ and, assuming as before that $p=5b=5/2$, the lifetime of the target asteroid varies as $D^{13/6}$. However, while noting that this dependence on $D$ is closely similar to the dependence on $D$ of the YORP rotational disruption timescale, which proves to be useful in Section \ref{section7} where we discuss the SFD of the Vesta family, we also note that we have little confidence in this relation because of our lack of knowledge of the SFD of the small asteroid bullets or the size variation of the factor $f$.

Given that our aim is to place observational constraints on the various asteroid loss timescales, all of these loss mechanisms should be modelled separately. However, in any given size range one mechanism tends to dominate and we are able to show, from the observations, that for asteroids with diameters in the range $\sim$1 to $\sim$10 $km$ orbital evolution due to Yarkovsky radiation forces could be the dominant loss mechanism. It is therefore expedient to reduce the number of variables by bundling all the loss mechanisms that do not depend on the inclination into one other group and writing 
\begin{equation}
    \frac{1}{N(D)} \frac{dN(D)}{dt}= -\big(\frac{1}{T_\mathrm{L}}\big) \big( \frac{1\,km}{D} \big)^{\beta},
	\label{eq:14}
\end{equation}
where $T_\mathrm{L}$ is the timescale of these other loss mechanisms and $N(D)dD$ is the number of asteroids with diameters in the range $D$ to $D+dD$.

\section{Models and results}

\begin{figure*}
    \centering  
	\includegraphics[width=0.355\textwidth]{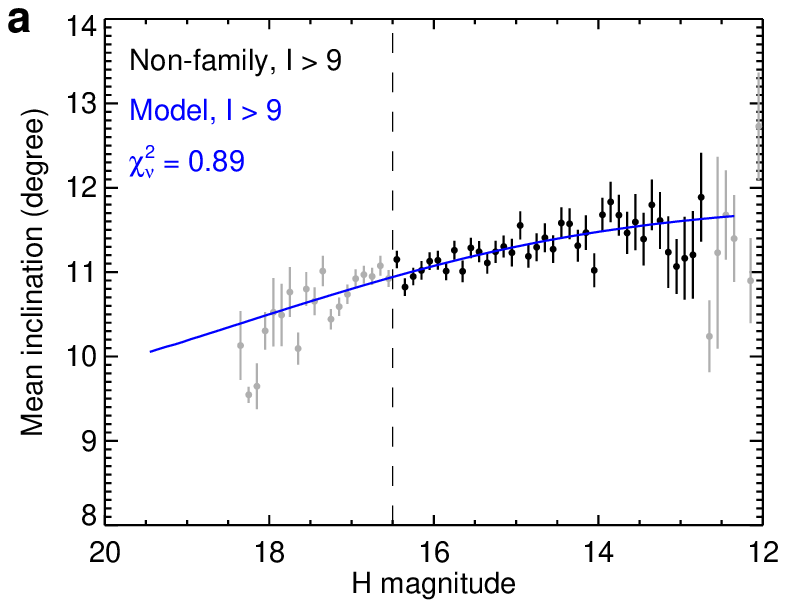}
	\includegraphics[width=0.355\textwidth]{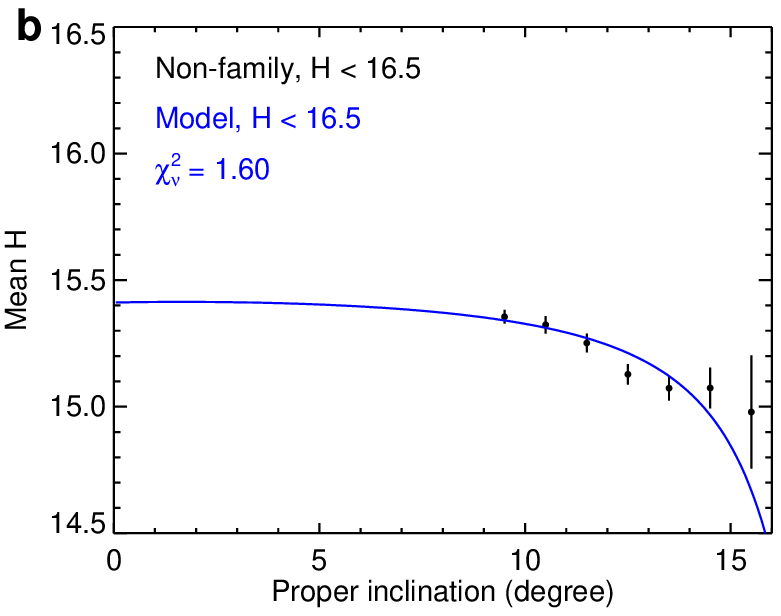}
	\includegraphics[width=0.355\textwidth]{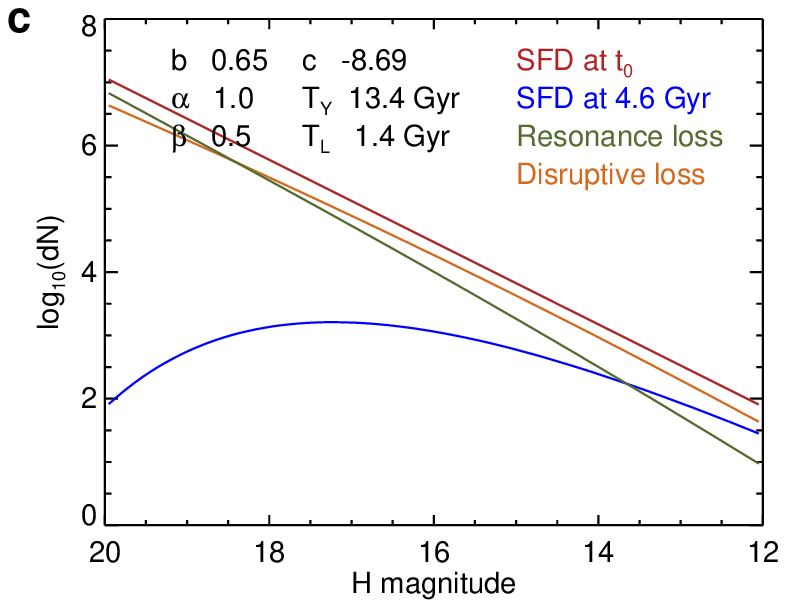}
	\includegraphics[width=0.355\textwidth]{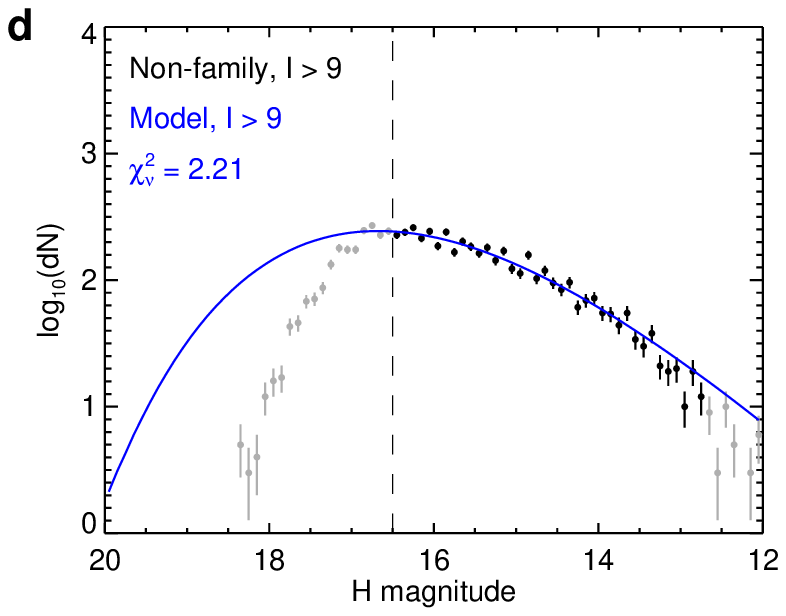}
	\includegraphics[width=0.355\textwidth]{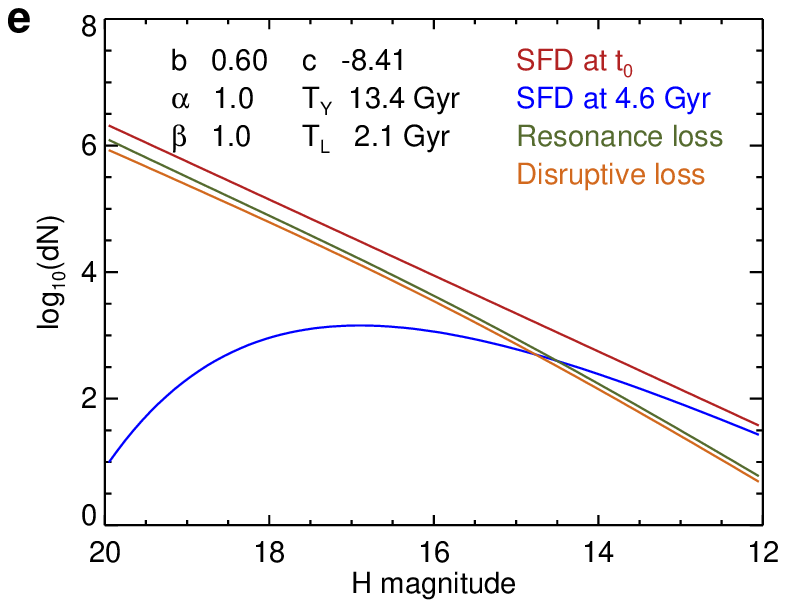}
	\includegraphics[width=0.355\textwidth]{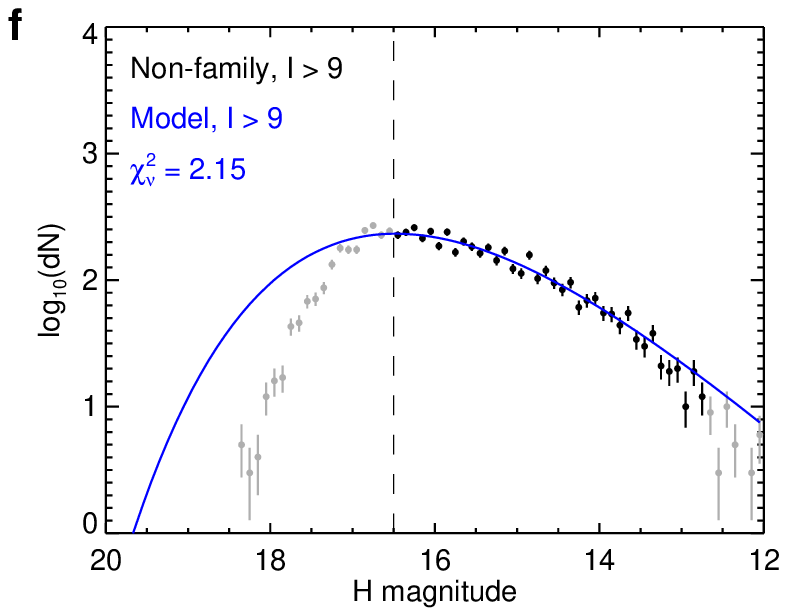}
	\includegraphics[width=0.355\textwidth]{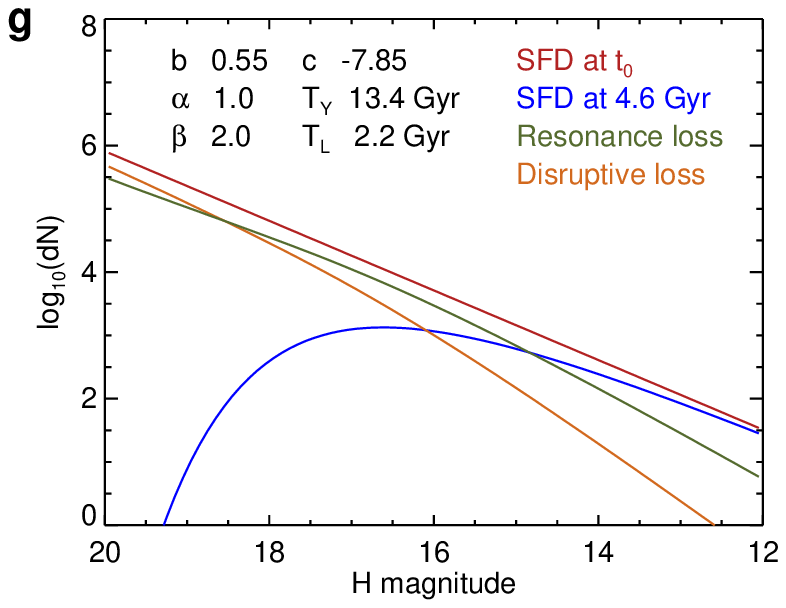}
	\includegraphics[width=0.355\textwidth]{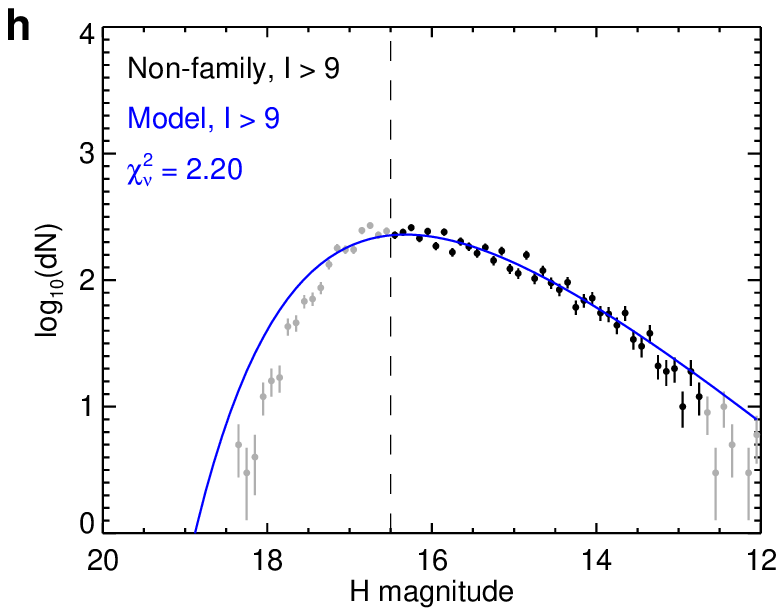}
    \caption{Models for the depletion of all the high-inclination, non-family asteroids in the IMB due to (a) a Yarkovsky force that changes the semimajor axes on a timescale $T_\mathrm{Y}$, and (b) all  other loss mechanisms that do not depend on the proper inclination, $I$ and result in loss of asteroids on a timescale $T_\mathrm{L}$. $\alpha$ and $\beta$ describe the dependence of these timescales on the asteroid diameter, $D$, $b$ is the slope of the initial SFD and $c$ is a normalizing constant that determines the total number of asteroids in the distribution. Panel A: Variation of the mean proper inclination of the high inclination ($I>9$ $deg$), non-family asteroids with absolute magnitude, $H$. The data is shown binned in $H$, but the slope has been determined from the individual points in the range $16.5<H<13.5$. Panel B: Variation of the mean absolute magnitude, $H$ of all the non-family asteroids in the IMB with proper inclination, $I$. In contrast to panel A, the data in panel B are shown binned in proper inclination.  The curves in panels A and B are the best-fit curves with $\alpha=1$ and the Yarkovsky timescale, $T_\mathrm{Y}=13.4$ $Gyr$. Panels C-H use the same values of $T_\mathrm{Y}$ and $\alpha$, but allow $\beta$ to take the values of 0.5 (panel C), 1.0 (panel E) and 2.0 (panel G). The values of $T_\mathrm{L}$ are the best-fit values found from the models. Panels C, E and G show the fractional losses due to the separate loss mechanisms as a function of $H$. The magenta curve shows the SFD at time zero and the blue curve the SFD after 4.6 $Gyr$ of evolution. The green and orange curves show, respectively, the total number of asteroids lost through the resonance escape hatches and the total number lost through all other loss mechanisms. Panels D, F and H compare the model SFDs with the observations with $\beta$ increasing from 0.5 to 2.0. The grey points in panels D, F and H are not used in the fits. }
    \label{fig:8}
\end{figure*}

\renewcommand{\arraystretch}{1.2}

\begin{table}
	\centering
	\caption{Best fit for the Yarkovsky loss timescale, $T_\mathrm{Y}$.}
	\label{tab:3}
	\begin{tabular}{ccc}
		\hline
		$\alpha$ & $T_\mathrm{Y}$ ($Gyr$) & $\chi_{\nu}^2$ \\
		\hline
		0.6	& $20.3^{+2.0} _{-1.8} $ & 1.01\\
		0.7	& $18.2^{+1.8} _{-1.5} $ & 0.94\\
		0.8	& $16.4^{+1.6} _{-1.4} $ & 0.90\\
		0.9	& $14.8^{+1.5} _{-1.3} $ & 0.88\\
		1.0	& $13.4^{+1.4} _{-1.2} $ & 0.89\\
		1.1	& $12.1^{+1.2} _{-1.0} $ & 0.91\\
		1.2	& $11.0^{+1.1} _{-0.9} $ & 0.95\\
		1.3	& $10.0^{+1.1} _{-0.9} $ & 1.01\\
		1.4	& $9.1^{+1.0} _{-0.8} $ & 1.08\\
		1.5	& $8.3^{+0.9} _{-0.8} $ & 1.17\\
		1.6	& $7.6^{+0.9} _{-0.8} $ & 1.26\\
		1.7	& $6.9^{+0.9} _{-0.7} $ & 1.37\\
		1.8	& $6.4^{+0.8} _{-0.7} $ & 1.49\\
		1.9	& $5.8^{+0.8} _{-0.6} $ & 1.61\\
		2.0	& $5.3^{+0.8} _{-0.6} $ & 1.74\\
		\hline
	\end{tabular}
	\vspace{1ex}
	
     {Note: $\chi_\nu^2$ is calculated using the plot of <$I$> vs. $H$, Fig.~\ref{fig:8}c, and the calculated value of $T_\mathrm{Y}$ assumes that $t_{\mathrm{evol}}=4.6$ $Gyr$.}
\end{table}

\begin{table}
	\centering
	\caption{Best fit for the other loss timescale, $T_\mathrm{L}$.}
	\label{tab:4}
	\begin{tabular}{cccccc}
		\hline
		$\alpha$ & $T_\mathrm{Y}$ ($Gyr$) & $\beta$ & $b$ & $T_\mathrm{L}$ ($Gyr$) & $\chi_{\nu}^2$ \\
		\hline
	     \multirow{7}{*}{1.0} & \multirow{7}{*}{13.4} & \multirow{7}{*}{2.0} & 0.5 & $7.4^{+5.5} _{-2.2} $ & 2.01\\
		 & & & 0.55 & $2.2^{+0.3} _{-0.2} $ & 1.87\\
		 & & & 0.6 & $1.3^{+0.1} _{-0.1} $ & 2.00\\
		 & & & 0.65 & $1.2^{+0.1} _{-0.0} $ & 5.58\\
		 & & & 0.7 & $1.1^{+0.1} _{-0.0} $ & 14.3\\
		 & & & 0.75 & $1.0^{+0.1} _{-0.0} $ & 26.5\\
		 & & & 0.8 & $0.9^{+0.1} _{-0.0} $ & 40.3\\
		\hline
		\multirow{7}{*}{1.0} & \multirow{7}{*}{13.4} & \multirow{7}{*}{1.0} & 0.5 & $12.5^{+10.1} _{-4.9} $ & 2.07\\
		& & & 0.55 & $3.6^{+0.4} _{-0.4} $ & 1.93\\
		 & & & 0.6 & $2.1^{+0.1} _{-0.1} $ & 1.86\\
		 & & & 0.65 & $1.5^{+0.1} _{-0.1} $ & 1.88\\
		 & & & 0.7 & $1.1^{+0.1} _{-0.0} $ & 2.10\\
		 & & & 0.75 & $0.9^{+0.1} _{-0.0} $ & 2.29\\
		 & & & 0.8 & $0.8^{+0.1} _{-0.0} $ & 2.31\\
		 \hline
		 \multirow{7}{*}{1.0} & \multirow{7}{*}{13.4} & \multirow{7}{*}{0.5} & 0.5 & $11.9^{+10.2} _{-3.7} $ & 2.11\\
		& & & 0.55 & $3.4^{+0.4} _{-0.4} $ & 2.02\\
		 & & & 0.6 & $2.0^{+0.1} _{-0.1} $ & 1.96\\
		 & & & 0.65 & $1.4^{+0.1} _{-0.1} $ & 1.90\\
		 & & & 0.7 & $1.1^{+0.1} _{-0.1} $ & 1.93\\
		 & & & 0.75 & $0.9^{+0.1} _{-0.0} $ & 1.98\\
		 & & & 0.8 & $0.7^{+0.1} _{-0.0} $ & 2.56\\
		 \hline
	\end{tabular}
	\vspace{1ex}
	
     {Note: $\chi_\nu^2$ is the mean $\chi_\nu^2$ of the SFD plot and the plot of <$H$> vs. $I$, Fig.~\ref{fig:8}b, d, and the calculated value of $T_\mathrm{Y}$ assumes that $t_{\mathrm{evol}}=4.6$ $Gyr$.}
\end{table}

Our models depend on four assumptions. The first assumption is that asteroid loss is driven by several mechanisms, but only one of these loss mechanisms depends on the orbital inclination. The second assumption is that, as suggested by the present distribution of high-inclination ($I>9$ $deg$) orbits shown in Fig. \ref{fig:2}f, the initial asteroid distribution in $a-I$ space was uniform. The third assumption is that the SFD of the high-inclination asteroids was initially linear on a log-log scale and that this SFD had some unknown slope, $b$. The assumption of uniformity is key to our analysis as it allows us to separate the effects of the inclination-dependent loss mechanism from those of the other loss mechanisms. Given these three assumptions, our models depend on the five parameters $\alpha$, $T_\mathrm{Y}$, $\beta$, $T_\mathrm{L}$, and $b$ defined by equations (\ref{eq:2}), (\ref{eq:6}), and (\ref{eq:14}). We constrain our models by the three sets of observations shown in Fig. \ref{fig:8}: (1) <$I$> vs. $H$, (2) the SFD, and (3) <$H$> vs. $I$. We also assume (and this is the fourth assumption) that the asteroids are collisionally evolved. However, here we allow for any changes in size by simply assuming that the asteroids in the model have evolved for an average time $t_{\mathrm{evol}}$ that may be less than the age of the solar system.  In the Tables \ref{tab:3} and \ref{tab:4}, we list values of $T_\mathrm{Y}$ and $T_\mathrm{L}$ assuming that $t_{\mathrm{evol}}=4.6$ $Gyr$. However, it should be noted that our models only calculate ratios of periods and that all the modeled values of $T_\mathrm{Y}$ and $T_\mathrm{L}$ scale with $t_{\mathrm{evol}}$.  More details of our methods are given in Appendix \ref{appendixB}. 
   
Binning the data in H and plotting <$I$> vs. $H$, implies that we are effectively tracking the inclinations of asteroids that have one particular size. Given the assumption that the initial asteroid distribution was uniform in $a-I$ space, that is, that both the initial SFD and the initial number density in $a-I$ space were independent of both $a$ and $I$, it follows that the modeled variation of <$I$> with $H$ (with the data binned in $H$ – Fig. \ref{fig:8}a) and the modeled variation of <$H$> with $I$ (with the data binned in $I$ – Fig. \ref{fig:8}b) are independent of the SFD and depend only on the Yarkovsky timescale $T_\mathrm{Y}$ and the parameter $\alpha$ shown in equation~(\ref{eq:6}). By minimizing $\chi_\nu^2$ in the plot of <$I$> vs. $H$, (Fig. \ref{fig:8}a), we obtain the best-fit $T_\mathrm{Y}$ as a function of $\alpha$ and find from the observations that $\alpha$ is close to unity (Table \ref{tab:3}). Assuming that $\alpha$ is unity, we find that $T_\mathrm{Y}/t_{\mathrm{evol}}= 2.9\pm0.3$, and that if $t_{\mathrm{evol}}=4.6$ $Gyr$, then the best-fit $T_\mathrm{Y}=13.4\pm1.3$ $Gyr$. The quoted uncertainty is that which makes $\chi_\nu^2$ 10\% greater than the minimum. 

The plots of the modeled SFD in Fig. \ref{fig:8} depend on both Yarkovsky orbital evolution and the other loss mechanisms. By setting $\alpha=1.0$, $t_{\mathrm{Y}}=13.4$ $Gyr$, $t_{\mathrm{evol}}=4.6$ $Gyr$ (that is, $T_\mathrm{Y}/t_{\mathrm{evol}}= 2.9$) and using various values of $\beta$, we can remove the <$I$> vs. $H$ plot from our $\chi_\nu^2$ analysis and focus on the roles of $T_\mathrm{L}$ and $b$. In Table \ref{tab:4} and Fig. \ref{fig:8} we show our model results for $\beta=0.5$, 1.0 and 2.0. The fit for the SFD for $H<16.5$ is largely insensitive to the value of $\beta$. However, after more asteroid discoveries have been made and the completeness limit has been extended to, say, $H=18$, that will not be the case and information will then be available on the other loss mechanisms. Assuming that $\beta=2$ gives the best fit, the minimum $\chi_\nu^2$ occurs for $b=0.55$ and $T_\mathrm{L}/T_{\mathrm{Y}}= 0.16\pm0.03$. If $t_{\mathrm{evol}}=4.6$ $Gyr$, then the best-fit $T_\mathrm{L}=2.2\pm0.3$ $Gyr$ (Table \ref{tab:4}). Again, the uncertainty is that which makes $\chi_\nu^2$ 10\% greater than the minimum. The model results in Fig. \ref{fig:8}g show that for $H<16.5$, Yarkovsky-driven loss through the resonant escape hatches is the dominant asteroid loss mechanism.  We also note that the value of $b$ ($=0.55$) that minimizes $\chi_\nu^2$ is close to the critical value of 0.6 for which the SFD has equal mass in each size interval, $dD$ \citep{DurdaDermott1997}. 

The footprints of Yarkovsky forces in the asteroid belt produced by the transport and loss of asteroids are (1) the V-shaped distribution of family asteroids in semimajor axis and inverse diameter ($1/D$) space and (2) the sharp truncation of some of the families at the major Kirkwood gaps \citep{FarinellaVokrouhlicky1999,Bottke.et.al2002a}. However, these pivotal observations do not establish an orbital evolution timescale and the ages of the families can only be inferred by using estimates of the strengths of the Yarkovsky forces or by extrapolating from the NEA orbital evolution observations. Our observation that the mean size of the surviving asteroids in the IMB increases with increasing inclination is therefore useful because we are able to interpret this size-inclination correlation in terms of the ratios of timescales: the Yarkovsky timescale, $T_{\mathrm{Y}}$ defined by equation (\ref{eq:6}), the other timescale, $T_{\mathrm{L}}$ defined by equation (\ref{eq:14}), and the time of orbital evolution, $t_{\mathrm{evol}}$. 

\begin{figure*}
    \centering
	\includegraphics[width=0.32\textwidth]{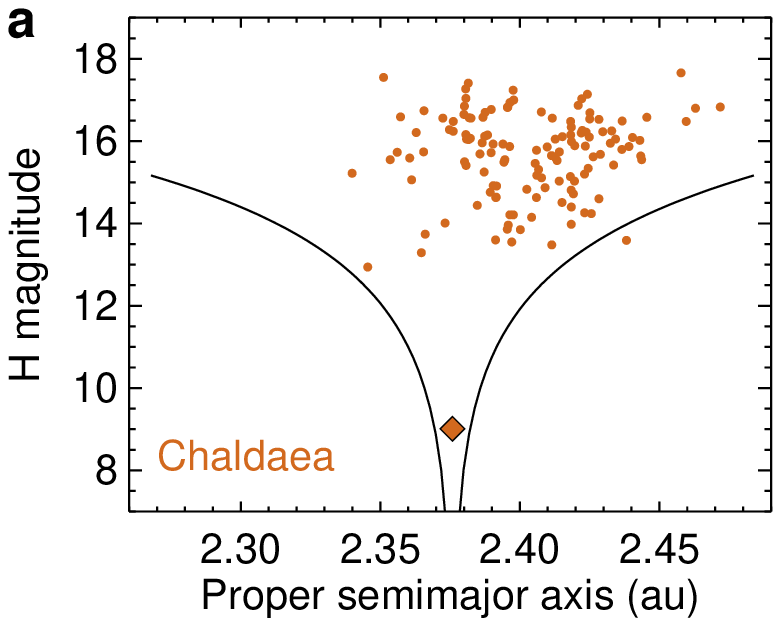}
	\includegraphics[width=0.32\textwidth]{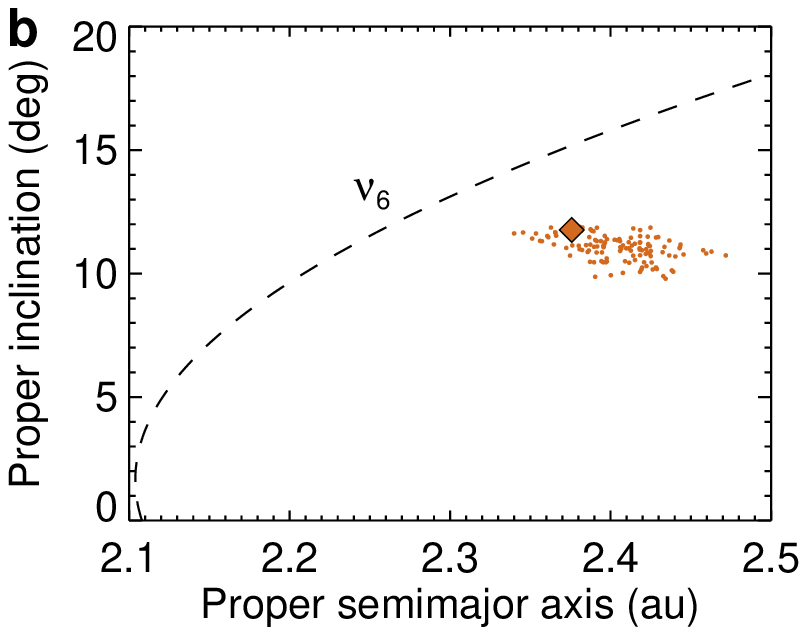}
	\includegraphics[width=0.32\textwidth]{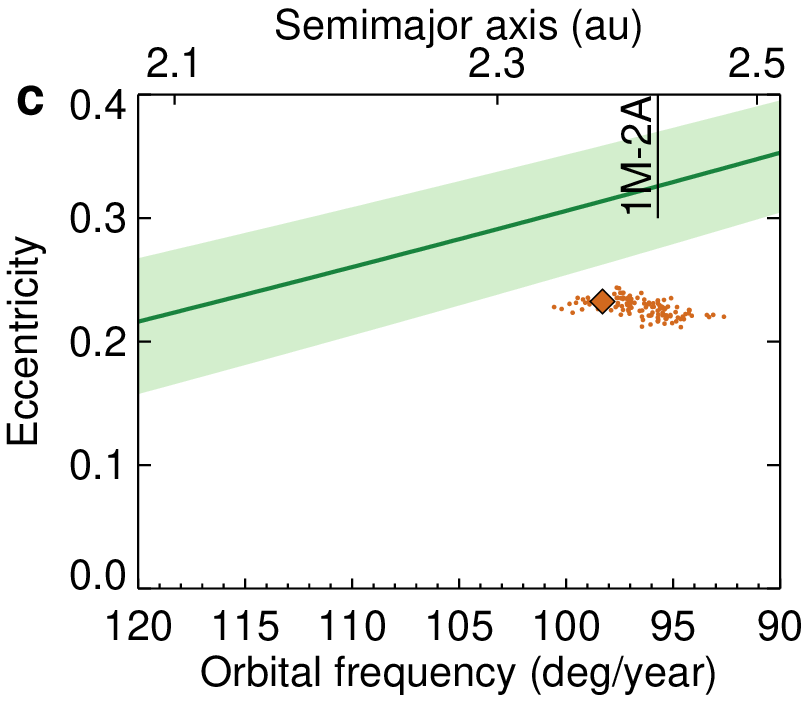}
    \caption{Panels A, B and C: Distributions of the absolute magnitude, proper inclination, and proper eccentricity of the asteroids in the Chaldea family. }
    \label{fig:9}
\end{figure*}

If we now assume that the asteroids in our data set are primordial with sizes unchanged since the time of their formation, then $t_{\mathrm{evol}}=4.6$ $Gyr$ and $T_\mathrm{Y}=13.4\pm1.3$ $Gyr$. This timescale is 3.4 times larger than the 4 $Gyr$ timescale derived from observations of the NEAs \citep{Greenberg.et.al2020}. The discrepancy is even larger if we reduce our estimates of the lengths of the asteroid escape routes. We have assumed that the length, $L$ of the escape route is determined by the exact location of the $\nu_6$ resonance. In Fig. \ref{fig:9}, we show the asymmetrical distribution of the asteroids in the IMB Chaldaea family. \citet{Morate2019} suggest that this marked asymmetry is due to the loss of those asteroids that while not located at the exact location of the $\nu_6$ resonance were sufficiently close to that resonance to ensure their removal. This is consistent with the estimate of \citet{MorbidelliVokrouhlicky2003} that asteroids are lost from this secular resonance at locations given by
\begin{equation}
    \nu_6(a)=2.12+6.003\mathrm{sin}(I)^{2.256} \,au.
	\label{eq:15}
\end{equation}
cf. equation (\ref{eq:6}). Use of equation (\ref{eq:15}) in the determination of $L$ results in a reduction in L of $\sim$0.2 and a corresponding increase in $T_\mathrm{Y}$. Thus, our estimate of $T_\mathrm{Y}$ based on $t_{\mathrm{evol}}=4.6$ $Gyr$ could be $\sim$4 times larger than the NEA estimate. 

One possibility for this unacceptably large difference is that the NEA timescale describes changes over short periods of time and assumes, appropriately, that the orbital evolution rates are constant. However, over times longer than $\sim$10$^8$ $yr$ we must allow for changes in the spin directions of the asteroids and the consequent changes in the signs of $da/dt$. Orbital evolution over long timescales could be more like a random walk and this would result in a large increase in our estimate of the orbital evolution timescale \citep{Bottke2015}. A second possibility is that the small asteroids have experienced significant mass loss due to erosion and therefore their masses have steadily decreased with time \citep{Holsapple2020}. A third possibility is that the time of evolution, $t_{\mathrm{evol}}$ is effectively less than the age of the solar system. This would be the case if the majority of asteroids in our data set are collision products. In our models, the time of evolution, $t_{\mathrm{evol}}$ would then refer to the average time over which the asteroids have had their current sizes. The most numerous asteroids in our data set have $H=16.5$, corresponding to diameters of 2.5 $km$ and 1.4 $km$ for asteroids with albedos of, respectively, 0.07 (CC) and 0.24 (NC). The collision lifetimes of the asteroids are uncertain, but given that small asteroids with $1<D<10$ $km$ are probably rubble-piles it is unlikely that these small asteroids are primordial \citep{Holsapple2020}. While we do not discount the first two possibilities, and indeed consider that all three possibilities are probably operative, we note that the third possibility implies that the small asteroids in our data set must be members of ghost families.

\section{Evidence for ghost families}

\begin{figure*}
    \centering
	\includegraphics[width=0.32\textwidth]{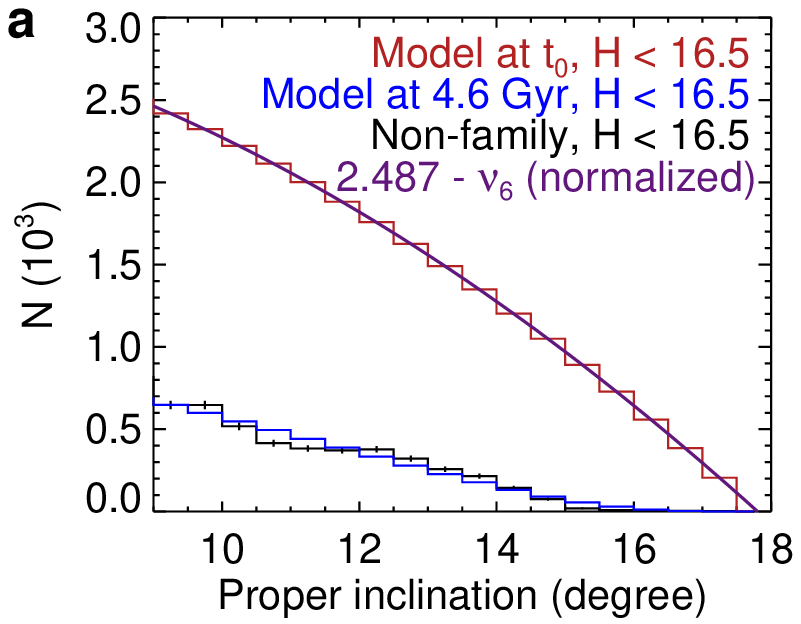}
	\includegraphics[width=0.32\textwidth]{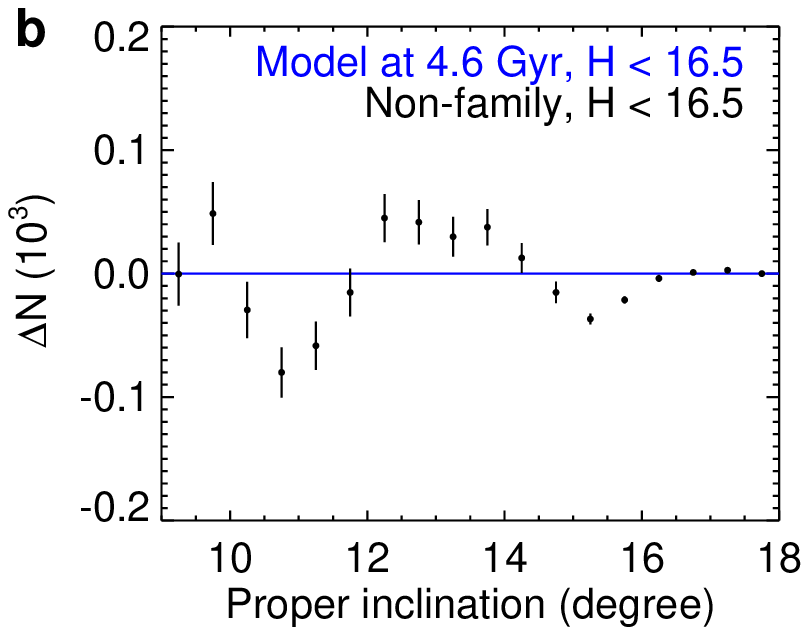}
	\includegraphics[width=0.32\textwidth]{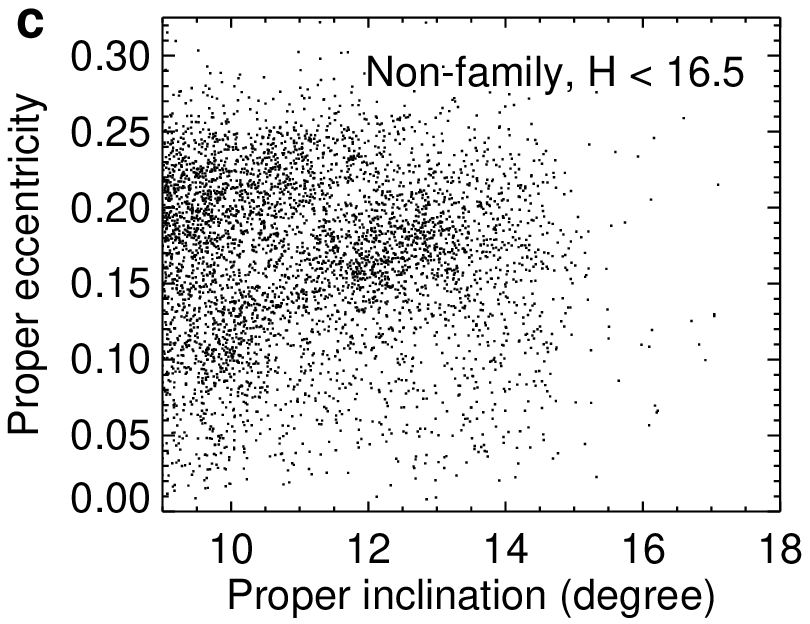}
    \caption{Panel A: The black histogram shows the observed distribution of the high-inclination ($I>9$ $deg$) non-family asteroids with $H<16.5$. The red histogram shows the initial distribution of the high-inclination asteroids in the model shown in Fig. \ref{fig:8}a. The initial distribution of the model is set by the shape of the $\nu_6$ secular resonance (the smooth, normalized curve). Specifically, the total number of asteroids at a given inclination is proportional to the distance between the $\nu_6$ secular resonance and the 3:1 Jovian mean motion resonance. The blue histogram shows the distribution of the high-inclination ($I>9$ $deg$) non-family asteroids after 4.6 $Gyr$ of evolution. Panel B: The difference between the predicted and observed distributions of the high-inclination ($I>9$ $deg$) non-family asteroids after 4.6 $Gyr$ of evolution. The magnitude of the largest deviation is 16\% of the expected number. Panel C: A scatter plot of the proper $e$ and the proper $I$ of the high-inclination ($I>9$ $deg$) non-family asteroids. }
    \label{fig:10}
\end{figure*}

\begin{figure}
    \centering
	\includegraphics[width=1.0\columnwidth]{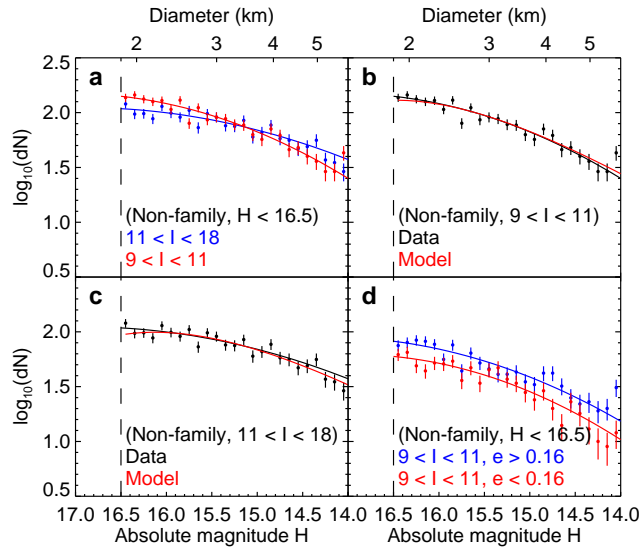}
    \caption{Panel A: SFDs of the high inclination non-family asteroid group are divided into two ranges of inclination: $9<I<11$ $deg$ and $11<I<18$ $deg$. Panels B and C: Comparison of the observed SFDs shown in Panel A with predictions of the model shown in Figure \ref{fig:8}a. Panel D: SFDs of the asteroids with inclinations in the range $9<I<11$ $deg$ divided into low and high eccentricity groups. }
    \label{fig:11}
\end{figure}

\begin{figure}
    \centering
	\includegraphics[width=0.8\columnwidth]{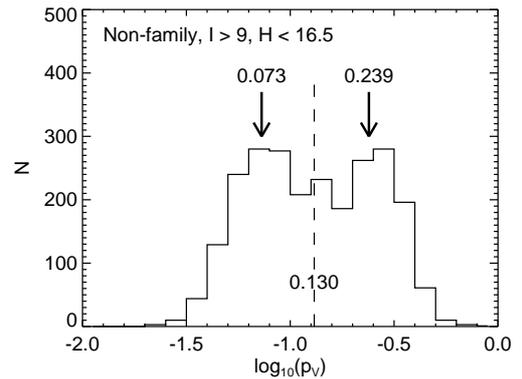}
    \caption{Distribution of the WISE albedos for 2,422 of the 4,400 asteroids in the IMB with $I>9$ $deg$ and $H<16.5$ \citep{Masiero.et.al2014}. The average measured albedo is 0.130. The 1,200 CC asteroids with albedo < 0.130 have an average albedo of 0.073. The 1,222 NC asteroids with albedo > 0.130 have an average albedo of 0.239. }
    \label{fig:12}
\end{figure}

\begin{figure*}
    \centering
	\includegraphics[width=0.36\textwidth]{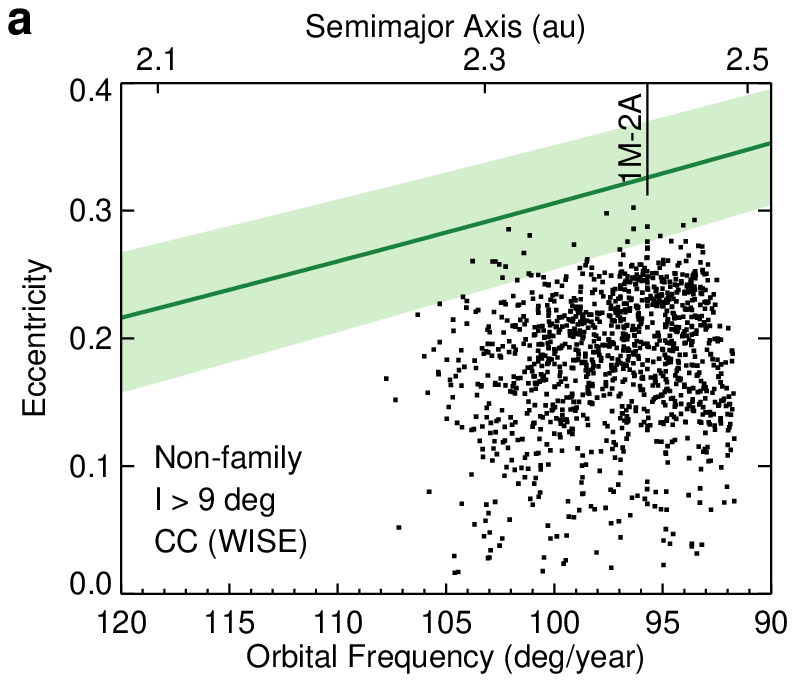}
	\includegraphics[width=0.36\textwidth]{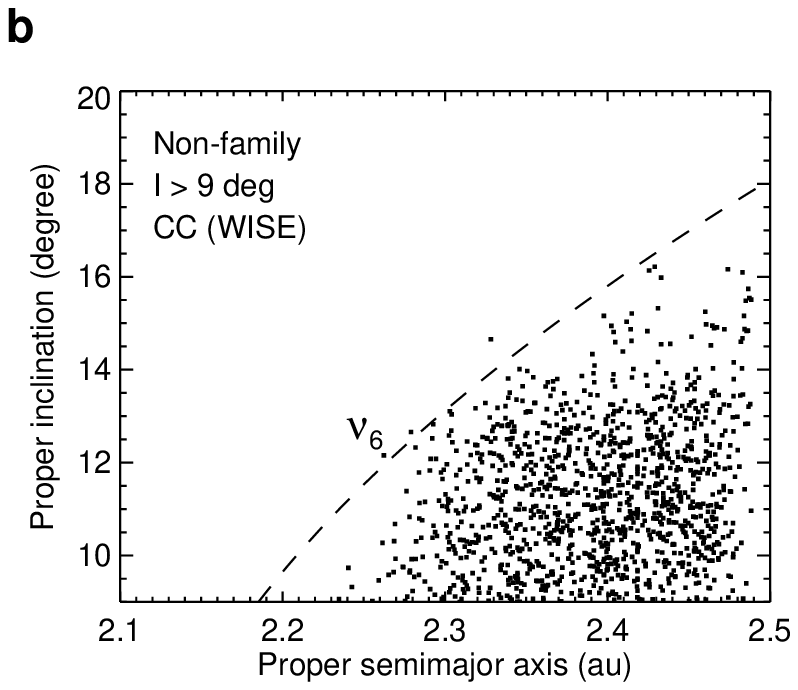}
	\includegraphics[width=0.36\textwidth]{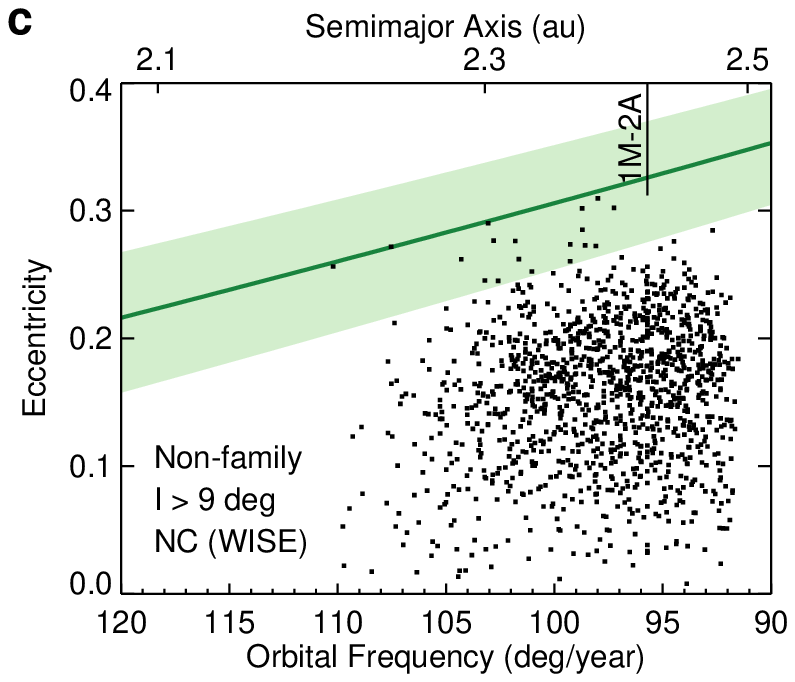}
	\includegraphics[width=0.36\textwidth]{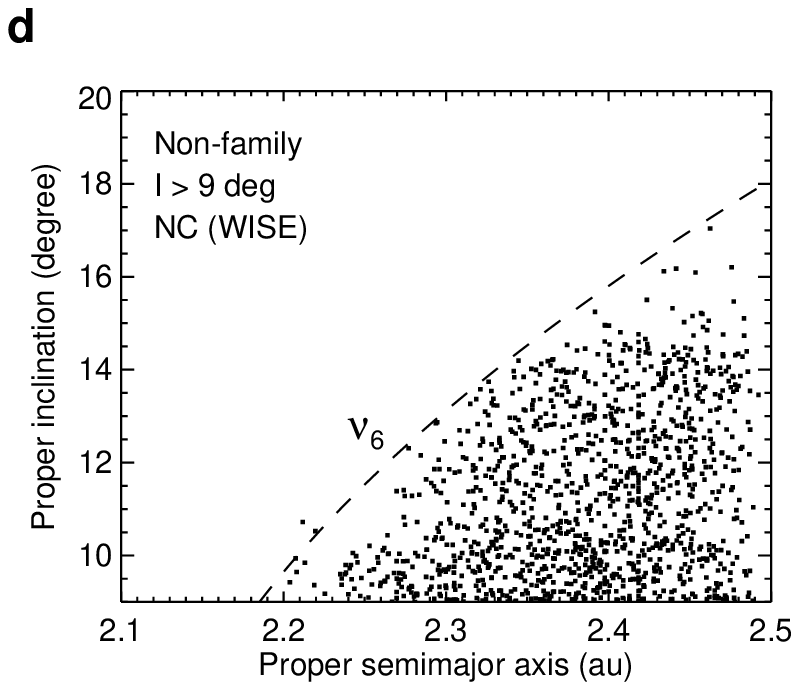}
    \caption{Scatter plots of all the non-family asteroids in the IMB with high inclinations ($I>9$ $deg$), $H<16.5$, and WISE albedos, $A$. Those asteroids with $A<0.13$ are designated as CC, those with $A>0.13$ are designated as NC. The green shaded region is the Mars-crossing zone and $\nu_6$ is the secular resonance. The WISE albedo data is incomplete. Of the 4,400 high inclination asteroids in the IMB, 1,200 are CC, 1,222 are NC and thus 45\% are undesignated. }
    \label{fig:13}
\end{figure*}

\begin{figure}
    \centering
	\includegraphics[width=1.0\columnwidth]{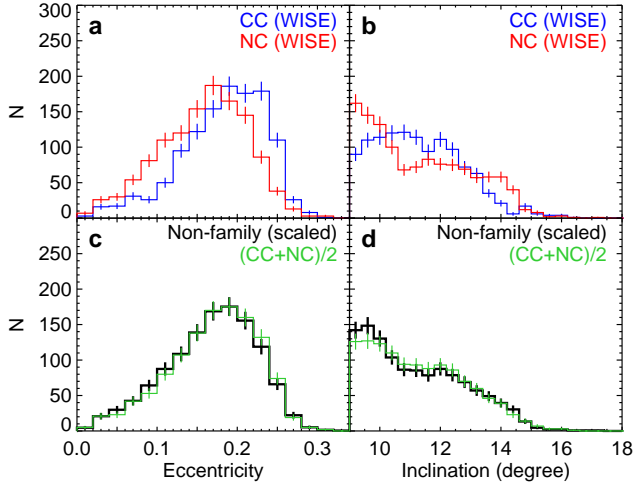}
    \caption{Panels A and B compare the $e$ and $I$ distributions of those high inclination ($I>9$ $deg$) asteroids with WISE albedos \citep{Masiero.et.al2014}. These 2,422 asteroids with $I>9$ $deg$ and $H<16.5$ have been divided by their albedos into two near-equal number groups: CC ($A<0.13$) and NC ($A>0.13$). Panels C and D compare the combined $e$ and $I$ distributions of these CC and NC asteroids with those distributions of the 4,400 asteroids in the observationally complete set, scaled so that the distributions have equal numbers. }
    \label{fig:14}
\end{figure}

\begin{figure}
    \centering
	\includegraphics[width=0.8\columnwidth]{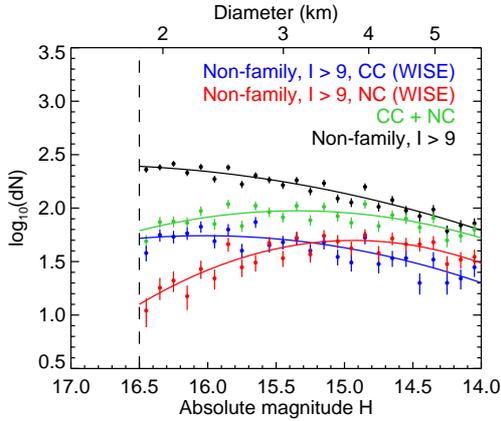}
    \caption{SFDs of the observationally complete set of high inclination, non-family  asteroids with $H<16.5$ and $I>9$ $deg$ and the SFDs of those asteroids in the set with WISE albedos. The blue curve is the SFD of the CC asteroids ($A<0.13$). The red curve is the SFD of the NC asteroids ($A>0.13$). The green curve shows the sum of the CC and NC data. }
    \label{fig:15}
\end{figure}

\begin{figure}
    \centering
	\includegraphics[width=0.49\columnwidth]{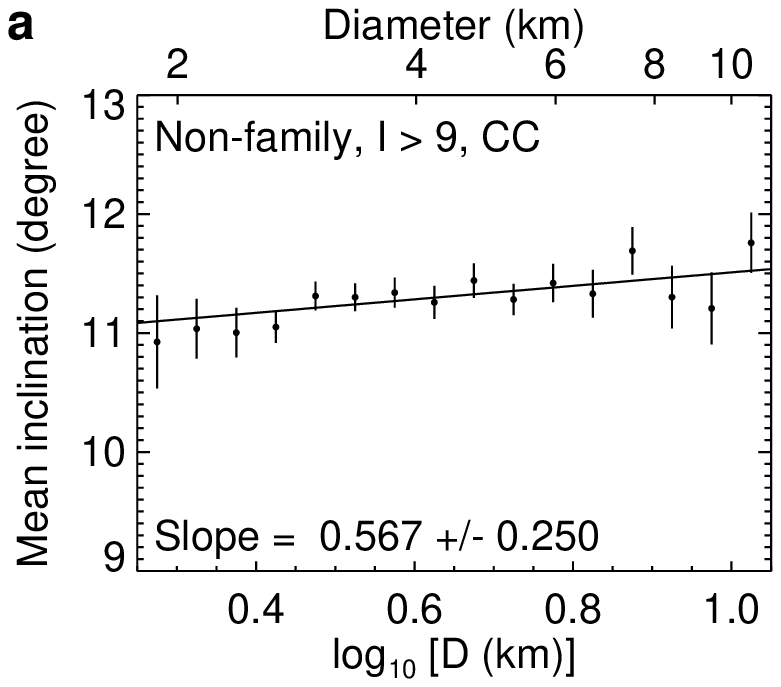}
	\includegraphics[width=0.49\columnwidth]{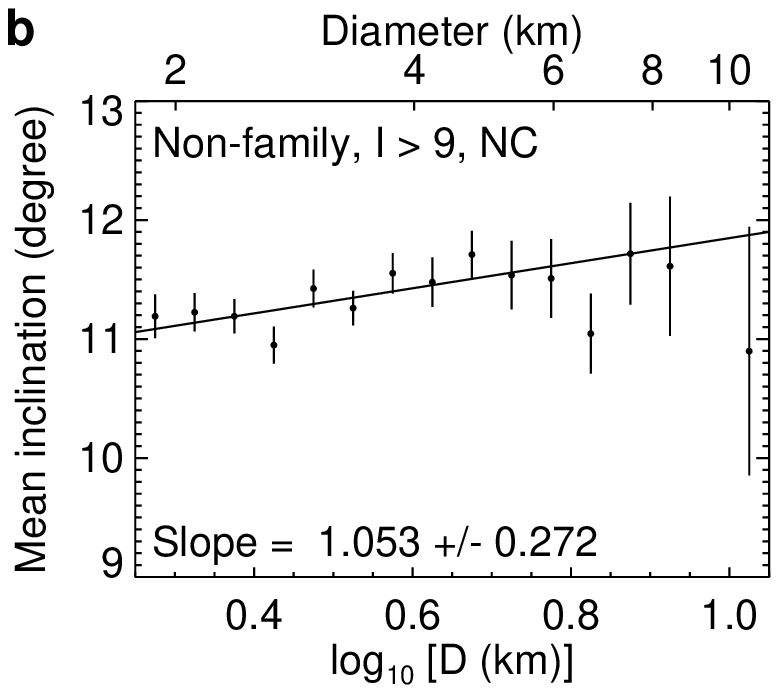}
	\includegraphics[width=0.49\columnwidth]{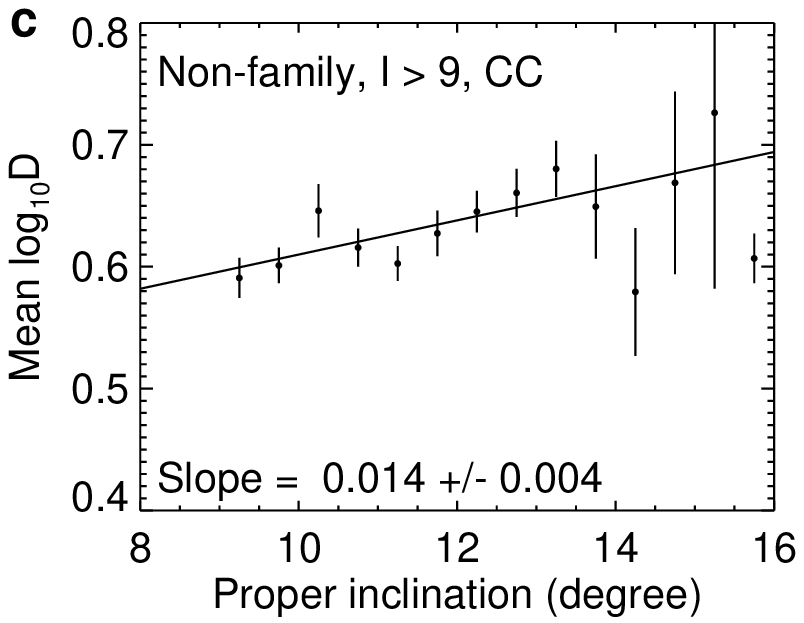}
	\includegraphics[width=0.49\columnwidth]{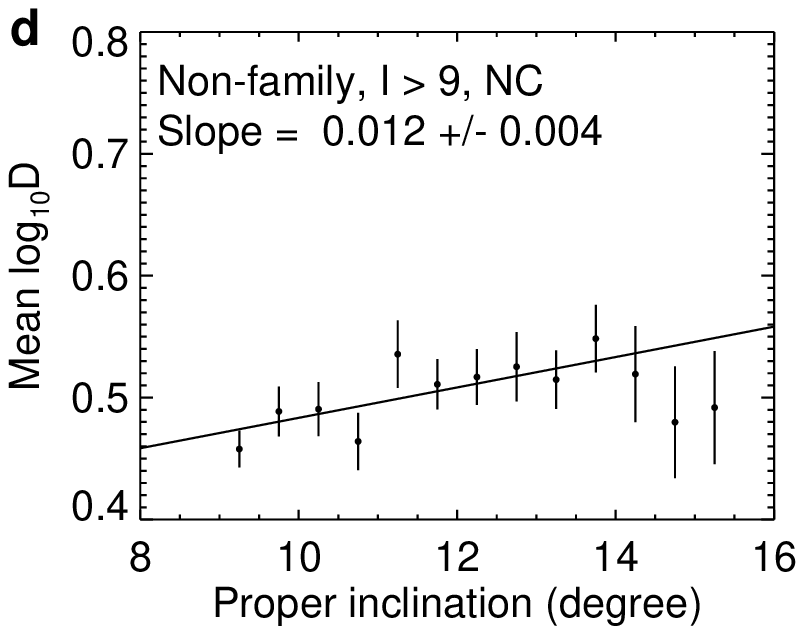}
    \caption{These plots are similar to those shown in Fig. \ref{fig:7}, but here we have divided the data for the high inclination, non-family asteroids into CC and NC groups as determined by their WISE albedos. Panels A and B: Variation of the mean proper inclination with asteroid diameter, $D$. The data is shown binned in log$D$, but the slope has been determined from the individual points in the range $16.5<H<13.5$. Panel A shows the CC data and Panel B shows the NC data. Panels C and D: Variation of the mean asteroid diameter, $D$ with proper inclination, $I$. In contrast to panels A and B, the data in panels C and D are shown binned in proper inclination.}
    \label{fig:16}
\end{figure}

\begin{figure}
    \centering      
	\includegraphics[width=1.0\columnwidth]{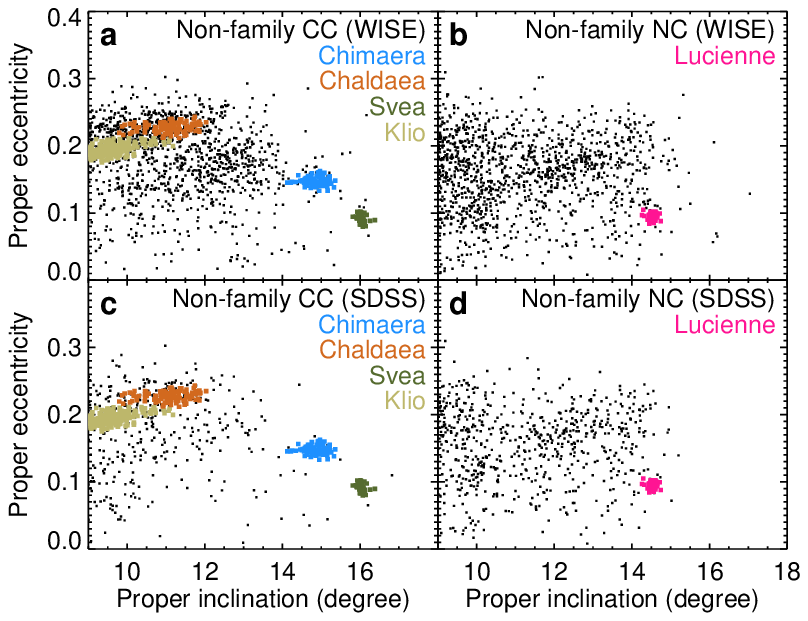}
    \caption{Panels A and B: $a,e$ and $a,I$ scatter plots of all the asteroids in the IMB with $H<16.5$ and $I>9$ $deg$. The family asteroids are colour-coded. The non-family asteroids have been divided into two near-equal number groups of CC and NC asteroids according to their WISE albedos \citep{Masiero.et.al2014}.  Panels C and D show similar scatter plots for asteroids divided into CC and NC groups according to their SDSS colours \citep{Parker2008}. }
    \label{fig:17}
\end{figure}

Our explanation for the size-inclination correlation observed in the IMB is based on the unique dynamics of the IMB and assumes an asteroid distribution that was initially uniform in $a-I$ space. In Fig. \ref{fig:10}a, we test this assumption by comparing the initial and final inclination distributions implied by the model shown in Fig. \ref{fig:8}g ($\beta=2$). We note that this model accounts for the near-total loss of those asteroids with $I>15$ $deg$ and that 80\% of the other asteroids in the initial distribution (with $H<16.5$) have also been lost. In Fig, \ref{fig:10}b, we show that the largest observed deviation from the predicted final distribution is 16\% and while this deviation is small, supporting the assumption of initial uniformity, it is not negligible. The scatter plot of the proper eccentricities and inclinations of the asteroids in our complete data set, Fig. \ref{fig:10}c, shows strong hints of non-uniformity, with possible ghost families located at $I$$\sim$12.5 $deg$ and $I$$\sim$9.5 $deg$ that could account for the deviations shown in Fig. \ref{fig:10}b. 

If these putative ghost families were initially clustered in $a-e-I$ space, then any clustering in the semimajor axes would have been lost through large-scale orbital evolution. Equation (\ref{eq:3}) shows that the eccentricities do not significantly affect the rates of change of the semimajor axes and therefore any eccentricity clustering cannot affect the size-inclination correlation. This is supported by the plots shown in Fig. \ref{fig:11}. In Fig. \ref{fig:11}a, we have divided our data set into two inclination ranges ($9<I<11$ $deg$) and ($11<I<18$ $deg$). The SFDs of both groups shown in Figs. \ref{fig:11}b and \ref{fig:11}c are well accounted for by the model shown in Fig. \ref{fig:8}g ($\beta=2$) and, as expected, the higher inclination group shows the largest loss of small asteroids. The asteroids in the lower inclination group have a wide range of eccentricities (see Fig. \ref{fig:10}c). In Fig. \ref{fig:11}d, we divide the asteroids in this group into low and high-eccentricity groups and show that the observed SFDs have no significant differences. Thus, while it is possible, and probably likely, that the asteroids in our data set are members of  ghost families, our model for the loss of the small asteroids from those families, that is based on an initially uniform distribution of inclinations, could be a good first-order approximation. 

If all the asteroids in the IMB that do not derive from the known families are members of old ghost families and originate from the catastrophic disruption of a few large asteroids, then separating the asteroids into CC and NC groups and analyzing their orbital element distributions could yield some supporting evidence for the existence of these ghost families. Approximately half (2,422 out of 4,400) of the asteroids in our data set have measured WISE albedos \citep{Masiero.et.al2014}. The distribution of these albedos shown in Fig. \ref{fig:12} is bimodal with a mean albedo, $A\simeq0.13$ and the asteroids can be divided into two numerically near-equal groups of CC asteroids (1,200 with $A<0.13$) and NC asteroids (1,222 with $A>0.13$). The distributions of the eccentricities and inclinations of these two groups are shown in Figs. \ref{fig:13} and \ref{fig:14}. The eccentricity and inclination distributions of the CC and NC groups are both significantly different. The difference in the mean eccentricities of the CC and NC asteroids shown in Fig. \ref{fig:14}a is $0.0249\pm0.0022$, a difference that is significant at the 11$\sigma$ level. However, because the WISE data are observationally incomplete, we must examine whether observational bias could be responsible for these observed orbital element differences. The asteroids in our data set have $H<16.5$ and therefore the data set, as a whole, is devoid of any observational bias. In Figs. \ref{fig:14}b and \ref{fig:14}d we compare the distributions of the eccentricities and inclinations of the combined CC and NC asteroids, that comprise about half of our data set, with those of the complete data set. The differences of the eccentricity and the inclination distributions of the complete and the incomplete data sets, when normalized to have equal weight, are negligible, implying that observational bias does not account for the large orbital element differences shown in Figs \ref{fig:14}a and \ref{fig:14}c. Given that the CC and the NC asteroids were scattered into the present asteroid belt from reservoirs with different heliocentric distances, we might expect the two groups to have different mean eccentricities and if that were the only observation to be considered we could argue that it supports the idea of formation in two separate reservoirs. However, the observed inclination distributions do not have a simple mean difference and we conclude from this that the observed orbital element differences favour a ghost family origin. Other evidence for the existence of old asteroid families in the IMB with highly dispersed orbital eccentricities and inclinations has been presented by \citet{Delbo2017} and \citet{Delbo.et.al2019}.  

Following \citet{Ivezic2002}, we have used the Sloane Digital Sky Survey (SDSS) colours, $a^*$ and $i-z$, to divide the asteroids into CC and NC groups. CC asteroids have $a^*<0$ and NC asteroids have $a^*>0$ and $i-z<-0.15$. However, the SDSS asteroid colour data are severely incomplete, only 364 CC and 649 NC asteroids of the 4,400 asteroids in our data set have SDSS colors and this incomplete data is not useful for statistical purposes. The SFDs of the CC and NC groups, as determined from the WISE albedo data alone, are shown in Fig. \ref{fig:15}. While both the CC and NC groups show a lack of small asteroids, both groups are also observationally incomplete and therefore, at present, we cannot make any useful deductions. In Fig. \ref{fig:16}, using the WISE albedo data alone both to separate the asteroids into CC and NC groups and to determine their diameters, we show the separate diameter-inclination correlations of the two groups. The observed correlations are consistent with those of the absolute magnitude-inclination correlations of the complete data set shown in Fig. \ref{fig:7}. However, given that the WISE CC and NC data are incomplete we cannot, at present, use these plots to determine either the ages or the thermal properties of the separate asteroid groups. In Fig. \ref{fig:17}, separate scatter plots of the eccentricities and inclinations of those asteroids with WISE albedos and those with SDSS colours show that the $I>15$ deg asteroid desert is clearly present in both the CC and NC groups. Most of the asteroids in the CC group with $I>14$ $deg$ reside in the Chimaera and Svea families. Given the low $e$ and $I$ dispersions of these two families, it is likely that most of the asteroids with these high inclinations are only present because they are recent ($<10^8$ $yr$) collision products. In Section \ref{section7}, we argue that the degree of loss of small asteroids is a measure of the family age. 

\section{Size-eccentricity correlations}

\begin{figure}
    \centering
	\includegraphics[width=0.8\columnwidth]{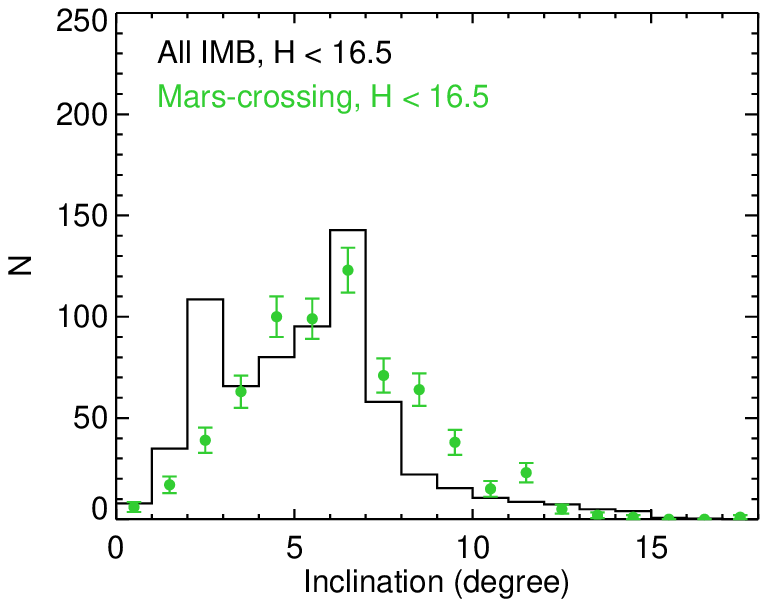}
    \caption{Histograms of the inclination distributions of (a) all asteroids in the IMB with $H<16.5$ and (b) the asteroids in the Mars-crossing zone shown in Fig. \ref{fig:2}d. }
    \label{fig:18}
\end{figure}

\begin{figure}
    \centering
	\includegraphics[width=0.8\columnwidth]{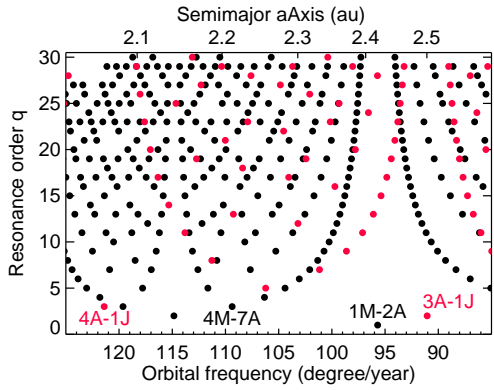}
    \caption{Distribution of all the 2-body mean motion resonances in the inner main belt with orders $\leq30$ involving an asteroid, Mars or Jupiter. A, M and J denote the mean motions of an asteroid, Mars and Jupiter. Red and black points refer, respectively, to Jovian and Martian resonances. }
    \label{fig:19}
\end{figure}

\begin{figure*}
    \centering
	\includegraphics{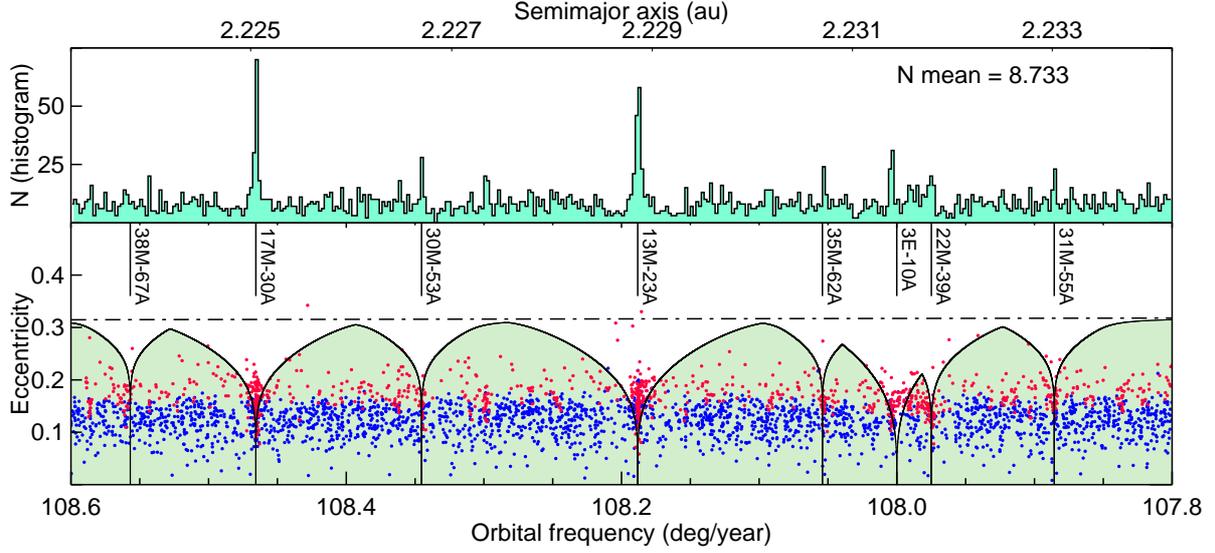}
    \caption{The top panel shows a high-resolution histogram of IMB asteroids with $H<16.5$ and semimajor axes spanning a narrow range of 0.011 $au$. The lower panel is an $e,a$ scatter plot of the asteroids in that range. The colours indicate the maximum Lyapunov Characteristic Exponents, mLCE of the orbits, as calculated by \citet{KnezevicMilani2007}, with red being the most chaotic (mLCE>0.00012 year$^{-1}$) and blue the most stable (mLCE<0.00012 year$^{-1}$). All the Martian mean motion resonances of order, $q\leq30$ are shown and labelled. For example, at $a=2.225$ $au$, we show the 30A-17M resonance, where A is the asteroid mean motion and M is the mean motion of Mars. We also show the location of the 10A-3E resonance involving the mean motion of Earth, E. The curves that bracket each of these resonances represent the maximum libration widths associated with the strongest eccentricity term in the full disturbing function. The dashed line denotes the asteroid eccentricity  ($e=0.31$) needed for the asteroid to cross the orbit of Mars when the eccentricity of Mars is zero. }
    \label{fig:20}
\end{figure*}

\begin{figure*}
    \centering
	\includegraphics[width=0.32\textwidth]{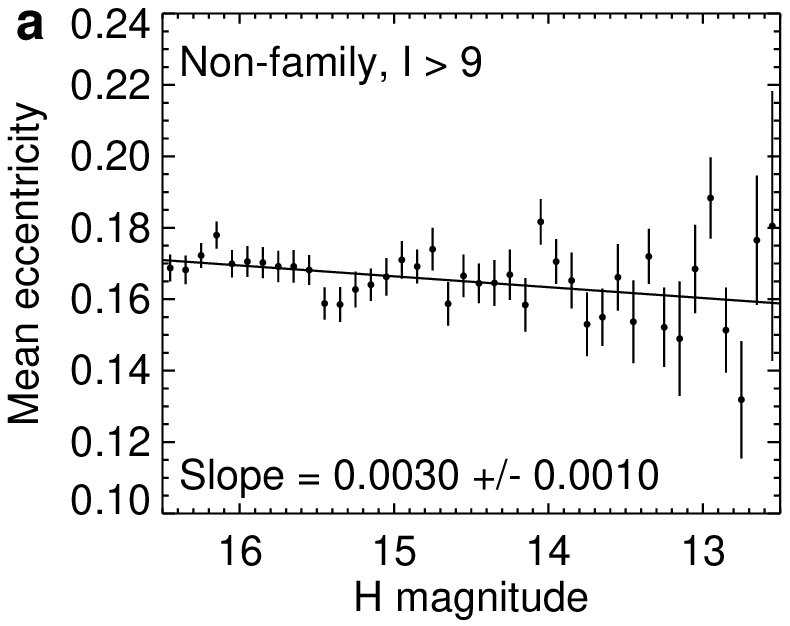}
	\includegraphics[width=0.32\textwidth]{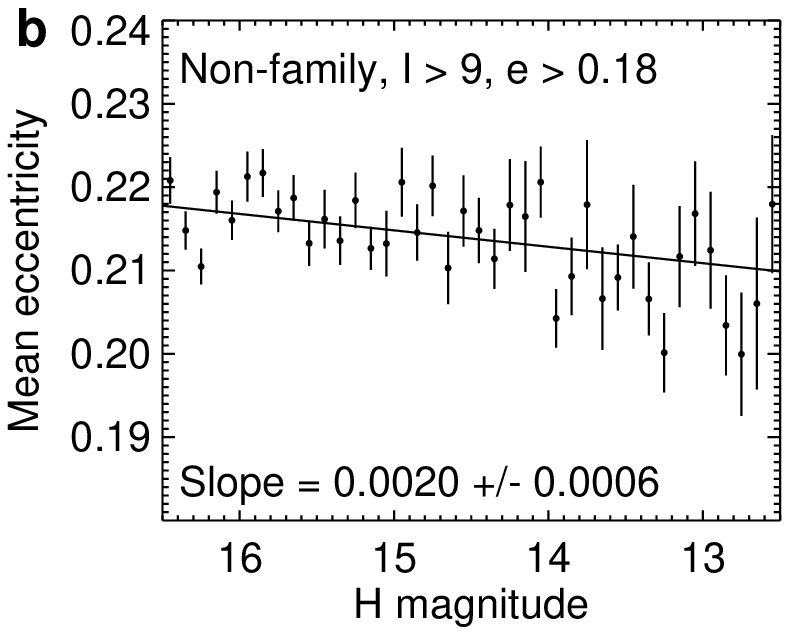}
	\includegraphics[width=0.32\textwidth]{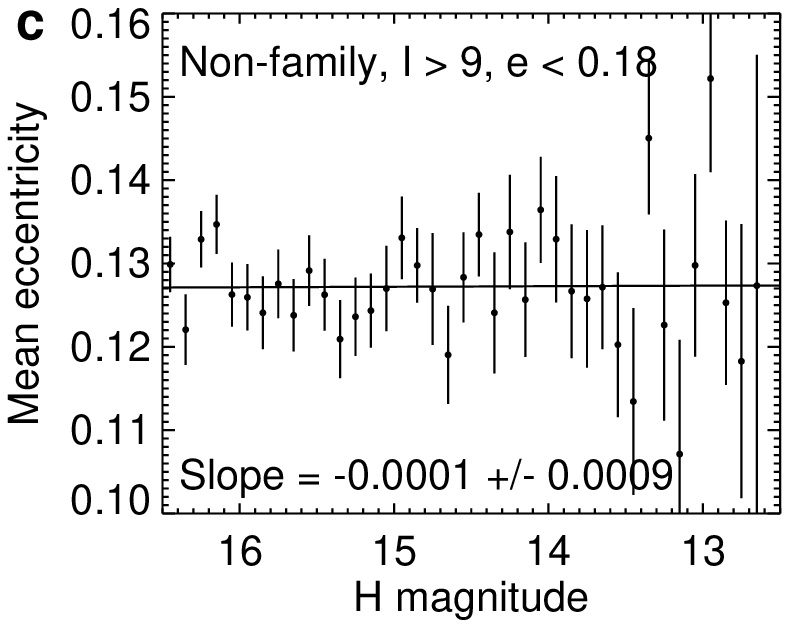}
    \caption{Panel A: Variation of the mean proper eccentricity of the high inclination ($I>9$ $deg$), non-family asteroids with absolute magnitude, $H$. The data is shown binned in $H$, but the slope has been determined from the individual points in the range $16.5<H<13.5$. Panels B and C: The data shown in Panel A have been divided into two groups according to their eccentricities with $e>0.18$ in Panel B and $e<0.18$ in Panel C. }
    \label{fig:21}
\end{figure*}

The observed size-inclination correlation of the asteroids in the IMB is evidence that small asteroids are transported towards the Sun by Yarkovsky forces. As a result of this evolution, asteroids entering the $\nu_6$ secular resonance experience an increase in their eccentricities \citep{Farinella.et.al1994} causing their pericentres to enter the Mars-crossing zone. After scattering by Mars this could result in some asteroids being fully transported to the Mars-crossing zone before eventually being scattered out of the main belt. This could account for the observations, shown in Fig. \ref{fig:3}b and Fig. \ref{fig:6}, that the Mars-crossing zone has an excess of small asteroids. However, other observations of the orbital elements of the asteroids in the Mars-crossing zone show that this may not be the only mechanism that feeds small asteroids into that zone. The distribution of the semimajor axes of the asteroids currently in the zone is shown in Fig. \ref{fig:2}d. The asteroids mostly have semimajor axes in the range $2.15<a<2.45$ $au$. If orbital evolution through the $\nu_6$ resonance was the only mechanism feeding asteroids into the Mars-crossing zone, then given the shape of the locus of the resonance in $a-I$ space (Fig. \ref{fig:2}b), we might expect most of the asteroids in the zone to have inclinations, $I>8$ $deg$. Comparison of the inclinations of the asteroids in the Mars-crossing zone with those in the IMB as a whole, after scaling for equal numbers, shows that this is not the case. In Fig. \ref{fig:18}, we observe that for asteroids in the IMB with $H<16.5$, the ratio of the number of asteroids with inclinations either greater or less than 8 degrees, $N(I>8\,deg)/N(I<8\,deg)$ is $0.124\pm0.013$ while for those asteroids in the Mars-crossing zone alone the ratio is $0.288\pm0.093$. Thus, the number of asteroids in the Mars-crossing zone with $I>8$ $deg$ is excessive and this supports the argument that the $\nu_6$ resonance has a significant role in replenishing the zone. However, Fig. \ref{fig:18} also shows that the majority of asteroids in the zone have inclinations, $I<8$ $deg$ indicating that there are other sources of replenishment. Because the majority of the asteroids in the zone have low inclinations, these other sources are probably associated with the major families and their halos.

In Fig. \ref{fig:2}b, we see that the shape of the distribution of asteroids in a-I space is close to triangular and we have shown that this results in the observed size-inclination correlation. In Fig. \ref{fig:2}d, we see a similar, but less pronounced, triangular distribution of asteroids in $a-e$ space caused by the positive slope of the boundaries of the Mars-crossing zone. Because of this positive slope, the length of the escape route between the 3:1 Jovian mean motion resonance and the Mars-crossing zone decreases with increasing eccentricity. Assuming an initially uniform distribution in $a-e$ space of those non-family asteroids with $I>9$ $deg$ and using the same argument that we used to account for the observed size-inclination correlation, we might expect the mean eccentricity of the remaining asteroids to increase with decreasing $H$. However, the distribution of the mean eccentricities shown in Fig. \ref{fig:21}a displays the opposite trend: the mean eccentricities decrease with decreasing $H$. This correlation is significant at the 3$\sigma$ level and, while not discounting this loss mechanism, we show here that resonant interactions of the asteroids with Mars and Jupiter may provide a more productive mechanism.

Chaos in the IMB arises from a dense web of high-order resonances \citep{Milani2014} and numerical integrations by \citet{MorbidelliNesvorny1999} suggest that these resonances could have a role in the delivery of asteroids to the Mars-crossing-zone. However, their numerical integrations do not include the effects of Yarkovsky forces and a resonance can only exist if it is strong enough to withstand those forces. A mean motion resonance arises if some argument, $\phi$ of the orbital elements librates rather than circulates. For 2-body resonance,
\begin{equation}
    \phi = i\lambda_A + j\lambda_P + r\varpi_A + s\Omega_A + t\varpi_P + u\Omega_P, 
    \label{eq:16}
\end{equation}
where $i$, $j$, $r$, $s$, $t$, $u$ are integers, and $\lambda$, $\varpi$, $\Omega$ denote the mean longitude, the longitude of pericentre and the longitude of node. The subscripts $A$, $P$ refer to an asteroid and to a particular planet ($E$, $M$, $J$ refer to Earth, Mars and Jupiter). Because the orbital frequencies, $\dot{\lambda}=n$ are much greater than the rates of change of the pericentres and nodes, the condition for resonance, $\dot{\phi}\simeq0$ requires that
\begin{equation}
    in_A+jn_P \simeq 0, 
    \label{eq:17}
\end{equation}
where $|i+j|=q$ is the order of the resonance. The distribution of the locations of all the 2-body Martian and Jovian resonances in the IMB with order $q\leq30$ is shown in Fig. \ref{fig:19}. Assuming that $\ddot{\lambda}_A$ is the dominant term in the equation for $\ddot{\phi}$, the equation of motion of the resonant argument reduces to the pendulum equation, 
\begin{equation}
    \ddot{\phi} =\pm\nu^2\mathrm{sin}\phi + \dot{n}_A,
    \label{eq:18}
\end{equation}
where $\nu$ is the libration frequency and $\dot{n}_A$ is the rate of change of the mean motion due to the Yarkovsky forces acting on the asteroid \citep{MurrayDermott1999}. For a resonance to exist, the resonant argument $\phi$ must librate and the sign of $\ddot{\phi}$ must change periodically. Therefore, the Yarkovsky force must satisfy the stability condition
\begin{equation}
    |\dot{n}_A| <\nu^2.
    \label{eq:19}
\end{equation}
The full libration width, $\Delta n_A$ of a resonance is defined by the maximum width of the separatrix  that separates libration from circulation and is given by 
\begin{equation}
    \Delta n_A=4\nu/i
    \label{eq:20}
\end{equation}
\citep{DermottMurray1983}. Using the expression for the Yarkovsky timescale given in equation (\ref{eq:6}), the stability condition, equation (\ref{eq:19}) can be written as
\begin{equation}
    \Delta n_A > \frac{4}{i} \big( \frac{3n_A}{2T_{\mathrm{Y}}(D/1\,km)} \big) ^{1/2}.
    \label{eq:21}
\end{equation}
The ability of a resonance to withstand the Yarkovsky forces is determined solely by the width of the resonance.

Fig. \ref{fig:2}d shows an $a,e$ scatter plot of the IMB non-family asteroids with the points colour-coded according to their maximum Lyapunov Characteristic Exponents as calculated by \citet{KnezevicMilani2007} with red being the most chaotic. Fig. \ref{fig:20}, a high-resolution version of Fig. \ref{fig:2}d, shows that the regions of enhanced chaos are mostly associated with 2-body Martian resonances and that the widths of the chaotic zones expand as the eccentricities increase and the asteroid pericentre distances approach the Mars-crossing zone. The condition for resonance, equation (\ref{eq:17}) implies that in the presence of a drag force the perturbing planet and the asteroid must exchange energy, $E$ and angular momentum, $L$ such that
\begin{equation}
    \frac{1}{n_A} \big( \frac{dn_A}{dt} \big) _{res} = \frac{1}{n_P} \big( \frac{dn_P}{dt} \big) _{res}
    \label{eq:22}
\end{equation}
and the resonance is maintained (we use a subscript $res$ to denote bodies in resonance). High-order resonances are multiplets of many arguments and the disturbing function describing their perturbing potential consists of a large number of terms of comparable strength. For $q=30$, this number is $\sim$$q^{2.3}\approx2,500$. Consequently, analytical methods that use a truncated disturbing function containing a single resonant argument, that may be appropriate for the analysis of resonances involving, for example, the satellites of the major planets, cannot be used for the analysis of the high-order resonances in the IMB. Nevertheless, we can use a truncated disturbing function to estimate the magnitudes of the perturbing forces. 

For a 2-body resonance between an asteroid and a planet, the strongest term in the planet's disturbing function is usually associated with an e-type resonance with an argument
\begin{equation}
    \phi = i\lambda_A + j\lambda_P -(i+j) \varpi_A. 
    \label{eq:23}
\end{equation}
For this resonant argument, the leading term in the disturbing function has the form $S\mathrm{cos}\phi$. To lowest order in eccentricity, 
\begin{equation}
    S = (Gm_p/a_p)\beta f(\alpha) e_A^{|i+j|}
    \label{eq:24}
\end{equation}
and the associated libration frequency is given by
\begin{equation}
    \nu^2 = 3i^2 (\alpha/\beta) |f(\alpha)| (m_P/M_{\odot}) e_A^{|i+j|} n_A^2
    \label{eq:25}
\end{equation}
where $G$ is the gravitational constant, $M_{\odot}$ is the mass of the Sun, $m_P$ is the mass of the perturbing planet, $\alpha(<1)$ is the ratio of the semimajor axes and $f(\alpha)$ is a function of Laplace coefficients. If the orbit of the perturbing planet is external to that of the asteroid, then $\beta=1$, otherwise $\beta=\alpha$. The total energy of this 2-body system is given by
\begin{equation}
    E = -\frac{GM_{\odot}m_P}{2a_P}-\frac{GM_{\odot}m_A}{2a_A}.
    \label{eq:26}
\end{equation}
Before capture into resonance (subscript $nonres$), the Yarkovsky force on the asteroid results in a rate of change of energy of the asteroid, 
\begin{equation}
    \big( \frac{dE}{dt}\big)_{nonres}= -\frac{GM_{\odot}}{3}\frac{m_A}{a_A}\frac{1}{n_A}\big(\frac{dn_A}{dt}\big)_{nonres}.
    \label{eq:27}
\end{equation}
After capture into resonance, energy is exchanged between the asteroid and the planet to maintain the resonance and 
\begin{equation}
    \big( \frac{dE}{dt}\big)_{res}= -\frac{GM_{\odot}}{3}\big(\frac{m_P}{a_P}+\frac{m_A}{a_A}\big)\frac{1}{n_A}\big(\frac{dn_A}{dt}\big)_{res}.
    \label{eq:28}
\end{equation}
Given that rate of change of $E$ before and after capture into resonance is unchanged, and assuming that $m_A/m_P\ll 1$,
\begin{equation}
    \big( \frac{dn_A}{dt}\big)_{res}= \frac{m_A}{m_P}\frac{a_P}{a_A}\big(\frac{dn_A}{dt}\big)_{nonres}.
    \label{eq:29}
\end{equation}
If, for simplicity, we now consider a coplanar system, then given that the rate of change of the total angular momentum of the system,  
\begin{equation}
    L=m_A a_A^2 n_A (1-e_A^2)^{1/2} + m_P a_P^2 n_P (1-e_P^2)^{1/2}
    \label{eq:30}
\end{equation}
before and after capture into resonance is also unchanged, and assuming that $e_A^2\ll 1$, we have
\begin{equation}
    \frac{1}{a_A}\big(\frac{da_A}{dt}\big)_{nonres}\big(1-\frac{n_A}{n_P}\big)=-2e_A^2\frac{1}{e_A}\big(\frac{de_A}{dt}\big)_{res}.
    \label{eq:31}
\end{equation}
This expression depends only on the existence of the resonance and not on the order of the resonance. For an asteroid with an orbit external to that of Mars, $n_A/n_P<1$, hence
\begin{equation}
    sign\big(\frac{de_A}{dt}\big)_{res}=-sign\big(\frac{da_A}{dt}\big)_{nonres}
    \label{eq:32}
\end{equation}
and the eccentricity evolution timescale, $T_e$ is given by
\begin{equation}
    |T_e| = \big(\frac{1}{e_A}\big(\frac{de_A}{dt}\big)_{res}\big)^{-1}=\frac{2e_A^2}{(1-n_A/n_P)}|T_\mathrm{Y}(D/1\,km)|.
    \label{eq:33}
\end{equation}
For $n_A/n_P=13/23$ (see Fig. \ref{fig:20}) and $e_A=0.15$, $|T_e|\approx0.1|T_\mathrm{Y} (D/1\,km)|$. This timescale has been estimated for a coplanar system and is therefore a minimum. For the more realistic non-coplanar case, as long as the system is stable against the action of Yarkovsky forces, the system will hop chaotically between resonant states that involve both the eccentricities and the inclinations \citep{Dermott1988} and this chaotic hopping increases both the eccentricity and the inclination evolution timescales. 

The variations of the libration widths with eccentricity, $e_A$ up to those eccentricities at which neighbouring resonances overlap are shown in Fig. \ref{fig:20}. This figure includes all the 2-body resonances of order, $q\leq30$. The libration widths have been calculated for the coplanar case by numerically averaging the full disturbing function with Mars in a circular orbit. The magnitude of the leading term $S$ in the expansion of the full disturbing function depends on $e_A^q$ and if $q$ is high one might expect the high-order resonances shown in Fig. \ref{fig:20} to be of little dynamical significance. However, this argument neglects the fact that the coefficient $f(\alpha)$ in $S$ increases exponentially with $q$ with the result that the libration widths increase markedly with increasing $e_A$ and become comparable for resonances of all orders as the Mars-crossing zone is approached. Fig. \ref{fig:2}d shows that the orbits just below the Mars-crossing zone in $a-e$ space are highly chaotic. If an asteroid that is evolving either towards or away from the Sun becomes trapped in a mean motion resonance, then the action of the gravitational forces alone results in an increase in the dispersion of the orbital eccentricities (see Appendix \ref{appendixA}). In addition, the Yarkovsky forces driving the small asteroids into the mean motion resonances act, depending on the asteroid spin directions, to either increase or decrease the eccentricities. The asteroid spin directions are critical because if the eccentricity decreases, then the strength of the resonance also decreases leading to the loss of the asteroid from the resonance. For that reason, Yarkovsky forces could have a role in driving small asteroids into high-order mean motion resonances where, on average, their eccentricities increase resulting in transport to the Mars-crossing zone. 

Are the resonances strong enough to withstand the Yarkovsky forces, that is, are the libration widths large enough to satisfy the stability condition given by equation (\ref{eq:21})? If we assume, based on the NEA observations, that the Yarkovsky timescale is $4\times10^9$ $yr$, then a stable resonance could withstand the Yarkovsky force on a 1 $km$ asteroid if it has a libration width $>$0.00027 $deg/yr$ or $4\times 10^{-6}$ $au$. For $e_A\gtrsim0.2$, these widths are small compared to the libration widths shown in Fig. \ref{fig:20}. Meteorite-sized bodies with diameters as small as 1 $m$ would need libration widths larger by a factor $\sqrt{1,000}$, that is widths $\sim$0.0001 $au$, but even these widths are comparable to those shown in Fig. \ref{fig:20}. It may be possible that some of these resonances have had a role in the delivery of meteorites to Earth. A meteorite-sized body with $D$$\sim$1 $m$ would have an eccentricity evolution timescale, $T_e$$\sim$$4\times10^5$ $yr$ which is shorter than the Cosmic Ray Exposure ages of some chondritic meteorites \citep{Eugster2006}. However, we must emphasize that all of these high-order resonances are not stable and an analysis of the dynamics based on a truncated disturbing function is of limited value.  Rather than asking if a resonance can withstand a Yarkovsky force, we need to determine if the time spent in a resonant multiplet is significantly longer than the time needed to traverse that multiplet in the absence of any resonant interactions. If the time spent in the multiplet is longer than the Yarkovsky transit time, then this could result in an increase in the dispersions of the eccentricities and inclinations. Hence, those asteroids with orbits close to the Mars-crossing zone could traverse a large number of high-order resonances before diffusing into the Mars-crossing zone and eventually leaving the main belt. A full numerical analysis of this dynamical flow problem is beyond the scope of this paper. Here, we support the above discussion concerning the possible role of high-order mean motion resonances, and the critical role of the asteroid eccentricity, by showing that the asteroid size-eccentricity correlation shown in Fig. \ref{fig:21}a and Fig. \ref{fig:21}b is a property of the high eccentricity asteroids alone (those with $e_A>0.18$) and does not exist in the low-eccentricity asteroid group (Fig. \ref{fig:21}c). 

\section{The age of the Vesta family}\label{section7}

\begin{table}
	\centering
	\caption{Age of the Vesta family from the two-mechanism model.}
	\label{tab:5}
	\begin{tabular}{ccccccc}
		\hline
		$b$ & $\alpha$ & $T_\mathrm{Y}$ ($Gyr$) & $\beta$ & $T_\mathrm{L}$ ($Gyr$) & $T_{\mathrm{Vesta}}$ ($Gyr$) & $\chi_{\nu}^2$ \\
		\hline
	     0.5 & \multirow{7}{*}{1.0} & \multirow{7}{*}{13.4} & \multirow{7}{*}{2.0} & 7.4 & $3.2^{+0.5} _{-0.5} $ & 14.3\\
		 0.55 & & & & 2.2 & $3.5^{+0.3} _{-0.3} $ & 10.8\\
		 0.6 & & & & 1.3 & $3.7^{+0.2} _{-0.2} $ & 8.15\\
		 0.65 & & & & 1.2 & $4.4^{+0.2} _{-0.1} $ & 6.55\\
		 0.7 & & & & 1.1 & $4.7^{+0.0} _{-0.1} $ & 7.34\\
		 0.75 & & & & 1.0 & $4.8^{+0.0} _{-0.1} $ & 14.9\\
		 0.8 & & & & 0.9 & $4.7^{+0.0} _{-0.1} $ & 35.4\\
		\hline
	\end{tabular}
	\vspace{1ex}
	
     {Note: $\chi_\nu^2$ is calculated in the SFD plot, Fig.~\ref{fig:23}b.}
\end{table}

\begin{figure}
    \centering  
	\includegraphics[width=1.0\columnwidth]{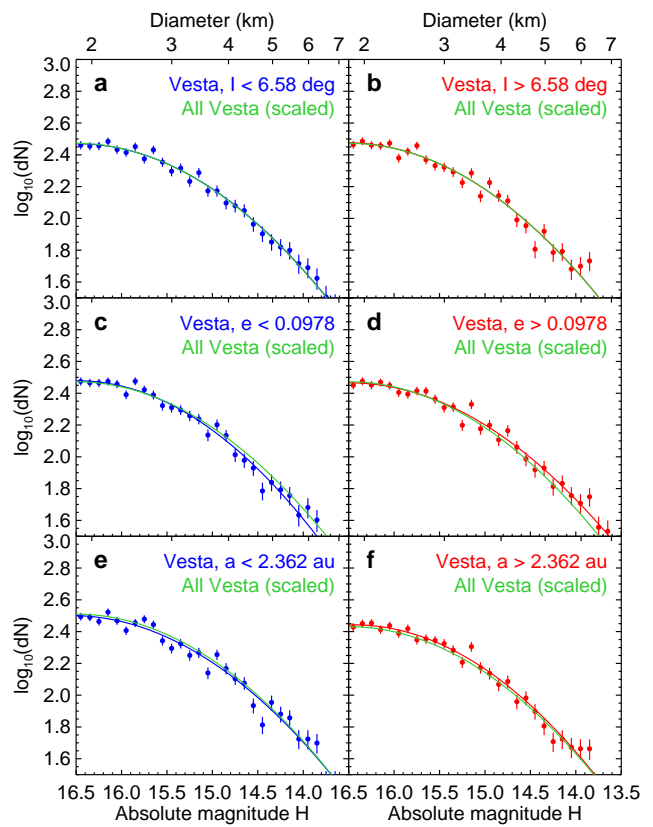}
    \caption{Panel A: Quadratic polynomial fits to the SFDs (with $dH=0.1$) for the low-inclination ($I<6.58$ $deg$) asteroids in the Vesta family compared with the SFD of the Vesta family as a whole after scaling to allow for the different numbers of asteroids in the two groups. Two curves are shown but they are indistinguishable. Panel B: The same plots for the high-inclination ($I>6.58$ $deg$) asteroids in the Vesta family. Panels C and D: Similar plots for the high-eccentricity and the low-eccentricity asteroids in the Vesta family. Panels E and F: Similar plots for the high-semimajor axis and the low-semimajor axis asteroids in the Vesta family. In each case two curves are shown, but they are nearly indistinguishable. }
    \label{fig:22}
\end{figure}

\begin{figure*}
    \centering
	\includegraphics[width=0.72\textwidth]{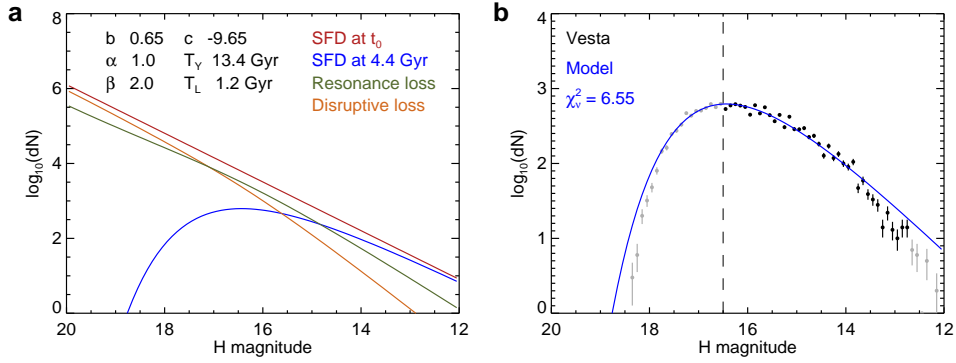}
    \caption{Panel A: Model for the loss of the Vesta family asteroids due to (a) a Yarkovsky force that changes the semimajor axis on a timescale $T_\mathrm{Y}$ and (b) all other loss mechanisms that do not depend on the proper inclination with timescale $T_\mathrm{L}$. $\alpha$ and $\beta$ describe the dependence of these timescales on the asteroid diameter, $D$, $b$ is the slope of the initial SFD and $c$ is a normalizing constant that determines the total number of asteroids in the distribution. The magenta curve shows the SFD at time zero and the blue curve the SFD after 4.4 $Gyr$ of evolution. The green and orange curves show, respectively, the total number of asteroids lost through the resonances and the total number lost through the other loss mechanisms. Panel B: Comparison of the model SFD with the observed SFD. The grey points in panel B are not used in the fit. }
    \label{fig:23}
\end{figure*}

\begin{figure*}
    \centering
	\includegraphics[width=0.36\textwidth]{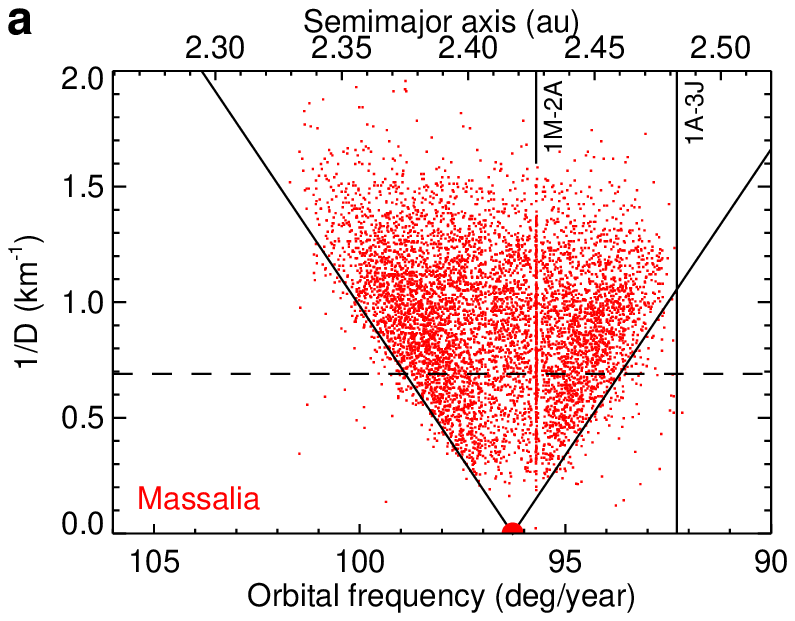}
	\includegraphics[width=0.36\textwidth]{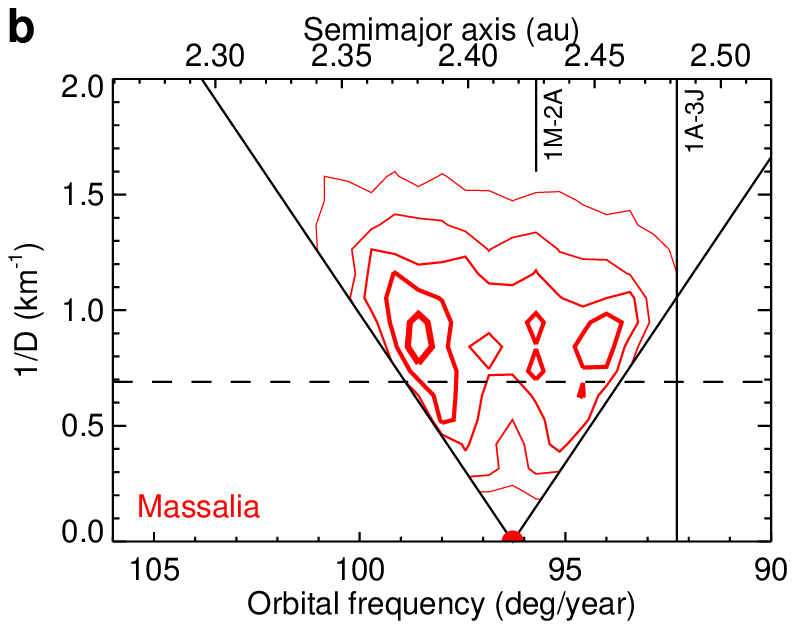}
	\includegraphics[width=0.36\textwidth]{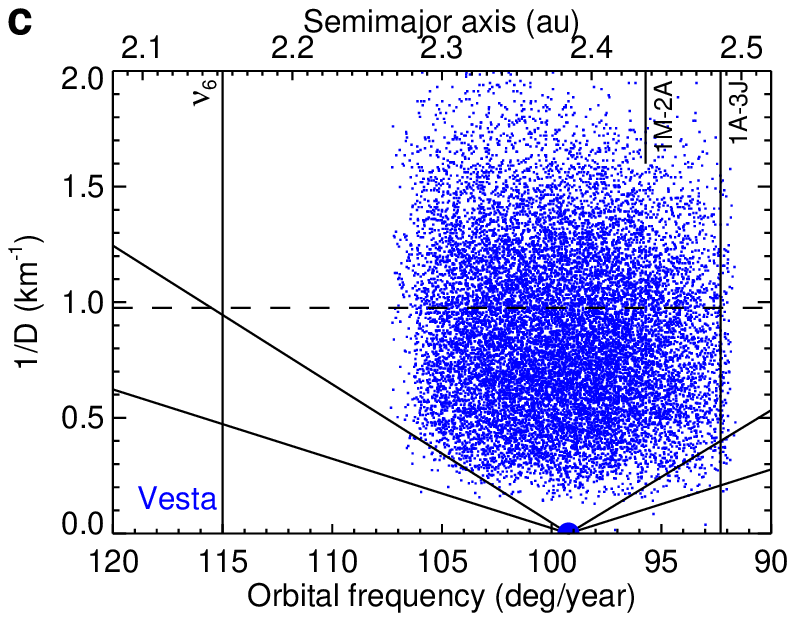}
	\includegraphics[width=0.36\textwidth]{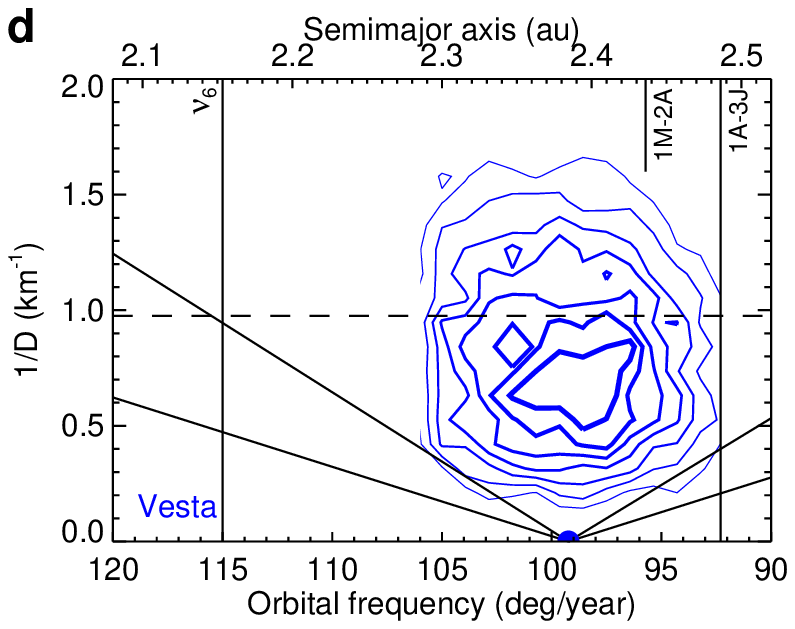}
    \caption{Panels A and C: Distributions in $1/D-a$ space of the asteroids in the Massalia and Vesta families with no limit on $H$. Diameters, $D$ have been calculated assuming albedos of 0.210 (Massalia) and 0.423 (Vesta). The V-shaped lines for Massalia correspond to the fit determined by \citet{Milani2014}. Panels B and D are the corresponding number density plots with linear spacing of the contour plots. In panels C and D, we have drawn in two sets of V-shaped lines. The lower set corresponds to an age, $t_\mathrm{Vesta}=1.3\times10^9$ $yr$, the upper set corresponds to an age half as short, $t_\mathrm{Vesta}=6.2\times10^8$ $yr$. }
    \label{fig:24}
\end{figure*}

\begin{figure}
    \centering
	\includegraphics[width=0.8\columnwidth]{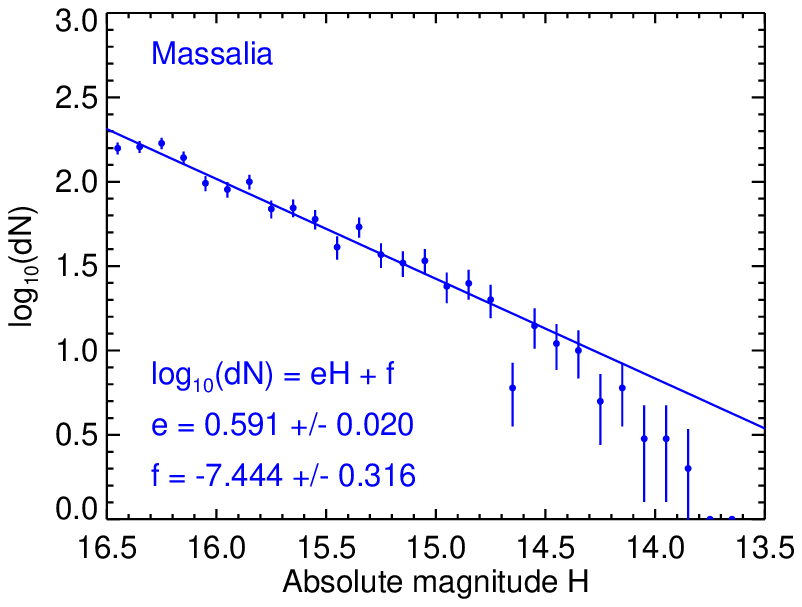}
    \caption{A linear fit to the SFD of the Massalia family. The data is shown binned in $H$, but the slope has been determined from the individual points in the range $16.5<H<13.5$. }
    \label{fig:25}
\end{figure}

Having placed constraints on the timescale of Yarkovsky-driven orbital evolution and on the timescales of the other asteroid loss mechanisms, we now analyze the effects of these same forces on the SFD of an asteroid family. If we assume that the SFD of the small asteroids in a given family was initially linear on a log-log scale, then modeling leads to a determination of the initial slope, $b$ of the SFD and allows us to place constraints on the age of the family. This analysis is not useful if a family is a conglomerate of several families with different ages. This is the case for the Nysa-Polana-Eulalia family complex and probably the case for the Flora family \citep{Brouwer1951,Tedesco1979,Zappala.et.al1990}. This analysis is also not useful if a family is young and the Yarkovsky forces have not had sufficient time to change the SFD, but it is useful if the family is not young and has a single, dominant source. 

Fig. \ref{fig:6} shows that the SFD of the small asteroids ($H>14$) in the Vesta family is well-defined and quite distinct from the SFDs of other IMB major families: the Nysa-Polana-Eulalia complex and the Flora family, and also from the SFDs of the high-inclination non-family asteroids and the asteroids in the Mars-crossing zone. The Vesta family as defined by \citet{Nesvorny.et.al2015} has 9,631 members with $H<16.5$ (Table \ref{tab:1}). Given that in the IMB the ratio of family+halo asteroids to family members alone is 1.7 (Table \ref{tab:2}), the Vesta members listed by \citet{Nesvorny.et.al2015} are probably a fraction $\lesssim0.6$ of the complete family. This leaves open the possibility that the halo asteroids that are more displaced from the family centre in $a-e-I$ space may have an SFD that is different from the core SFD. To explore that possibility, in Fig. \ref{fig:22} we split the asteroids in the Nesvorn\'y family into groups and compare the SFDs of (a) the low-inclination and the high-inclination family members, (b) the low-eccentricity and the high-eccentricity family members, and (c) the family asteroids with low-semimajor axis (asteroids moving, on average, towards the Sun) and those with high-semimajor axis (asteroids moving, on average, away from the Sun). We observe that all groups, regardless of their orbital elements, have SFDs that are close to indistinguishable from the SFD of the family as a whole.

The Vesta family was probably formed by the impact that created the giant $\sim$500 $km$ diameter and $\sim$20 $km$ deep Rheasilvia basin that overlies and partially obscures the smaller ($\sim$400 $km$ diameter) and older Veneneia crater \citep{Marchi2012,Burbine2017}. To constrain the age of the Vesta family from the observed shape of the SFD of the small asteroids ($H>14$), we assume that the impact that created the Rheasilvia basin was the dominant event that created the family and we introduce the age of the basin, $t_\mathrm{Vesta}$ as a model parameter. We keep the initial slope of the SFD, $b$ as a free parameter, but set $\alpha=1$, $\beta=2$ and $T_\mathrm{Y}$, $T_\mathrm{L}$ to the best-fitting values derived from the models for the high-inclination ($I>9$ $deg$), non-family asteroids shown in Fig. \ref{fig:8}g. The best-fitting model for the Vesta family has $b=0.65$ and $t_\mathrm{Vesta}/T_\mathrm{Y}=0.33\pm0.015$ (Fig. \ref{fig:23} and Table \ref{tab:5}). Using the value of $T_\mathrm{Y}=4$ $Gyr$ derived from the NEA observations, we estimate that $t_\mathrm{Vesta}=1.3\pm0.1$ $Gyr$. This age is consistent with the age, $\sim$1 $Gyr$, obtained from Vesta surface crater counts \citep{Marchi2012}. However, we argue here that our age estimate is likely to be a significant underestimate. The NEA observations do not allow for collisional evolution or for changes in spin directions and therefore these observations are not applicable to the small asteroids in a family that is old.

We do not have a direct measurement or a theoretical prediction of the size distribution of the rocks in the interior of a rubble-pile asteroid or of those rocks produced by a cratering event. In addition, we must allow that the size distribution of a family of rubble-pile asteroids that may have been destroyed and reaccumulated many times since their initial formation may differ from the initial SFD acquired on formation. \citet{SanchezScheeres2014} adopt in their models of rubble-pile interiors, partly on the basis of the Hayabusa spacecraft observations of the distribution of the rocks on surface of the asteroid Itokawa \citep{Mazrouei2014,Michikami2008}, a cumulative size distribution, $N=Cd^{-p}$ with $p=5b=3$. A distribution with $b=0.6$ corresponds to an SFD that has equal mass in each size interval, $dD$. In Fig. \ref{fig:24}a, we show the distribution in $a-1/D(km)$ space of the asteroids in the Massalia family. Massalia has an albedo, $A=0.210$ \citep{Tedesco.et.al2004} and an observational completeness limit of $H=16.5$ corresponds to a diameter, $D=1.45$ $km$ or $1/D=0.7$ $km^{-1}$. Therefore, the asteroids in Fig. \ref{fig:24}a are observationally incomplete. Nevertheless, the figure has several features of interest. The analysis of \citet{Milani2014} shows that the 1 $km$ diameter asteroids have a spread in semimajor axes, $\Delta a=\pm0.063$ $au$ and we have used that result to plot the V-shaped lines shown in Fig. \ref{fig:24}a. The density of 20 Massalia is 3,540 $kg/m^3$, but we allow for some porosity in the family members and assume a density of $\sim$3,000 $kg/m^3$. Then, using the NEA observations, we calculate that for the Massalia family asteroids, $T_\mathrm{Y}\sim$6 $Gyr$ and the age of the family is $\sim1.6\times10^8$ $yr$. Thus, the family is young and in calculating this age it may be appropriate not to make any allowance for changes in the spin directions or for the collisional evolution of the asteroids. \citet{Hanus2018} have shown that the larger asteroids in the Eos family, those with diameters $\gtrsim7-10$ $km$, have spin directions that are probably largely unchanged since the time of family formation. \citet{Nesvorny.et.al2015} and \citet{BrozMorbidelli2013} estimate that the age of the Eos family is between 1.5 and 1.9 $Gyr$ which is about 10$\times$ older than the Massalia family and, on that basis, we would expect the young Massalia family members with $D\gtrsim2$ $km$ to have largely unchanged spin directions. This is partly confirmed by the distinctive distribution of asteroids shown in Fig. \ref{fig:24}a that is clearly V-shaped and has a central depletion of asteroids. Fig. \ref{fig:25} shows that this young family, which may have experienced minimal collisional and spin direction evolution, has a linear log-log SFD with a slope consistent with $b=0.6$ and close to the initial slope that we have deduced for the Vesta family. 

Both the Massalia family and the Vesta family were created by cratering events \citep{Milani2019}. However, the present distribution in $a-1/D(km)$ space of the asteroids in the Vesta family shown in Fig. \ref{fig:24}c is quite different from that of the young Massalia family. The central depletion in the Massalia distribution decreases as $1/D$ increases and is absent for $1/D\gtrsim1.2$ $km^{-1}$. In the Vesta family, the central depletion is absent at all diameters and there is only weak evidence of a V-shaped distribution. The Vesta distribution suggests that most of the Vesta family asteroids have experienced collisional evolution and significant changes in spin directions. It should be noted that changes in spin direction can occur without asteroid disruption, but asteroid disruption always results in changes in the spin directions. Consider an asteroid evolving towards the Sun. If that asteroid is disrupted, then immediately after disruption the collision products will be distributed on a near-vertical line in $a-1/D(km)$ space. However, these collision products will not all have the same spin direction as that of the precursor asteroid. Thus, these collision products will then evolve both towards and away from the Sun, suggesting that the spread of small asteroids in old families should resemble a random walk. We do not have observations of the spin directions of the small asteroids to confirm this suggestion, but we do observe in Figs. \ref{fig:22}c and \ref{fig:22}f that the SFDs of the family asteroids with $a<a_{\mathrm{Vesta}}$ and those with those with $a>a_{\mathrm{Vesta}}$ are indistinguishable. We also note that the V-shaped distribution of the Massalia family asteroids in $a-1/D(km)$ space implies, as seen in Fig. \ref{fig:24}b, that $<1/D(km)>$ increases with increasing distance from the family centre. In contrast, the distribution of $<1/D(km)>$ of the Vesta family shown in Fig. \ref{fig:24}d has a strong central concentration, implying that over the full semimajor axis range of the family the asteroids could be evolving both towards and away from the Sun without regard to their distances from the family centre. The Vesta family shown in Fig. \ref{fig:24} is truncated on the left by overlap with members of the Flora family and on the right by the 3:1 Jovian resonance. For reference, in Figs. \ref{fig:24}c and \ref{fig:24}d we have drawn in two sets of V-shaped lines. The lower set corresponds to an age, $t_\mathrm{Vesta}=1.3\times10^9$ $yr$, the upper set corresponds to an age half as short, $t_\mathrm{Vesta}=6.2\times10^8$ $yr$ If we ignore the size and spin direction evolution of the Vesta asteroids, then we estimate from the shape of the SFD that the age of this family is $>1.3\pm0.1$ $Gyr$. However, given that it is likely that both the size and the spin directions of the small asteroids have evolved, it follows that this age is an underestimate. We note that for both the high-inclination, non-family asteroids and the Vesta family, $t_\mathrm{evol}/T_\mathrm{Y}\simeq1/3$, consistent with our argument that both groups of asteroids have had similar evolutionary histories and that both have ages comparable to the age of the solar system.

\section{Conclusions}

Using a data set that is devoid of observational selection effects and the biases introduced by the major asteroid families and their halos, we have shown that there are statistically significant correlations between the sizes, inclinations and eccentricities of IMB asteroids. We have also shown that for asteroids in the diameter range $\sim$1 $km$ to $\sim$10 $km$, these correlations are determined by the action of Yarkovsky forces driving small asteroids into the $\nu_6$ secular resonance, the Jovian 3:1 mean motion resonance, and into numerous high-order Martian mean motion resonances. Evidence for the $\nu_6$ resonance driving asteroids into the Mars-crossing zone is provided by the observed size-inclination correlation and it follows that an equal number of IMB asteroids have been driven into the Jovian 3:1 mean motion resonance. The observations that the asteroids in the Mars-crossing zone have predominantly low inclinations ($I<8$ $deg$), and that the mean size of the IMB asteroids with $e>0.18$, but not those with $e<0.18$, decreases with increasing eccentricity suggest that asteroids are also transported into the Mars-crossing zone by Yarkovsky forces driving small asteroids into numerous high-order Martian mean motion resonances. Mechanisms other than Yarkovsky-driven orbital evolution may also have a role in the loss of small asteroids from the IMB. These mechanisms include collisional and rotational disruption and the continuous loss of mass through erosion. It is likely that both collisions and YORP forces have changed the spin directions of the small asteroids resulting in a random walk of the semimajor axes driven by Yarkovsky forces. 

If the asteroids in our high-inclination, non-family data set are primordial, with sizes unchanged since the time of their formation, then $t_{\mathrm{evol}}=4.6$ $Gyr$ and $T_{\mathrm{Y}}=13.4\pm1.3$ $Gyr$, a value that is unacceptably larger than the 4 $Gyr$ timescale derived from observations of the orbital evolution of near-Earth asteroids \citep{Greenberg.et.al2020}. There are at least three explanations for this large discrepancy and all three explanations could be valid and operative. The first explanation is that the asteroids are collisionally evolved. If we ignore spin reversal and use the orbital evolution timescale derived from the NEA observations, then $T_{\mathrm{Y}}=4$ $Gyr$ and the average age of the asteroids, $t_{\mathrm{evol}}=1.4$ $Gyr$, a value very much less than the age of the solar system. The second explanation is that the small asteroids have experienced spin reversal with consequent increases in their orbital evolution timescales. We have not modeled the effects of stochastic changes in the spin directions, but the implication of these two explanations, acting together, is that the time over which, on average, the  orbits of the small asteroids have evolved is $1.4<t_{\mathrm{evol}}<4.6$ $Gyr$. The third explanation is that the asteroids have experienced gradual mass loss due to erosion and the orbital evolution timescale is underestimated because, on average, over the time of their orbital evolution, the masses of the small asteroids were greater than their present masses. 

By applying these loss mechanisms to the evolution of the SFD the Vesta asteroid family, we estimate that the age of this family is $>1.3\pm0.1$ $Gyr$ and is likely to be comparable with the age of the solar system. Previous methods of dating the age of the Vesta family have used chronologies based on the counting of craters. These methods rely on the assumption that the impact rates on the surfaces of bodies in the inner solar system are time invariant. Our models of the evolution of an asteroid family indicates that this assumption may not be justified. It is clear from the observed SFDs of the major families in the IMB that these SFDs now have a peak, implying that the SFD of a given family, particularly with respect to the small family members, is strongly time dependent. If the initial SFD of a major family was linear on a log-log scale, then the number of asteroids in that family would have been determined by the number of the smallest members, but that number would have decayed on timescale $\lesssim10^8$ $yr$ and much less than the age of the solar system. If the total number of asteroids in the main belt has been determined by the destruction of a small number of large asteroids, then it likely that the impact rates on the surfaces of the asteroids has varied with time and cannot be used as standard clocks. This argument is similar to the argument that has been pursued with respect to the origin of the major solar system dust bands \citep{DurdaDermott1997,Dermott2002} and is consistent with the observations and conclusions of \citet{Heck2017} that the meteorite flux has varied over geological time as asteroid disruptions create new fragment populations that then slowly fade away due to collisional and dynamical evolution.

The smallest and most abundant asteroids in our data set have $H<16.5$ and, depending on the assumed asteroid albedo, their diameters range from 1.4 to 2.5 $km$. That these small asteroids are unlikely to be primordial is supported by the observation that the distributions of the eccentricities and the inclinations of the CC and NC asteroids in our high-inclination, non-family data set are significantly different. We conclude that it is most likely that these asteroids are members of ghost families originating from the catastrophic disruption of a small number of large asteroids. This conclusion is further supported by the non-random distribution of the CC and NC asteroids in $e-I$ space. However, we note that while the current asteroids in the IMB are probably derived from a small number of large asteroids, it does not follow that all asteroids were formed big. Any asteroids that were formed small would have been lost shortly after their transfer to the main belt and these asteroids may have left no trace of their existence (Milani, A. 2018, private discussion). 

Our analysis of the dynamical evolution of the IMB could shed some further light on the origins of meteorites and NEAs. Given that the IMB is the major source of both chondritic meteorites and NEAs, a key question is how many asteroids were originally in the IMB? Inspection of the groupings in $e-I$ space of CC and NC asteroids shown in Fig. \ref{fig:17} suggest that originally there were at least 3 NC and 3 CC asteroids with $I>9$ $deg$. Given that the area in $a-I$ space for asteroids with inclinations in the range $0<I<17.8$ $deg$ is a factor of 2.3 greater than that available to asteroids in the range $9<I<17.8$ $deg$, and given that, in addition, the major families in the IMB originate from 6 asteroids, we estimate the initial number of asteroids in the IMB that are the root sources of both the meteorites and the NEAs that originate from the IMB, was $\sim$20.  

\section*{Acknowledgements}
Astronomical Research at the Armagh Observatory and Planetarium is grant-aided by the Northern Ireland Department for Communities (DfC). CM thanks the UK Science and Technology Facilities Council (Grant No. ST/M001202/1) for financial assistance. This publication uses data products from NEOWISE, a project of the Jet Propulsion Laboratory/California Institute of Technology, funded by the Planetary Science Division of the NASA. We made use of the NASA/IPAC Infrared Science Archive, which is operated by the Jet Propulsion Laboratory, California Institute of Technology, under contract with the NASA.

\section*{Data availability}
The data underlying this article will be shared on reasonable request to the corresponding author. The datasets were derived from sources in the public domain: 
https://newton.spacedys.com/$\sim$astdys2/propsynth/all.syn (AstDys - Asteroid Proper Elements) and https://pds.nasa.gov/ds-view/pds/viewDataset.jsp?dsid=EAR-A-VARGBDET-5-NESVORNYFAM-V3.0 (Nesvorn\'{y} HCM Asteroid Families V3.0). The {\it orbit9} code used for the simulations reported in Appendix A is also available in the public domain at http://adams.dm.unipi.it/$\sim$orbmaint/orbfit/ (The OrbFit Software Package).
%%%%%%%%%%%%%%%%%%%%%%%%%%%%%%%%%%%%%%%%%%%%%%%%%%


\bibliographystyle{mnras}
\bibliography{myBibtex} % if your bibtex file is called example.bib

\begin{thebibliography}{}
\makeatletter
\relax
\def\mn@urlcharsother{\let\do\@makeother \do\$\do\&\do\#\do\^\do\_\do\%\do\~}
\def\mn@doi{\begingroup\mn@urlcharsother \@ifnextchar [ {\mn@doi@}
  {\mn@doi@[]}}
\def\mn@doi@[#1]#2{\def\@tempa{#1}\ifx\@tempa\@empty \href
  {http://dx.doi.org/#2} {doi:#2}\else \href {http://dx.doi.org/#2} {#1}\fi
  \endgroup}
\def\mn@eprint#1#2{\mn@eprint@#1:#2::\@nil}
\def\mn@eprint@arXiv#1{\href {http://arxiv.org/abs/#1} {{\tt arXiv:#1}}}
\def\mn@eprint@dblp#1{\href {http://dblp.uni-trier.de/rec/bibtex/#1.xml}
  {dblp:#1}}
\def\mn@eprint@#1:#2:#3:#4\@nil{\def\@tempa {#1}\def\@tempb {#2}\def\@tempc
  {#3}\ifx \@tempc \@empty \let \@tempc \@tempb \let \@tempb \@tempa \fi \ifx
  \@tempb \@empty \def\@tempb {arXiv}\fi \@ifundefined
  {mn@eprint@\@tempb}{\@tempb:\@tempc}{\expandafter \expandafter \csname
  mn@eprint@\@tempb\endcsname \expandafter{\@tempc}}}

\bibitem[\protect\citeauthoryear{{Bolin}, {Morbidelli}  \& {Walsh}}{{Bolin}
  et~al.}{2018}]{bolin2018a}
{Bolin} B.~T.,  {Morbidelli} A.,   {Walsh} K.~J.,  2018, \mn@doi [\aap]
  {10.1051/0004-6361/201732079}, \href
  {https://ui.adsabs.harvard.edu/abs/2018A&A...611A..82B} {611, A82}

\bibitem[\protect\citeauthoryear{{Bottke}, {Vokrouhlick{\'y}}, {Rubincam}  \&
  {Broz}}{{Bottke} et~al.}{2002}]{Bottke.et.al2002a}
{Bottke} W.~F. J.,  {Vokrouhlick{\'y}} D.,  {Rubincam} D.~P.,   {Broz} M.,
  2002, {The Effect of Yarkovsky Thermal Forces on the Dynamical Evolution of
  Asteroids and Meteoroids}.
The University of Arizona Press, pp 395--408

\bibitem[\protect\citeauthoryear{{Bottke} et~al.,}{{Bottke}
  et~al.}{2015}]{Bottke2015}
{Bottke} W.~F.,  et~al., 2015, \mn@doi [\icarus]
  {10.1016/j.icarus.2014.09.046}, \href
  {https://ui.adsabs.harvard.edu/abs/2015Icar..247..191B} {247, 191}

\bibitem[\protect\citeauthoryear{{Brouwer}}{{Brouwer}}{1951}]{Brouwer1951}
{Brouwer} D.,  1951, \mn@doi [\aj] {10.1086/106480}, \href
  {https://ui.adsabs.harvard.edu/abs/1951AJ.....56....9B} {56, 9}

\bibitem[\protect\citeauthoryear{{Bro{\v{z}}} \& {Morbidelli}}{{Bro{\v{z}}} \&
  {Morbidelli}}{2013}]{BrozMorbidelli2013}
{Bro{\v{z}}} M.,  {Morbidelli} A.,  2013, \mn@doi [\icarus]
  {10.1016/j.icarus.2013.02.002}, \href
  {https://ui.adsabs.harvard.edu/abs/2013Icar..223..844B} {223, 844}

\bibitem[\protect\citeauthoryear{Burbine, DeMeo, Rivkin  \& Reddy}{Burbine
  et~al.}{2017}]{Burbine2017}
Burbine T.~H.,  DeMeo F.~E.,  Rivkin A.~S.,   Reddy V.,  2017, Evidence for
  Differentiation among Asteroid Families.
Cambridge University Press, p. 298–320, \mn@doi{10.1017/9781316339794.014}

\bibitem[\protect\citeauthoryear{{Carry}}{{Carry}}{2012}]{Carry2012}
{Carry} B.,  2012, \mn@doi [\planss] {10.1016/j.pss.2012.03.009}, \href
  {https://ui.adsabs.harvard.edu/abs/2012P&SS...73...98C} {73, 98}

\bibitem[\protect\citeauthoryear{{Davis}, {Chapman}, {Weidenschilling}  \&
  {Greenberg}}{{Davis} et~al.}{1985}]{Davis.et.al1985}
{Davis} D.~R.,  {Chapman} C.~R.,  {Weidenschilling} S.~J.,   {Greenberg} R.,
  1985, \mn@doi [\icarus] {10.1016/0019-1035(85)90170-8}, \href
  {https://ui.adsabs.harvard.edu/abs/1985Icar...62...30D} {62, 30}

\bibitem[\protect\citeauthoryear{{DeMeo} \& {Carry}}{{DeMeo} \&
  {Carry}}{2013}]{DeMeoCarry2013}
{DeMeo} F.~E.,  {Carry} B.,  2013, \mn@doi [\icarus]
  {10.1016/j.icarus.2013.06.027}, \href
  {https://ui.adsabs.harvard.edu/abs/2013Icar..226..723D} {226, 723}

\bibitem[\protect\citeauthoryear{Delbo{\textquoteright}, Walsh, Bolin,
  Avdellidou  \& Morbidelli}{Delbo{\textquoteright} et~al.}{2017}]{Delbo2017}
Delbo{\textquoteright} M.,  Walsh K.,  Bolin B.,  Avdellidou C.,   Morbidelli
  A.,  2017, \mn@doi [Science] {10.1126/science.aam6036}, 357, 1026

\bibitem[\protect\citeauthoryear{{Delbo{\textquoteright}}, {Avdellidou}  \&
  {Morbidelli}}{{Delbo{\textquoteright}} et~al.}{2019}]{Delbo.et.al2019}
{Delbo{\textquoteright}} M.,  {Avdellidou} C.,   {Morbidelli} A.,  2019,
  \mn@doi [\aap] {10.1051/0004-6361/201834745}, \href
  {https://ui.adsabs.harvard.edu/abs/2019A&A...624A..69D} {624, A69}

\bibitem[\protect\citeauthoryear{{Dermott} \& {Murray}}{{Dermott} \&
  {Murray}}{1983}]{DermottMurray1983}
{Dermott} S.~F.,  {Murray} C.~D.,  1983, \mn@doi [\nat] {10.1038/301201a0},
  \href {https://ui.adsabs.harvard.edu/abs/1983Natur.301..201D} {301, 201}

\bibitem[\protect\citeauthoryear{{Dermott}, {Malhotra}  \& {Murray}}{{Dermott}
  et~al.}{1988}]{Dermott1988}
{Dermott} S.~F.,  {Malhotra} R.,   {Murray} C.~D.,  1988, \mn@doi [\icarus]
  {10.1016/0019-1035(88)90074-7}, \href
  {https://ui.adsabs.harvard.edu/abs/1988Icar...76..295D} {76, 295}

\bibitem[\protect\citeauthoryear{{Dermott}, {Kehoe}, {Durda}, {Grogan}  \&
  {Nesvorn{\'y}}}{{Dermott} et~al.}{2002}]{Dermott2002}
{Dermott} S.~F.,  {Kehoe} T. J.~J.,  {Durda} D.~D.,  {Grogan} K.,
  {Nesvorn{\'y}} D.,  2002, in {Warmbein} B.,  ed.,  ESA Special Publication
  Vol. 500, Asteroids, Comets, and Meteors: ACM 2002. pp 319--322

\bibitem[\protect\citeauthoryear{{Dermott}, {Christou}, {Li}, {Kehoe}  \&
  {Robinson}}{{Dermott} et~al.}{2018}]{Dermott.et.al2018}
{Dermott} S.~F.,  {Christou} A.~A.,  {Li} D.,  {Kehoe} T. J.~J.,   {Robinson}
  J.~M.,  2018, \mn@doi [Nature Astronomy] {10.1038/s41550-018-0482-4}, \href
  {https://ui.adsabs.harvard.edu/abs/2018NatAs...2..549D} {2, 549}

\bibitem[\protect\citeauthoryear{{Dohnanyi}}{{Dohnanyi}}{1969}]{Dohnanyi1969}
{Dohnanyi} J.~S.,  1969, \mn@doi [\jgr] {10.1029/JB074i010p02531}, \href
  {https://ui.adsabs.harvard.edu/abs/1969JGR....74.2531D} {74, 2531}

\bibitem[\protect\citeauthoryear{{Durda} \& {Dermott}}{{Durda} \&
  {Dermott}}{1997}]{DurdaDermott1997}
{Durda} D.~D.,  {Dermott} S.~F.,  1997, \mn@doi [\icarus]
  {10.1006/icar.1997.5803}, \href
  {https://ui.adsabs.harvard.edu/abs/1997Icar..130..140D} {130, 140}

\bibitem[\protect\citeauthoryear{{Eugster}, {Herzog}, {Marti}  \&
  {Caffee}}{{Eugster} et~al.}{2006}]{Eugster2006}
{Eugster} O.,  {Herzog} G.~F.,  {Marti} K.,   {Caffee} M.~W.,  2006,
  {Irradiation Records, Cosmic-Ray Exposure Ages, and Transfer Times of
  Meteorites}.
p.~829

\bibitem[\protect\citeauthoryear{{Farinella} \& {Vokrouhlicky}}{{Farinella} \&
  {Vokrouhlicky}}{1999}]{FarinellaVokrouhlicky1999}
{Farinella} P.,  {Vokrouhlicky} D.,  1999, \mn@doi [Science]
  {10.1126/science.283.5407.1507}, \href
  {https://ui.adsabs.harvard.edu/abs/1999Sci...283.1507F} {283, 1507}

\bibitem[\protect\citeauthoryear{{Farinella}, {Froeschl{\'e}}, {Froeschl{\'e}},
  {Gonczi}, {Hahn}, {Morbidelli}  \& {Valsecchi}}{{Farinella}
  et~al.}{1994}]{Farinella.et.al1994}
{Farinella} P.,  {Froeschl{\'e}} C.,  {Froeschl{\'e}} C.,  {Gonczi} R.,  {Hahn}
  G.,  {Morbidelli} A.,   {Valsecchi} G.~B.,  1994, \mn@doi [Nature]
  {10.1038/371314a0}, \href
  {https://ui.adsabs.harvard.edu/abs/1994Natur.371..314F} {371, 314}

\bibitem[\protect\citeauthoryear{{Gladman} et~al.,}{{Gladman}
  et~al.}{1997}]{Gladman1997}
{Gladman} B.~J.,  et~al., 1997, \mn@doi [Science]
  {10.1126/science.277.5323.197}, \href
  {https://ui.adsabs.harvard.edu/abs/1997Sci...277..197G} {277, 197}

\bibitem[\protect\citeauthoryear{{Gradie} \& {Tedesco}}{{Gradie} \&
  {Tedesco}}{1982}]{gradieTedesco1982}
{Gradie} J.,  {Tedesco} E.,  1982, \mn@doi [Science]
  {10.1126/science.216.4553.1405}, \href
  {https://ui.adsabs.harvard.edu/abs/1982Sci...216.1405G} {216, 1405}

\bibitem[\protect\citeauthoryear{{Granvik}, {Morbidelli}, {Vokrouhlick{\'y}},
  {Bottke}, {Nesvorn{\'y}}  \& {Jedicke}}{{Granvik} et~al.}{2017}]{Granvik2017}
{Granvik} M.,  {Morbidelli} A.,  {Vokrouhlick{\'y}} D.,  {Bottke} W.~F.,
  {Nesvorn{\'y}} D.,   {Jedicke} R.,  2017, \mn@doi [\aap]
  {10.1051/0004-6361/201629252}, \href
  {https://ui.adsabs.harvard.edu/abs/2017A&A...598A..52G} {598, A52}

\bibitem[\protect\citeauthoryear{{Granvik} et~al.,}{{Granvik}
  et~al.}{2018}]{Granvik2018}
{Granvik} M.,  et~al., 2018, \mn@doi [\icarus] {10.1016/j.icarus.2018.04.018},
  \href {https://ui.adsabs.harvard.edu/abs/2018Icar..312..181G} {312, 181}

\bibitem[\protect\citeauthoryear{{Greenberg}, {Margot}, {Verma}, {Taylor}  \&
  {Hodge}}{{Greenberg} et~al.}{2020}]{Greenberg.et.al2020}
{Greenberg} A.~H.,  {Margot} J.-L.,  {Verma} A.~K.,  {Taylor} P.~A.,   {Hodge}
  S.~E.,  2020, \mn@doi [\aj] {10.3847/1538-3881/ab62a3}, \href
  {https://ui.adsabs.harvard.edu/abs/2020AJ....159...92G} {159, 92}

\bibitem[\protect\citeauthoryear{{Hanu{\v{s}}} et~al.,}{{Hanu{\v{s}}}
  et~al.}{2018}]{Hanus2018}
{Hanu{\v{s}}} J.,  et~al., 2018, \mn@doi [\icarus]
  {10.1016/j.icarus.2017.07.007}, \href
  {https://ui.adsabs.harvard.edu/abs/2018Icar..299...84H} {299, 84}

\bibitem[\protect\citeauthoryear{{Heck} et~al.,}{{Heck}
  et~al.}{2017}]{Heck2017}
{Heck} P.~R.,  et~al., 2017, \mn@doi [Nature Astronomy]
  {10.1038/s41550-016-0035}, \href
  {https://ui.adsabs.harvard.edu/abs/2017NatAs...1E..35H} {1, 0035}

\bibitem[\protect\citeauthoryear{{Hendler} \& {Malhotra}}{{Hendler} \&
  {Malhotra}}{2020}]{hendler.et.al2020}
{Hendler} N.~P.,  {Malhotra} R.,  2020, \mn@doi [The Planetary Science Journal]
  {10.3847/PSJ/abbe25}, \href
  {https://ui.adsabs.harvard.edu/abs/2020PSJ.....1...75H} {1, 75}

\bibitem[\protect\citeauthoryear{{Holsapple}}{{Holsapple}}{2020}]{Holsapple2020}
{Holsapple} K.~A.,  2020, arXiv e-prints, \href
  {https://ui.adsabs.harvard.edu/abs/2020arXiv201215300H} {p. arXiv:2012.15300}

\bibitem[\protect\citeauthoryear{{Holsapple}, {Giblin}, {Housen}, {Nakamura}
  \& {Ryan}}{{Holsapple} et~al.}{2002}]{holsapple2002}
{Holsapple} K.,  {Giblin} I.,  {Housen} K.,  {Nakamura} A.,   {Ryan} E.,  2002,
  {Asteroid Impacts: Laboratory Experiments and Scaling Laws}.
pp 443--462

\bibitem[\protect\citeauthoryear{{Ivezi{\'c}} et~al.,}{{Ivezi{\'c}}
  et~al.}{2002}]{Ivezic2002}
{Ivezi{\'c}} {\v{Z}}.,  et~al., 2002, \mn@doi [\aj] {10.1086/344077}, \href
  {https://ui.adsabs.harvard.edu/abs/2002AJ....124.2943I} {124, 2943}

\bibitem[\protect\citeauthoryear{{Jacobson}, {Marzari}, {Rossi}, {Scheeres}  \&
  {Davis}}{{Jacobson} et~al.}{2014}]{Jacobson.et.al2014}
{Jacobson} S.~A.,  {Marzari} F.,  {Rossi} A.,  {Scheeres} D.~J.,   {Davis}
  D.~R.,  2014, \mn@doi [\mnras] {10.1093/mnrasl/slu006}, \href
  {https://ui.adsabs.harvard.edu/abs/2014MNRAS.439L..95J} {439, L95}

\bibitem[\protect\citeauthoryear{{Kne\v{z}evi\'{c}} \&
  {Milani}}{{Kne\v{z}evi\'{c}} \& {Milani}}{2007}]{KnezevicMilani2007}
{Kne\v{z}evi\'{c}} Z.,  {Milani} A.,  2007, {Synthetic proper elements of
  Jupiter Trojans computed numerically (superseded in May 2012 by a new version
  containing 4030 entries)}, \url
  {http://hamilton.dm.unipi.it/~astdys2/propsynth/tro.syn}

\bibitem[\protect\citeauthoryear{{Kruijer}, {Burkhardt}, {Budde}  \&
  {Kleine}}{{Kruijer} et~al.}{2017}]{Kruijer.et.al2017}
{Kruijer} T.~S.,  {Burkhardt} C.,  {Budde} G.,   {Kleine} T.,  2017, \mn@doi
  [Proceedings of the National Academy of Science] {10.1073/pnas.1704461114},
  \href {https://ui.adsabs.harvard.edu/abs/2017PNAS..114.6712K} {114, 6712}

\bibitem[\protect\citeauthoryear{{Marchi} et~al.,}{{Marchi}
  et~al.}{2012}]{Marchi2012}
{Marchi} S.,  et~al., 2012, \mn@doi [Science] {10.1126/science.1218757}, \href
  {https://ui.adsabs.harvard.edu/abs/2012Sci...336..690M} {336, 690}

\bibitem[\protect\citeauthoryear{{Masiero}, {Grav}, {Mainzer}, {Nugent},
  {Bauer}, {Stevenson}  \& {Sonnett}}{{Masiero}
  et~al.}{2014}]{Masiero.et.al2014}
{Masiero} J.~R.,  {Grav} T.,  {Mainzer} A.~K.,  {Nugent} C.~R.,  {Bauer} J.~M.,
   {Stevenson} R.,   {Sonnett} S.,  2014, \mn@doi [\apj]
  {10.1088/0004-637X/791/2/121}, \href
  {https://ui.adsabs.harvard.edu/abs/2014ApJ...791..121M} {791, 121}

\bibitem[\protect\citeauthoryear{{Mazrouei}, {Daly}, {Barnouin}, {Ernst}  \&
  {DeSouza}}{{Mazrouei} et~al.}{2014}]{Mazrouei2014}
{Mazrouei} S.,  {Daly} M.~G.,  {Barnouin} O.~S.,  {Ernst} C.~M.,   {DeSouza}
  I.,  2014, \mn@doi [\icarus] {10.1016/j.icarus.2013.11.010}, \href
  {https://ui.adsabs.harvard.edu/abs/2014Icar..229..181M} {229, 181}

\bibitem[\protect\citeauthoryear{{Michikami} et~al.,}{{Michikami}
  et~al.}{2008}]{Michikami2008}
{Michikami} T.,  et~al., 2008, \mn@doi [Earth, Planets, and Space]
  {10.1186/BF03352757}, \href
  {https://ui.adsabs.harvard.edu/abs/2008EP&S...60...13M} {60, 13}

\bibitem[\protect\citeauthoryear{{Migliorini}, {Michel}, {Morbidelli},
  {Nesvorny}  \& {Zappala}}{{Migliorini} et~al.}{1998}]{Migliorini.et.al1998}
{Migliorini} F.,  {Michel} P.,  {Morbidelli} A.,  {Nesvorny} D.,   {Zappala}
  V.,  1998, \mn@doi [Science] {10.1126/science.281.5385.2022}, \href
  {https://ui.adsabs.harvard.edu/abs/1998Sci...281.2022M} {281, 2022}

\bibitem[\protect\citeauthoryear{{Milani} \& {Nobili}}{{Milani} \&
  {Nobili}}{1988}]{MilaniNobili1988}
{Milani} A.,  {Nobili} A.~M.,  1988, \mn@doi [Celestial Mechanics]
  {10.1007/BF01234550}, \href
  {https://ui.adsabs.harvard.edu/abs/1988CeMec..43....1M} {43, 1}

\bibitem[\protect\citeauthoryear{{Milani}, {Cellino}, {Kne{\v{z}}evi{\'c}},
  {Novakovi{\'c}}, {Spoto}  \& {Paolicchi}}{{Milani} et~al.}{2014}]{Milani2014}
{Milani} A.,  {Cellino} A.,  {Kne{\v{z}}evi{\'c}} Z.,  {Novakovi{\'c}} B.,
  {Spoto} F.,   {Paolicchi} P.,  2014, \mn@doi [\icarus]
  {10.1016/j.icarus.2014.05.039}, \href
  {https://ui.adsabs.harvard.edu/abs/2014Icar..239...46M} {239, 46}

\bibitem[\protect\citeauthoryear{{Milani}, {Kne{\v{z}}evi{\'c}}, {Spoto}  \&
  {Paolicchi}}{{Milani} et~al.}{2019}]{Milani2019}
{Milani} A.,  {Kne{\v{z}}evi{\'c}} Z.,  {Spoto} F.,   {Paolicchi} P.,  2019,
  \mn@doi [\aap] {10.1051/0004-6361/201834056}, \href
  {https://ui.adsabs.harvard.edu/abs/2019A&A...622A..47M} {622, A47}

\bibitem[\protect\citeauthoryear{{Minton} \& {Malhotra}}{{Minton} \&
  {Malhotra}}{2010}]{MintonMalhotra2010}
{Minton} D.~A.,  {Malhotra} R.,  2010, \mn@doi [\icarus]
  {10.1016/j.icarus.2009.12.008}, \href
  {https://ui.adsabs.harvard.edu/abs/2010Icar..207..744M} {207, 744}

\bibitem[\protect\citeauthoryear{{Morate} et~al.,}{{Morate}
  et~al.}{2019}]{Morate2019}
{Morate} D.,  et~al., 2019, \mn@doi [\aap] {10.1051/0004-6361/201935992}, \href
  {https://ui.adsabs.harvard.edu/abs/2019A&A...630A.141M} {630, A141}

\bibitem[\protect\citeauthoryear{{Morbidelli}}{{Morbidelli}}{2002}]{Morbidelli2002}
{Morbidelli} A.,  2002, {Modern celestial mechanics : aspects of solar system
  dynamics}

\bibitem[\protect\citeauthoryear{{Morbidelli} \& {Nesvorn{\'y}}}{{Morbidelli}
  \& {Nesvorn{\'y}}}{1999}]{MorbidelliNesvorny1999}
{Morbidelli} A.,  {Nesvorn{\'y}} D.,  1999, \mn@doi [\icarus]
  {10.1006/icar.1999.6097}, \href
  {https://ui.adsabs.harvard.edu/abs/1999Icar..139..295M} {139, 295}

\bibitem[\protect\citeauthoryear{{Morbidelli} \&
  {Vokrouhlick{\'y}}}{{Morbidelli} \&
  {Vokrouhlick{\'y}}}{2003}]{MorbidelliVokrouhlicky2003}
{Morbidelli} A.,  {Vokrouhlick{\'y}} D.,  2003, \mn@doi [\icarus]
  {10.1016/S0019-1035(03)00047-2}, \href
  {https://ui.adsabs.harvard.edu/abs/2003Icar..163..120M} {163, 120}

\bibitem[\protect\citeauthoryear{{Murray} \& {Dermott}}{{Murray} \&
  {Dermott}}{1999}]{MurrayDermott1999}
{Murray} C.~D.,  {Dermott} S.~F.,  1999, {Solar System Dynamics}.
Cambridge University Press, Cambridge

\bibitem[\protect\citeauthoryear{{Nesvorny}}{{Nesvorny}}{2015}]{Nesvorny2015}
{Nesvorny} D.,  2015, NASA Planetary Data System, \href
  {https://ui.adsabs.harvard.edu/abs/2015PDSS..234.....N} {pp
  EAR--A--VARGBDET--5--NESVORNYFAM--V3.0}

\bibitem[\protect\citeauthoryear{{Nesvorn{\'y}}, {Morbidelli},
  {Vokrouhlick{\'y}}, {Bottke}  \& {Bro{\v{z}}}}{{Nesvorn{\'y}}
  et~al.}{2002}]{Nesvorny.et.al2002}
{Nesvorn{\'y}} D.,  {Morbidelli} A.,  {Vokrouhlick{\'y}} D.,  {Bottke} W.~F.,
  {Bro{\v{z}}} M.,  2002, \mn@doi [\icarus] {10.1006/icar.2002.6830}, \href
  {https://ui.adsabs.harvard.edu/abs/2002Icar..157..155N} {157, 155}

\bibitem[\protect\citeauthoryear{{Nesvorn{\'y}}, {Bro{\v{z}}}  \&
  {Carruba}}{{Nesvorn{\'y}} et~al.}{2015}]{Nesvorny.et.al2015}
{Nesvorn{\'y}} D.,  {Bro{\v{z}}} M.,   {Carruba} V.,  2015, {Identification and
  Dynamical Properties of Asteroid Families}.
pp 297--321, \mn@doi{10.2458/azu_uapress_9780816532131-ch016}

\bibitem[\protect\citeauthoryear{{Parker}, {Ivezi{\'c}}, {Juri{\'c}}, {Lupton},
  {Sekora}  \& {Kowalski}}{{Parker} et~al.}{2008}]{Parker2008}
{Parker} A.,  {Ivezi{\'c}} {\v{Z}}.,  {Juri{\'c}} M.,  {Lupton} R.,  {Sekora}
  M.~D.,   {Kowalski} A.,  2008, \mn@doi [\icarus]
  {10.1016/j.icarus.2008.07.002}, \href
  {https://ui.adsabs.harvard.edu/abs/2008Icar..198..138P} {198, 138}

\bibitem[\protect\citeauthoryear{{Rubincam}}{{Rubincam}}{2000}]{Rubincam2000}
{Rubincam} D.~P.,  2000, \mn@doi [\icarus] {10.1006/icar.2000.6485}, \href
  {https://ui.adsabs.harvard.edu/abs/2000Icar..148....2R} {148, 2}

\bibitem[\protect\citeauthoryear{{S{\'a}nchez} \& {Scheeres}}{{S{\'a}nchez} \&
  {Scheeres}}{2014}]{SanchezScheeres2014}
{S{\'a}nchez} P.,  {Scheeres} D.~J.,  2014, \mn@doi [Meteoritics and Planetary
  Science] {10.1111/maps.12293}, \href
  {https://ui.adsabs.harvard.edu/abs/2014M&PS...49..788S} {49, 788}

\bibitem[\protect\citeauthoryear{{Spoto}, {Milani}  \&
  {Kne{\v{z}}evi{\'c}}}{{Spoto} et~al.}{2015}]{Spoto.et.al2015}
{Spoto} F.,  {Milani} A.,   {Kne{\v{z}}evi{\'c}} Z.,  2015, \mn@doi [\icarus]
  {10.1016/j.icarus.2015.04.041}, \href
  {https://ui.adsabs.harvard.edu/abs/2015Icar..257..275S} {257, 275}

\bibitem[\protect\citeauthoryear{{Tedesco}}{{Tedesco}}{1979}]{Tedesco1979}
{Tedesco} E.~F.,  1979, \mn@doi [\icarus] {10.1016/0019-1035(79)90030-7}, \href
  {https://ui.adsabs.harvard.edu/abs/1979Icar...40..375T} {40, 375}

\bibitem[\protect\citeauthoryear{{Tedesco}, {Noah}, {Noah}  \&
  {Price}}{{Tedesco} et~al.}{2004}]{Tedesco.et.al2004}
{Tedesco} E.~F.,  {Noah} P.~V.,  {Noah} M.,   {Price} S.~D.,  2004, NASA
  Planetary Data System, \href
  {https://ui.adsabs.harvard.edu/abs/2004PDSS...12.....T} {pp
  IRAS--A--FPA--3--RDR--IMPS--V6.0}

\bibitem[\protect\citeauthoryear{{Vokrouhlick{\'y}} \&
  {Farinella}}{{Vokrouhlick{\'y}} \&
  {Farinella}}{2000}]{VokrouhlickyFarinella2000}
{Vokrouhlick{\'y}} D.,  {Farinella} P.,  2000, \mn@doi [\nat]
  {10.1038/35036528}, \href
  {https://ui.adsabs.harvard.edu/abs/2000Natur.407..606V} {407, 606}

\bibitem[\protect\citeauthoryear{{Walsh}, {Morbidelli}, {Raymond}, {O'Brien}
  \& {Mandell}}{{Walsh} et~al.}{2011}]{Walsh.et.al2011}
{Walsh} K.~J.,  {Morbidelli} A.,  {Raymond} S.~N.,  {O'Brien} D.~P.,
  {Mandell} A.~M.,  2011, \mn@doi [\nat] {10.1038/nature10201}, \href
  {https://ui.adsabs.harvard.edu/abs/2011Natur.475..206W} {475, 206}

\bibitem[\protect\citeauthoryear{{Wisdom}}{{Wisdom}}{1985}]{Wisdom1985}
{Wisdom} J.,  1985, \mn@doi [\nat] {10.1038/315731a0}, \href
  {https://ui.adsabs.harvard.edu/abs/1985Natur.315..731W} {315, 731}

\bibitem[\protect\citeauthoryear{{Zappala}, {Cellino}, {Farinella}  \&
  {Knezevic}}{{Zappala} et~al.}{1990}]{Zappala.et.al1990}
{Zappala} V.,  {Cellino} A.,  {Farinella} P.,   {Knezevic} Z.,  1990, \mn@doi
  [\aj] {10.1086/115658}, \href
  {https://ui.adsabs.harvard.edu/abs/1990AJ....100.2030Z} {100, 2030}

\makeatother
\end{thebibliography}

%%%%%%%%%%%%%%%%% APPENDICES %%%%%%%%%%%%%%%%%%%%%

\appendix

\section{Chaotic evolution of the orbital inclinations}\label{appendixA}

\begin{figure}
    \centering  
	\includegraphics[width=1.0\columnwidth]{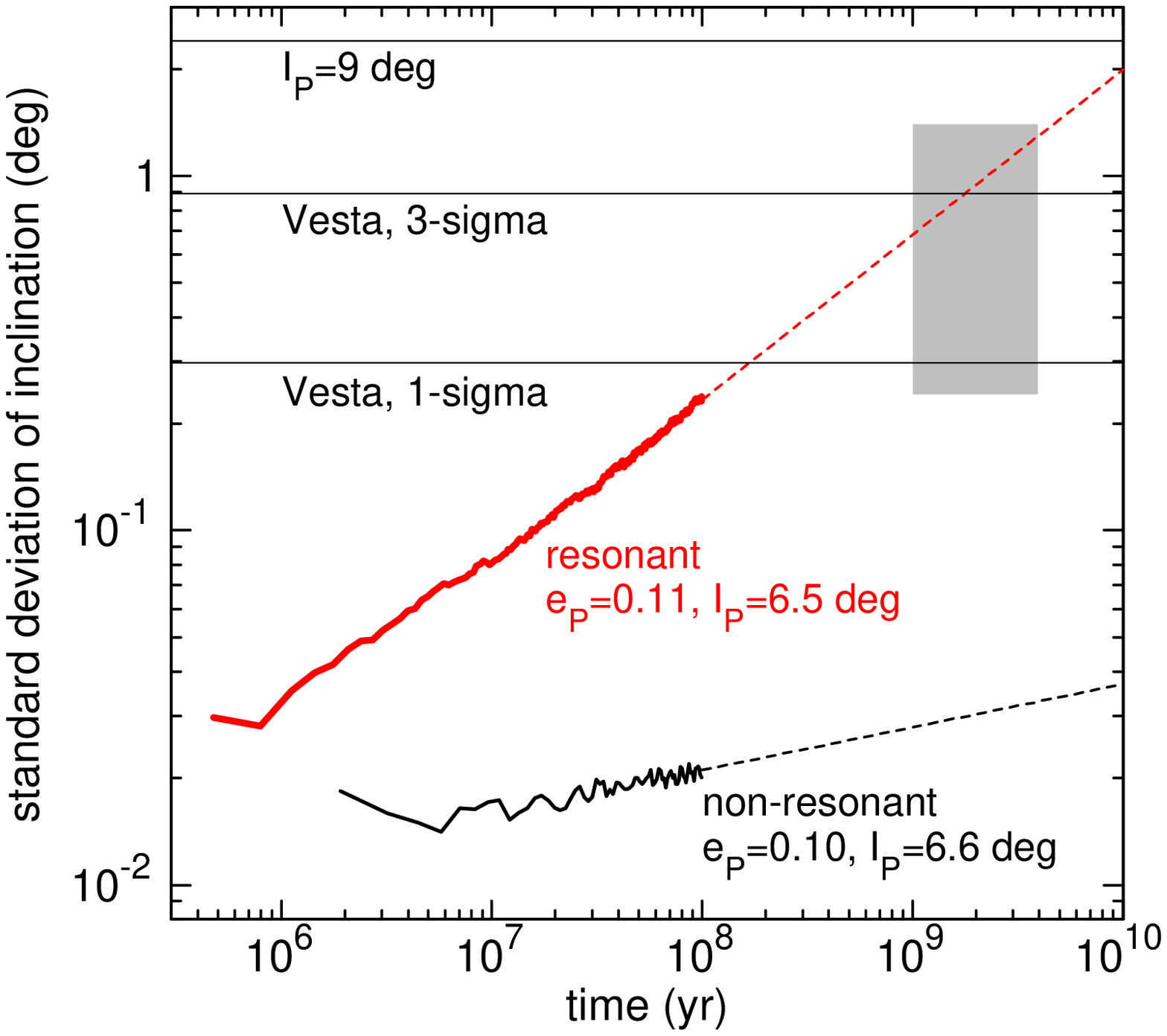}
    \caption{Evolution of the standard deviation of the proper inclination for two sets of test particles within the Vesta family, one started inside (red curve) and the other outside (black curve) the 1M-2A mean motion resonance and integrated for $10^8$ $yr$ from the present ($t=0$). The dashed lines represent fits to the data from $t=10^7$ $yr$ until $t=10^8$ $yr$. The fitted functions were of the form $\log_{10}  \sigma (t)=c+b \log_{10} (t)$. For the resonant group, $b=0.466$ and $c=-4.36$. For the non-resonant set, $b=0.121$ and $c=-2.64$. Horizontal lines represent: $1\times$ and $3\times$ multiples of the population standard deviation of asteroid orbits, calculated over all Vesta family asteroids \citep{Dermott.et.al2018}; and the nominal value of $\sigma(I)$ where a Vesta family asteroid, starting from $I_p=6.6$ $deg$ at $t=0$, will reach $I_p=9$ $deg$. The shaded box has been added to highlight the dispersion of resonant asteroids at $t=1-4\times10^9$ $yr$ ($0.7-1.3$ $deg$), compared to the 1$\sigma$ width of the Vesta family ($\sim0.3$ $deg$). }
    \label{fig:a1}
\end{figure}

The orbital elements of main belt asteroids osculate due to the gravitational pull of the planets. These short-term variations are periodic and produce no net changes in the orbits. Chaotic evolution induces slower, gradual changes in the proper semimajor axis $a_p$, eccentricity $e_p$ and inclination $I_p$ of the orbits, which can be thought of as the long-term averages of the respective osculating elements. Here, we show that asteroids with $I_p>9$ $deg$ do not originate from the Nesvorn\'y major families. Specifically, we show that asteroids from the Vesta family, that has the highest $I_p$ of all the major inner belt families, have not diffused into the $I_p>9$ $deg$ group.

We generated osculating elements for two sets of test particles with user-defined $e_p$ \& $I_p$. Set \#1 was placed just outside the 1M-2A mean motion resonance with $e_p=0.10$, $I_p=6.6$ $deg$, values that correspond to the averages of the asteroids in the Nesvorn\'y Vesta family. M and A are, respectively, the mean motions of Mars and an asteroid. Set \#2 was placed inside the resonance with $e_p=0.10$ \& $I_p=6.6$ $deg$. Each of the two sets of 400 particles were integrated for $10^8$ $yr$ from the present using the orbit9 \citep{Nesvorny.et.al2015,MilaniNobili1988}, code with a maximum step-size of 0.2 $yr$ and an output step of $10^4$ $yr$. Starting conditions for the particles and the solar system model that determines the gravitational perturbations of the particles were chosen as in our previous work (dermott2018). The inclination dispersion was evaluated from the simulation output using the time-dependent quantity
\begin{equation}
    \sigma  \big(I(t)\big)= \sqrt{ \big( \frac{1}{N-1} \big)  \sum  \big(I_{p,i}(t)-I_{p,0}(t)\big)^2  }  
    \label{eq:a1}
\end{equation}
where $N$ is the number of particles within each set and $i$ is an index running through the particles. We calculate $I_{p,i} (t)$ by averaging the integration output every 32 samples or $3.2\times10^5$ $yr$ for the resonant set and 128 samples or $1.3\times10^6$ $yr$ for the non-resonant set. 

Supplementary Fig.~\ref{fig:a1} shows $\sigma  \big(I(t)\big)$ for these two sets. The dashed lines represent extrapolations of power-law fits to the simulation data, omitting the first $10^7$ $yr$. The extrapolations predict values of 0.68 $deg$ after $10^9$ $yr$, 0.94 $deg$ after $2\times10^9$ $yr$ and 1.30 $deg$ after $4\times10^9$ $yr$ for the resonant particles. In contrast, the dispersion of non-resonant particles after $4\times10^9$ $yr$ is 0.033 $deg$.

The dispersion for resonant and non-resonant particles should bracket the growth of the standard deviation of $I_p$ for the Vesta family asteroids. We assume here that the proper inclinations $I_p (t)$ of each set of asteroids follow a gaussian distribution. Then, if the Vesta family is $\sim4$ $Gyr$, 3.2\% of family asteroids will achieve $I_p>9$ $deg$. For an age of $\sim2$ $Gyr$, 0.5\% will have $I_p>9$ $deg$ and, finally, for an age of $\sim1$ $Gyr$, only 0.02\% will have $I_p>9$ $deg$. The actual dispersion will generally be different, because the family had a finite width in $I_p$ to start with and the actual growth of the standard deviation will lie somewhere between the red and black curves. It seems likely, however, that no more than a few \% of Vesta family members have diffused to orbits with $I_p>9$ $deg$. 

To estimate the degree of contamination of high-$I_p$ asteroids by Vesta family asteroids, we consider that the 1M-2A resonance extends across $4\times10^{-3}$ $au$ or $\sim2$\% of the width of the Vesta family. Therefore, the number of family asteroids injected into high-$I_p$ orbits over 4 $Gyr$ is $0.02\times0.032\times9,667\sim6$ or $\sim0.1$\% (6/4414) of non-family $I_p>9$ $deg$. This is a lower limit because Yarkovsky radiation forces continuously inject new asteroids into the resonance, however the observed dispersion of the family (see Fig.~\ref{fig:3}) suggests that resonances alone do not drive $I_p$ evolution for Vesta family asteroids.

\section{The models}\label{appendixB}

We compute how many asteroids are lost by the two loss mechanisms over 4.6 $Gyr$ by a population of asteroids with a uniform distribution in the $a$-$I$ space at the initial time, $t_0$. To set up the model, consider a region in the $a$-$I$ space bounded by
\begin{align*}
    I_{max}&=17.8\,deg,\\
    I_{min}&=0\,deg,\\
    a_{max}&=2.487\,au,\\
    a_{min}&=\nu_6(I).
\end{align*}
Setting $a_{max}$ at 2.487 $au$ allows for the average width of the 3:1 Jovian e-type mean motion resonance.

Divide the above region into small horizontal strips of width $dI=0.1$ $deg$. Set the initial SFD of the top strip ($15.9 < I < 16.0$) to be
\begin{equation}
    \log dN=bH+c,
	\label{eq:b1}
\end{equation}
where $dN$ is the number of asteroids in a small bin, $dH=0.1$. All other strips at $I < 15.9$ $deg$, have the same initial SFDs with the same slope $b$, but the scaling factor $c$ varies so that the number density distribution is initially uniform in the $a$-$I$ space over the region defined above. In other words, the total number of asteroids in each horizontal strip (i.e., the integral of equation~\ref{eq:b1} over $12<H<20$) is proportional to the strip length $L$:
\begin{equation}
    L(au)=2.487-\nu_6(I).
	\label{eq:b2}
\end{equation}

We consider two loss mechanisms in our model. The first one is the Yarkovsky radiation force which, depending on the spin direction of the asteroid, expands or contracts the orbit. Once an asteroid reaches the $\nu_6$ secular resonance or the 3:1 Jovian mean motion resonance, it escapes from the IMB. The change in the semimajor axis, $\Delta a$, due to the Yarkovsky effect over a short period of time $\Delta t$, is described by
\begin{equation}
    \frac{1}{a} \frac{\Delta a}{\Delta t}= \frac{1}{T_\mathrm{Y}} \big( \frac{1\,km}{D} \big)^{\alpha},
	\label{eq:b3}
\end{equation}
where $D$ is the diameter of an asteroid given by
\begin{equation}
    D(km)= \frac{1329 \times 10^{-H/5}}{ \sqrt{A} },
	\label{eq:b4}
\end{equation}
and we assume an albedo, $A$ of 0.13. Consider a cell of $dI \times dH$, the number of asteroids escaping (in both directions) due to Yarkovsky forces during $\Delta t$ is
\begin{equation}
    \Delta N= \frac{N \times \Delta a}{ L },
	\label{eq:b5}
\end{equation}
where $N$ is the number of asteroids from the previous step. The second loss mechanism, that depends on $D$, but does not depend on the inclination, is described by
\begin{equation}
    \frac{1}{N(D)} \frac{\Delta N(D)}{\Delta t}= -\frac{1}{T_\mathrm{L}}\big( \frac{1\,km}{D} \big)^{\beta}.
	\label{eq:b6}
\end{equation}
To summarize, there are five free parameters in the model: 
\begin{align*}
    b, \alpha, T_\mathrm{Y}, \beta, \mathrm{and} \, T_\mathrm{L}.
\end{align*}
Note that the scaling factor $c$ in equation~\ref{eq:b1} is not a free parameter but a function of other parameters. Its value is determined by comparing the model with the data. Specifically, we adjust $c$ so that the final model SFD (for example, Fig.~\ref{fig:6}b in the main text) matches the observed SFD at $H=14$. Furthermore, notice that $\Delta N/N$ is a relative quantity in both equations~\ref{eq:b5} and \ref{eq:b6}, and therefore the evolution of $N$ does not depend on the absolute value of $c$. We can set up the model with any value of $c$ and calibrate it at the end of the simulation.

To find out how many asteroids are lost over 4.6 $Gyr$, we compute $\Delta N$ in a series of small steps ($\Delta t=0.1$ $Gyr$). In each step, we compute the Yarkovsky loss first, the result from which is then used as the input for computing the non-Yarkovsky loss. We find that using smaller steps ($\Delta t<0.1$ $Gyr$) or changing the order of computation in each step has no impact on the results. 

To quantify the goodness of fit, we compare the model with data in three different plots:

SFD: for example, Fig.~\ref{fig:8}b

$<I>$ vs. $H$: for example, Fig.~\ref{fig:8}c

$<H>$ vs. $I$: for example, Fig.~\ref{fig:8}d\\
and get three measures of $\chi_\nu^2$. However, to find the best fit model through a global $\chi_\nu^2$ minimization, one needs to weigh three measures of $\chi_\nu^2$, a process that is unavoidably arbitrary to some degree. Instead, we search for the best-fitting parameters by inspecting each plot separately.

% Don't change these lines
\bsp	% typesetting comment
\label{lastpage}
\end{document}